\documentclass[aps,floats]{revtex4-2}
\usepackage{amsmath,amssymb}
\usepackage{graphicx,epsfig}
\usepackage[greek,english]{babel}
\usepackage{bbold}

\usepackage{booktabs}
\usepackage{multirow}
\usepackage{floatrow}
\usepackage{mathtools}
\usepackage{makecell,tabularx}
\setcellgapes{1pt}
\makegapedcells
\usepackage[table]{xcolor}

\makeatletter\AtBeginDocument{\let\@elt\relax}\makeatother
 
\begin{document}
\bibliographystyle {plain}

\pdfoutput=1
\def\oppropto{\mathop{\propto}} 
\def\opsimeq{\mathop{\simeq}}
\def\opoverderline{\mathop{\overline}}
\def\operarrow{\mathop{\longrightarrow}}
\def\opsim{\mathop{\sim}}

\def\opmin{\mathop{\min}} 
\def\opmax{\mathop{\max}} 
\def\oplim{\mathop{\lim}}

%%%%%%%%%%%%%%%%%%%%%%%%%%%%%%%%%%%%%%%%%%%%%%%%%%%%%%%%%%%%%%%%%%%%%%%%%%%%
\title{ Conditioned diffusion processes with an absorbing boundary condition 
\\ for finite or infinite horizon  } 

%%%%%%%%%%%%%%%%%%%%%%%%%%%%%%%%%%%%%%%%%%%%%%%%%%%%%%%%%%%%%%%%%%%%%%%%%%%%
\author{C\'ecile Monthus}
\affiliation{Universit\'e Paris-Saclay, CNRS, CEA, Institut de Physique Th\'eorique, 91191 Gif-sur-Yvette, France}

\author{Alain Mazzolo}
\affiliation{Universit\'e Paris-Saclay, CEA, Service d'\'Etudes des R\'eacteurs et de Math\'ematiques Appliqu\'ees, 91191, Gif-sur-Yvette, France}

\begin{abstract}
When the unconditioned process is a diffusion living on the half-line $x \in ]-\infty,a[$ in the presence of an absorbing boundary condition at position $x=a$, we construct various conditioned processes corresponding to finite or infinite horizon. When the time horizon is finite $T<+\infty$, the conditioning consists in imposing the probability distribution $P^*(y,T ) $ to be surviving at time $T$ at the position $y \in ]-\infty,a[$, as well as the probability distribution $\gamma^*(T_a ) $ of the absorption time $T_a \in [0,T]$. When the time horizon is infinite $T=+\infty$, the conditioning consists in imposing the probability distribution $\gamma^*(T_a ) $ of the absorption time $T_a \in [0,+\infty[$, whose normalization $[1- S^*(\infty )]$ determines the conditioned probability $S^*(\infty ) \in [0,1]$ of forever-survival. This case of infinite horizon $T=+\infty$ can be thus reformulated as the conditioning of diffusion processes with respect to their first-passage-time properties at position $a$. This general framework is applied to the explicit case where the unconditioned process is the Brownian motion with uniform drift $\mu$ to generate stochastic trajectories satisfying various types of conditioning constraints. Finally, we describe the links with the dynamical large deviations at Level 2.5 and the stochastic control theory.
\end{abstract}

\maketitle

\section{ Introduction} 

\subsection{ Conditioned diffusion processes }

Diffusion processes describe the temporal evolution of a very large number of natural and artificial phenomena and have multiple applications in engineering, natural and social sciences, as well as finance. To analyze the conditioned processes that emerge when one imposes some constraints in the future, mathematicians have developed the so-called {\it{h}}-transform~\cite{refbookDoob,refbookRogers}, based on the pioneering work of Doob~\cite{refDoob}, which takes into account the desired constraint in a rigorous way. A gentle exposure of this method is given in Karlin and Taylor's book~\cite{refbookKarlin}. This technique is also exposed, from the physicist point of view, in the recent articles~\cite{refMajumdarOrland,refOrland}.
The most well-known example of conditioned process is the diffusion bridge,
where a one-dimensional diffusion process starting at position $x_0$ at the initial time $t=0$ is conditioned to end in configuration $x_f$ at the final time $t=t_f$ : 
for this bridge, the conditional probability distribution $B^*(x,t)$ to be at position $x$ at some internal time $t \in ]0,t_f[$ 
can be computed from the unconditioned propagator $P(x_2, t_2 \vert x_1, t_1)$ 
via the famous bridge formula
\begin{eqnarray}
B^*(x,t) =  \frac{P(x_f,t_f \vert x,t) P(x,t \vert x_0,0)}{P(x_f,t_f \vert x_0,0)}
\label{bridge}
\end{eqnarray}
which is normalized over the position $x$ as a consequence of the Chapman-Kolmogorov property.
The conditioned dynamics of this stochastic bridge can then be obtained from the backward dynamics of the unconditioned propagator $P(x_f,t_f \vert x,t) $ with respect to its initial variables $(x,t)$
and the forward dynamics of the unconditioned propagator $P(x,t \vert x_0,0) $ with respect to its final variables $(x,t)$.
In ecology, such bridges are standard processes for studying animal behaviors~\cite{refHorne}, while in mathematical finance they are employed as credit-risk models~\cite{refBrody}. 

More generally, depending on the physical applications, other constraints can be relevant. For example, in nuclear engineering, when the reactor is operating at the critical point, one should have a constant neutron population and a neutron flux as flat as possible (this critical regime is obtained thanks to the control rods)~\cite{refMulatier,refbookPazsit}. Among the many processes conditioned to satisfy certain constraints, let us quote the Brownian excursion, i.e. a Brownian bridge conditioned to be positive~\cite{refMajumdarExcursion,refChung}, the Brownian meander, i.e. a Brownian motion evolving under the condition that its minimum remains positive~\cite{refMajumdarMeander} and the taboo process i.e. a Brownian motion conditioned to stay in a prescribed (bounded) region~\cite{refKnight,refAlainTaboo}. For applications of such processes, we refer to the recent review~\cite{refMajumdarOrland}.
As can be already seen on the bridge formula of Eq. \ref{bridge},
the key ingredient of Doob's method is the finite-time propagator $P(x_2, t_2 \vert x_1, t_1)$
of the unconditioned process. Whenever this finite-time propagator is
known analytically, Doob's technique can be applied to construct various kinds of conditioned processes~\cite{refbookKarlin,refMajumdarOrland,refOrland,refSzavits,refBaudoin}. 
In particular, an important extension of the bridge formula of Eq. \ref{bridge} occurs when one imposes the joint probability distribution $E^*(x_f,t_f)$ of the final position $x_f$ and of the final time $t_f$
normalized over $x_f$ and over $t_f$
\begin{eqnarray}
\int_{-\infty}^{+\infty} dx_f \int_0^{+\infty} dt_f E^*(x_f,t_f) =1
\label{normaEstar}
\end{eqnarray}
while the initial position $x_0$ at the initial time $t=0$ is still fixed. 
The conditioned probability distribution $P^*(x,t) $ to be at position $x$ at time $t$ can then be reconstructed via an average of the bridge formula of Eq. \ref{bridge} over the final probability distribution $E^*(x_f,t_f)$ that one imposes 
\begin{eqnarray}
P^*(x,t)  =\int_{-\infty}^{+\infty} dx_f \int_t^{+\infty} dt_f E^*(x_f,t_f)   \frac{P(x_f,t_f \vert x,t) P(x,t \vert x_0,0)}{P(x_f,t_f \vert x_0,0)}
\label{sumoverbridge}
\end{eqnarray}
For instance, this formula has been applied
 to impose an arbitrary final distribution 
 of the final position $x_f$ at some fixed horizon $t_f=T$ \cite{refBaudoin,refMultiEnds}
or to analyze the conditioning with respect to the distribution of the
first-passage-time $t_f$ at the position $x_f=a$ \cite{refBaudoin,refMultiEnds}.
Other recent applications of Eq. \ref{sumoverbridge}
concern the conditioning of diffusion processes with killing rates \cite{us_DoobKilling}, 
and the conditioning of two diffusion processes
with respect to their first encounter properties \cite{us_FirstEncounter}.
Among the many other directions to extend the range of applicability of the Doob's method,
let us mention the discrete-time constrained random walks and L\'evy flights \cite{refGarbaczewski_Levy,bruyne_discrete}, run-and-tumble trajectories \cite{bruyne_run},
processes with resetting~\cite{refdeBruyne2022}, or non-intersecting Brownian bridges \cite{grela}.

Another recent generalization concerns the conditioning with respect to global dynamical constraints,
i.e. time-additive observables of the stochastic trajectories. In particular, the conditioning 
on the area has been studied via various methods
for Brownian processes or bridges \cite{refMazzoloJstat} and for Ornstein-Uhlenbeck bridges \cite{Alain_OU}.
The conditioning on the area and on other time-additive observables has been then analyzed 
both for the Brownian motion and for discrete-time random walks \cite{refdeBruyne2021},
while the conditioning with respect to one local time and two local times are studied in \cite{us_LocalTime} and \cite{us_TwoLocalTimes}.
This approach has been generalized recently \cite{c_microcanonical} to various types of discrete-time or continuous-time Markov processes, while the time-additive observable
can involve both the time spent in each configuration and the increments of the Markov process.
This general reformulation of the 'microcanonical conditioning', where the time-additive observable is constrained
to reach a given value after the finite time window $T$, allows one to make the link \cite{c_microcanonical} 
with the 'canonical conditioning' based on generating functions of additive observables
that has been much studied recently in the field of dynamical large deviations
of Markov processes for $T \to + \infty$ \cite{peliti,derrida-lecture,tailleur,sollich_review,lazarescu_companion,lazarescu_generic,jack_review,vivien_thesis,lecomte_chaotic,lecomte_thermo,lecomte_formalism,lecomte_glass,kristina1,kristina2,jack_ensemble,simon1,simon2,simon3,Gunter1,Gunter2,Gunter3,Gunter4,chetrite_canonical,chetrite_conditioned,chetrite_optimal,chetrite_HDR,touchette_circle,touchette_langevin,touchette_occ,touchette_occupation,derrida-conditioned,derrida-ring,bertin-conditioned,garrahan_lecture,Vivo,chemical,touchette-reflected,touchette-reflectedbis,c_lyapunov,previousquantum2.5doob,quantum2.5doob,quantum2.5dooblong,c_ruelle,lapolla,chabane}.
In these studies, as explained in detail in the two complementary papers \cite{chetrite_conditioned,chetrite_optimal}
and in the habilitation thesis \cite{chetrite_HDR},
the Doob conditioned processes correspond to the processes 
that optimize the dynamical large deviations in the presence of the imposed constraints,
showing the link with the field of stochastic control.
It should be stressed that the corresponding rate functions at Level 2.5 are explicit 
for many Markov processes, including discrete-time Markov chains
 \cite{fortelle_thesis,fortelle_chain,c_largedevdisorder,c_reset,c_inference},
continuous-time Markov jump processes
\cite{fortelle_thesis,fortelle_jump,maes_canonical,maes_onandbeyond,wynants_thesis,chetrite_formal,BFG1,BFG2,chetrite_HDR,c_ring,c_interactions,c_open,c_detailed,barato_periodic,chetrite_periodic,c_reset,c_inference,c_runandtumble,c_jumpdiff,c_skew,c_metastable,c_east,c_exclusion}
and Diffusion processes 
\cite{wynants_thesis,maes_diffusion,chetrite_formal,engel,chetrite_HDR,c_reset,c_lyapunov,c_inference,c_metastable}.

As incredible as it may seem,
the very deep connections between the field of Doob conditioning of large deviations and the field of stochastic control
are actually already present in the famous paper written in 1931 by E. Schr\"odinger \cite{Schrodinger},
as discussed in detail in the recent detailed commentary \cite{CommentSchrodinger} accompanying its english translation,
as well as in the two reviews \cite{ControlSchrodinger,MongeSchrodinger} written from the viewpoint of stochastic control and optimal transport.
The Schr\"odinger perspective that was introduced for the specific problem of the "Schr\"odinger bridge" between an arbitrary initial condition at time $t=0$ and an arbitrary final condition at time $t=T$ \cite{Schrodinger}
can be adapted to the present case of Eq. \ref{sumoverbridge} as follows.
The normalized distribution $E^*(x_f,t_f)$ of Eq. \ref{normaEstar}
is considered as the atypical empirical result measured in an experiment concerning a large number $N$ of unconditioned processes starting all at $x_0$ at time $t=0$. The goal is then to reconstruct a posteriori what is the most likely dynamics that has been able to produce this atypical result, 
via the optimization of the appropriate dynamical relative entropy.
So, this alternative Schr\"odinger construction of the conditioned process based on the notion of relative entropy
contains interesting new information with respect to the Doob construction, in particular the following two  points that will be useful in the present work :

(i) The relative entropy cost of the conditioning constraint $E^*(x_f,t_f) $ with respect to the corresponding typical result
allows one to measure how rare the conditioning event $E^*(x_f,t_f) $ is for the initial dynamics.

(ii) It becomes possible to construct the appropriate conditioned processes when the conditioning constraints are less detailed than the whole normalized joint distribution $E^*(x_f,t_f)$ of Eq. \ref{normaEstar} : one just needs to optimize the relative entropy in the presence of the remaining constraints that one imposes.

%%%%%%%%%%%%%%%%%%%%%%%%%%%%%%%%%%

\subsection{ Goals of the present work }

The conditioning of stochastic processes with respect to a random time
is also an important issue, especially for first passage times~\cite{refbookRedner,Bray,FirstPassage,Redner_2022}.
 Indeed, it is natural to try to modify a process so that it reaches a target faster, or at a given fixed time, or avoids it for a certain amount of time or even forever. However, despite the considerable amount of work mentioned before, very little is known for Brownian motion conditioned on the first passage time to level $a$, except for the pioneering work of Baudoin~\cite{refBaudoin} on the side of mathematics and the more recent work~\cite{refMultiEnds} on the side of physics.
The goal of the present paper is to revisit
 this conditioning with respect to first passage time properties
and to analyze the wealth of possibilities offered by Eq 3 for 
the conditioning of diffusion processes living on the half-line $x \in ]-\infty,a[$ in the presence 
of an absorbing boundary at position $x=a$, for a finite or infinite horizon.

More precisely, we will consider that the unconditioned process $X(t)$
satisfies the Ito Stochastic Differential Equation involving the drift $\mu(x)$, the diffusion coefficient $D(x)$,
and the Wiener process $W(t)$
\begin{eqnarray}
dX(t) = \mu(X(t)) dt + \sqrt{ 2 D(X(t)) } dW(t)
\label{ito}
\end{eqnarray}
in the region $X \in ]-\infty,a[$, while $x=a$ is an absorbing boundary.
As explained above on the examples of Eqs \ref{bridge} and \ref{sumoverbridge},
when one imposes some conditioning constraints,
one should first write
the corresponding conditioned probability distribution in the product form
\begin{eqnarray}
P^*(x,t)  =   Q(x,t) P(x,t \vert x_0,0) 
\label{conditional}
\end{eqnarray}
where $P(x,t \vert x_0,0) $ represents the unconditioned propagator,
while the remaining function $Q(x,t) $ has to be computed in terms of the precise conditioning constraints via Eq. \ref{sumoverbridge}.
One should then analyze the dynamics of $P^*(x,t) $ of Eq. \ref{conditional},
based on the forward Fokker-Planck dynamics satisfied by the unconditioned propagator
$P(x,t \vert x_0,0) $ and
on the backward Fokker-Planck dynamics satisfied the function $Q(x,t) $.
In the present setting, the conclusion of this dynamical analysis will be that
the function $Q(x,t) $ allows one to compute the conditioned drift 
\begin{eqnarray}
\mu^* (x,t) && =\mu(x) + 2 D(x) \partial_x \ln Q(x,t) 
\label{driftdoob}
\end{eqnarray}
which can be plugged into the Ito  analog to Eq. \ref{ito}
\begin{eqnarray}
dX^*(t) && = \mu^* (X^*(t),t) dt + \sqrt{ 2 D(X^*(t))} dW(t)
\label{itostar}
\end{eqnarray}
to generate stochastic trajectories of the conditioned process $X^*(t)$
with an absorbing boundary at $a$.

In summary, for each type of conditioning constraints that we will consider,
we will write the appropriate function $Q(x,t) $ of Eq. \ref{conditional}
to compute the corresponding conditioned drift $\mu^* (x,t) $ via Eq. \ref{driftdoob}.
For the Brownian motion of drift $\mu$, some examples of the conditioned drifts $\mu^* (x,t) $ that will be discussed are given in Table \ref{table1}.

\begin{table}[!h]
\begin{center}
\renewcommand{\arraystretch}{2}
%%\begin{tabular}{|p{7.5cm}||p{10.8cm}|} 
\begin{tabular}{|p{7.3cm}||p{10.4cm}|} 
\hline
\centering 
Conditioning the Brownian motion of drift $\mu$ \par
toward the distributions $\gamma^*(T_a ) $ and $P^*(y,T )$ \par
with the survival probability $S^*(T ) $ \par
at the time horizon $T$: \par 
$ \begin{aligned}
 &  
 \\
 &
\textstyle S^*(T ) = \int_{-\infty}^a dy P^*(y,T )   = 1- \int_{0}^T dT_a \gamma^*(T_a )
 \end{aligned} $
& \centering {\rm Conditioned \ drift } : 
$ \begin{aligned} 
 &  
 \\
 & \ \ \ \ \mu^* (x,t) 
=  \partial_x \ln  \bigg[
 \left( \frac{a-x}{a} \right)   
\int_t^{T} dT_a \gamma^*(T_a) 
  \left(\frac{T_a}{T_a-t}  \right)^{\frac{3}{2} }
e^{\frac{a^2}{2T_a} - \frac{(a-x)^2}{2(T_a-t)}}
\\ & 
+  e^{ \frac{a^2}{2 T}- \frac{(a-x)^2}{2(T-t)}} 
 \sqrt{ \frac{T}{T-t} } 
\int_{-\infty}^a dy P^*(y,T ) 
e^{ \frac{(a-y)^2}{2 T} - \frac{(a-y)^2}{2(T-t)}} 
\frac{ \sinh \left( \frac{(a-x) (a-y) }{T-t} \right)}
{\sinh \left( \frac{a (a-y) }{T} \right)} 
 \bigg]
 \end{aligned} $
\tabularnewline
\hline 
\centering 
  Conditioning toward absorption at $T^*$ at $a$ :\par  $ \gamma^*(T_a)=\delta(T_a-T^* ) $  
& \centering $ \mu^*(x,t) =    -\frac{1}{a-x} +  \frac{a-x}{T^*-t}  $ 
\tabularnewline  
\hline
\centering
Conditioning toward survival at $T$ at $y^*$ : \par $P^*(y,T )=\delta(y-y^*)   $  
& \centering   $\mu^*_T(x,t)    = \frac{a-x}{T-t} + \left( \frac{y^*-a}{T-t } \right) \coth \left( \frac{(a-x) (a-y^*) }{T-t} \right)$
\tabularnewline
\hline
\centering
$
\begin{aligned}
\centering
& {\rm Conditioning  \ toward  \ the  \ normalized  \ distribution \ }   \\
& \ \ \ \ \ \ \ \ \ \ \gamma^*(T_a ) {\rm \ for \  the  \ time \  horizon \ } T=\infty: \\  
 & \\
 & 
\textstyle
  \ \ \ \ \ \ \ \ \ \ \ \ \ \ \ \ \ \ \ \int_{0}^{+\infty} dT_a \gamma^*(T_a )=1  
\end{aligned} $
& 
\centering 
$ \begin{aligned} 
 &  
 \\
 &  
\mu^*_{\infty}(x,t)  =   
\frac{\int_t^{+\infty} dT_a \gamma^*(T_a)\left(\frac{T_a}{T_a-t}  \right)^{\frac{3}{2} }    
e^{\frac{a^2}{2T_a} - \frac{(a-x)^2}{2(T_a-t)}} \left[\frac{1}{x-a} +  \frac{a-x}{T_a-t}  \right]}
{\int_t^{+\infty} dT_a \gamma^*(T_a)\left(\frac{T_a}{T_a-t}  \right)^{\frac{3}{2} }   
e^{\frac{a^2}{2T_a} - \frac{(a-x)^2}{2(T_a-t)}} }  
 \\
 &
\end{aligned} $  
\tabularnewline 
\hline
\centering  Conditioning toward full survival $S^*(\infty )=1 $ \par
 at the infinite horizon $T=+\infty$ \par
 when the unconditioned drift is positive $\mu\geq 0$
& 
\centering 
$ \begin{aligned} 
 &  
 \\
 &
 \mu^*_{\infty}(x,t)  = - \frac{1}{a-x}   
\end{aligned} $
\tabularnewline  
\hline
\centering   Conditioning toward full survival $S^*(\infty )=1 $ \par
 at the infinite horizon $T=+\infty$ \par
 when the unconditioned drift is negative $\mu < 0$
 & 
\centering 
$ \begin{aligned} 
 &  
 \\
 &
 \mu^*_{\infty}(x,t)  = -\mu \coth(\mu (a-x))
\end{aligned} $
\tabularnewline 
\hline
\end{tabular}
\end{center}
\caption{Examples of conditioned drifts $\mu^* (x,t) $ for the Brownian motion of drift $\mu$
with an absorbing boundary condition at position $a$ : 
the first line contains the general formula for the finite time horizon $T$,
where the conditioned drift $\mu^* (x,t) $
is computed in terms of the absorbing distribution $\gamma^*(T_a ) $ for $T_a \in [0,T]$
and in terms the survival distribution $P^*(y,T )  $ for $y \in ]-\infty,a[$. The other lines display the applications to the simplest examples, either for finite horizon $T$ or for the limit of the infinite horizon $T=+\infty$. More details and other examples can be found in sections \ref{sec_finitehorizon} and \ref{sec_infinitehorizon}. } 
\label{table1}
\end{table}

%%%%%%%%%%%%%%%%%%%%%%%%%%%%%%%%%%

\subsection{ Organization of the paper }

The paper is organized as follows.
Section \ref{sec_general} explains the construction of the conditioned diffusion processes
for the different types of conditioning constraints that one wishes to consider. 
This general framework is then applied to the explicit case
where the unconditioned process is the Brownian motion with uniform drift $\mu$ 
starting at $x=0$ with absorbing condition at position $a>0$,
both for finite horizon $T<+\infty$ in section \ref{sec_finitehorizon}
and for infinite horizon $T=+\infty$ in section \ref{sec_infinitehorizon},
with many illustrative examples where
stochastic trajectories satisfying various types of conditioning constraints are generated.
Our conclusions are summarized in section \ref{sec_conclusion}.
Finally, the Appendices describe the links with the Schr\"odinger perspective
that involve the dynamical large deviations of the unconditioned process
and the stochastic control theory.

%%%%%%%%%%%%%%%%%%%%%%%%%%%%%%%%%%%%%%%%

\section{ Conditioned diffusion processes with absorption at position $x=a$} 

\label{sec_general}

In this section, we describe the general construction of the conditioned diffusion process $X^*(t)$
in the presence of an absorbing boundary at position $x=a$.

\subsection{ Unconditioned process : diffusion $X(t)$ on $]-\infty,a[$ with absorbing condition at position $a$ }

\label{subsec_unconditioned}

As explained in the Introduction and as can be seen on Eqs \ref{bridge} and \ref{sumoverbridge},
the essential building block of Doob's method 
is the finite-time propagator $P(x_2, t_2 \vert x_1, t_1)$
of the unconditioned process with its dynamics with respect to the final variables $(x_2,t_2)$
and with respect to the initial variables $(x_1,t_1)$.
In this subsection, we thus describe all the properties of the unconditioned process
that will be needed later to construct conditioned processes.

\subsubsection{ Forward and backward Fokker-Planck dynamics for the propagator $P(x_2,t_2 \vert x_1,t_1)$ }

The Fokker-Planck generator associated to the Ito Stochastic Differential Equation
of Eq. \ref{ito}
\begin{eqnarray}
{\cal F}_x = \mu(x) \partial_x + D(x) \partial^2_{x} 
\label{generator}
\end{eqnarray}
governs the backward dynamics of the propagator $P(x_2,t_2 \vert x_1,t_1)$
 with respect to its initial variables $(x_1,t_1)$ 
\begin{eqnarray}
-\partial_{t_1} P(x_2,t_2 \vert x_1,t_1) = {\cal F}_{x_1} P(x_2,t_2 \vert x_1,t_1) 
= \mu(x_1) \partial_{x_1} P(x_2,t_2 \vert x_1,t_1) + D(x_1) \partial^2_{x_1} P(x_2,t_2 \vert x_1,t_1)
\label{backward}
\end{eqnarray}
while the adjoint operator of the generator of Eq. \ref{generator}
\begin{eqnarray}
{\cal F}^{\dagger}_x = -  \partial_x \mu(x)+ \partial^2_{x}  D(x)
\label{adjoint}
\end{eqnarray}
governs the forward dynamics of the propagator $P(x_2,t_2 \vert x_1,t_1)$
 with respect to the its final variables $(x_2,t_2)$
\begin{eqnarray}
\partial_{t_2} P(x_2,t_2 \vert x_1,t_1) ={\cal F}^{\dagger}_{x_2}  P(x_2,t_2 \vert x_1,t_1)
= -  \partial_{x_2} \left[  \mu(x_2) P(x_2,t_2 \vert x_1,t_1) \right]+ \partial^2_{x_2} 
\left[ D(x_2) P(x_2,t_2 \vert x_1,t_1) \right]
\label{forward}
\end{eqnarray}
The absorbing boundary condition at position $a$
corresponds to the vanishing of the propagator $P(x_2,t_2 \vert x_1,t_1) $ at 
positions $x_2=a$ and $x_1=a$ at any time $t$
 \begin{eqnarray}
 P(x_2=a,t_2 \vert x_1,t_1) = 0 
 \nonumber \\
 P(x_2,t_2 \vert x_1=a,t_1) = 0 
\label{absorbing}
\end{eqnarray}
while the initial condition at coinciding times $t_2 =t_1$ reads
\begin{eqnarray}
 P(x_2,t_2=t_1 \vert x_1,t_1) = \delta(x_2-x_1)
\label{initial}
\end{eqnarray}

\subsubsection{ Survival probability $S(t_2 \vert x_1,t_1)  $ and probability distribution $\gamma(t_2 \vert x_1,t_1) $ of the absorption-time $t_2$ at position $a$ }

The total survival probability $S(t_2 \vert x_1,t_1) $ at time $t_2$ when starting at the position $x_1$ at time $t_1$
can be computed via the integration of the propagator $P(x_2,t_2 \vert x_1,t_1)  $ over all the possible positions $x_2 \in ]-\infty,a[$
\begin{eqnarray}
S(t_2 \vert x_1,t_1) = \int_{-\infty}^a dx_2 P(x_2,t_2 \vert x_1,t_1) 
\label{survival}
\end{eqnarray}
with the initial condition at coinciding times $t_2 =t_1$ inherited from Eq. \ref{initial}
\begin{eqnarray}
S(t_2=t_1 \vert x_1,t_1) = \int_{-\infty}^a dx_2  \delta(x_2-x_1) = 1 ~~ \mathrm{for~~} x_1<a
\label{survivalinitial}
\end{eqnarray}

The probability distribution $\gamma(t_2 \vert x_1,t_1)$ of the absorption-time $t_2$
can be obtained from the derivative of the survival probability of Eq. \ref{survival}
with respect to $t_2$
\begin{eqnarray}
\gamma(t_2 \vert x_1,t_1)  = - \partial_{t_2} S(t_2 \vert x_1,t_1) = - \int_{-\infty}^a dx_2 \partial_{t_2} P(x_2,t_2 \vert x_1,t_1) 
\label{gammadef}
\end{eqnarray}
Using the forward Fokker-Planck Eq. \ref{forward}
and the absorbing boundary condition of Eq. \ref{absorbing}, Eq. \ref{gammadef}
can be rewritten using integration by parts as
\begin{eqnarray}
\gamma(t_2 \vert x_1,t_1) 
&& = - \int_{-\infty}^a dx_2 \left( -  \partial_{x_2} \left[  \mu(x_2) P(x_2,t_2 \vert x_1,t_1) \right]
+ \partial^2_{x_2} \left[ D(x_2) P(x_2,t_2 \vert x_1,t_1) \right]\right)
\nonumber \\
&& =  \left[    \mu(x_2) P(x_2,t_2 \vert x_1,t_1) 
- \partial_{x_2} \left( D(x_2) P(x_2,t_2 \vert x_1,t_1) \right) \right]_{x_2=-\infty}^{x_2=a}
\nonumber \\
&& = - D(a)  \left( \partial_{x_2} P(x_2,t_2 \vert x_1,t_1) \right)\bigg\vert_{x_2=a}
\label{gammafirst}
\end{eqnarray}
where one recognizes the Fick diffusion current 
entering the absorbing boundary $x=a$.
The initial condition at $t_2=t_1$ reads using Eq. \ref{initial} for any $x_1<a$
\begin{eqnarray}
\gamma(t_2= t_1 \vert x_1,t_1) 
&& = - D(a)  \left( \delta'(x_2-x_1) \right)\bigg\vert_{x_2=a} =0 
\label{gammafirstinitial}
\end{eqnarray}

Using Eq. \ref{survivalinitial},
the normalization over the possible finite times $t_2 \in [t_1,+\infty[$ 
\begin{eqnarray}
\int_{t_1}^{+\infty} d t_2  \gamma(t_2 \vert x_1,t_1) = - \int_{t_1}^{+\infty} d t_2  \partial_{t_2} S(t_2 \vert x_1,t_1)
= - \left[ S(t_2 \vert x_1,t_1) \right]_{t_2=t_1}^{t_2=+\infty} =1 - S(\infty \vert x_1,t_1) = 1 - S(\infty \vert x_1)
\label{gammanormalization}
\end{eqnarray}
involves the probability $S(\infty \vert x_1) \in [0,1]$ to survive forever, i.e. to never touch the boundary $x=a$
when starting at position $x_1$.

Both the survival probability $S(t_2 \vert x_1,t_1) $ and the
 probability distribution $\gamma(t_2 \vert x_1,t_1) $ inherit from the propagator 
 $P(x_2,t_2 \vert x_1,t_1)  $
 the backward dynamics of Eq. \ref{backward} with respect to the initial variables $(x_1,t_1)$
\begin{eqnarray}
-\partial_{t_1} S(t_2  \vert x_1,t_1) && = {\cal F}_{x_1} S(t_2  \vert x_1,t_1)
= \mu(x_1) \partial_{x_1} S(t_2   \vert x_1,t_1) + D(x_1) \partial^2_{x_1} S(t_2  \vert x_1,t_1)
\nonumber \\
-\partial_{t_1} \gamma(t_2 \vert x_1,t_1) && = {\cal F}_{x_1} \gamma(t_2 \vert x_1,t_1)
= \mu(x_1) \partial_{x_1} \gamma(t_2 \vert x_1,t_1) + D(x_1) \partial^2_{x_1} \gamma(t_2 \vert x_1,t_1)
\label{gammabackward}
\end{eqnarray}
In particular, the forever-survival $S(\infty \vert x_1) $ satisfies
\begin{eqnarray}
0 = {\cal F}_{x_1} S(\infty \vert x_1)
= \left[ \mu(x_1) + D(x_1) \partial_{x_1}\right]\partial_{x_1} S(\infty \vert x_1) 
\label{survivalbackward}
\end{eqnarray}

\subsubsection{ Probability $P(a-\epsilon ,t_2\vert x_1,t_1) $ to be near the absorbing boundary at $x_2=a-\epsilon$ in terms of the absorption-time distribution}

For later purposes, it is also useful to evaluate 
the probability to be near the absorbing boundary at position $x_2=a-\epsilon$ 
via the Taylor expansion at first order in $\epsilon$ around $P(x_2=a ,t_2\vert x_1,t_1)=0 $ of Eq.  \ref{absorbing}
\begin{eqnarray}
P(a-\epsilon ,t_2\vert x_1,t_1) && =P(a ,t_2\vert x_1,t_1) - \epsilon  \left( \partial_{x_2} P(x_2,t_2 \vert x_1,t_1) \right)\bigg\vert_{x_2=a} + O(\epsilon^2)
\nonumber \\
&&   = \epsilon \frac{1}{ D(a) } \gamma(t_2 \vert x_1,t_1) + O(\epsilon^2)
\label{Absgamma}
\end{eqnarray}
where Eq. \ref{gammafirst} was used to rewrite
 the leading contribution at first order in $\epsilon$ in terms of the absorption-time distribution $\gamma(t_2 \vert x_1,t_1) $ of Eq. \ref{gammafirst}.

%%%%%%%%%%%%%%%%%%%%%%%%%%%%%%%%%%%%%%%%%%

\subsection{ Conditioned process $X^*(t)$ with respect to some finite horizon $T<+\infty$ }

\subsubsection{ Conditioning toward the distributions $P^*(y,T ) $ for $y \in ]-\infty,a[$ and $\gamma^*(T_a ) $  for $T_a \in [0,T]$ }

For the unconditioned diffusion process $X(t)$ starting at position $X(0)=x_0$ at time $t=0$ :

(i) the probability distribution $P(y,T \vert x_0, 0)$ to be surviving at the position $y$
is normalized over $y \in ]-\infty,a[$ to the survival probability at time $T$
\begin{eqnarray}
S(T \vert x_0,0) = \int_{-\infty}^a dy P(y,T \vert x_0, 0) 
\label{survivalT}
\end{eqnarray}

 (ii) the probability distribution $\gamma(T_a \vert x_0,0) $
 of the absorption time $T_a$ 
 is normalized over $T_a \in [0,T]$,
 to the probability to be already dead at time $T$
\begin{eqnarray}
 \int_{0}^T dT_a \gamma(T_a \vert x_0,0)   = 1- S(T \vert x_0,0) 
\label{deadT}
\end{eqnarray}
and is thus complementary to the survival probability of Eq. \ref{survivalT}.

Now we construct the conditioned diffusion process $X^*(t)$ by imposing instead 
the following other properties: 

(i) another probability distribution $P^*(y,T )$ to be surviving at position $y$ at time $T$,
whose normalization over $y \in ]-\infty,a[$ will be the conditioned survival probability 
$S^*(T )$ at time $T$
\begin{eqnarray}
\int_{-\infty}^a dy P^*(y,T )    = S^*(T )  
\label{survivalTstar}
\end{eqnarray}

 (ii) another probability distribution $\gamma^*(T_a ) $
 of the absorption time $T_a$, 
  whose normalization over $T_a \in [0,T]$ is the conditioned probability 
  to be already dead at time $T$, and is thus complementary to Eq. \ref{survivalTstar}
\begin{eqnarray}
 \int_{0}^T dT_a \gamma^*(T_a)   = 1- S^*(T ) 
\label{deadTstar}
\end{eqnarray}
 The conditioned survival probability $S^*(t)$ at any intermediate time $t \in ]0,T[$ can be computed via
\begin{eqnarray}
S^*(t ) = 1-  \int_{0}^t dT_a \gamma^*(T_a )    
\label{survivalstarinter}
\end{eqnarray}

In summary, for the time horizon $T$, we impose
the following joint distribution $E^*(x_f,t_f)$ for the final end-point $(x_f,t_f)$ for the 
stochastic trajectories 
\begin{eqnarray}
 E^*(x_f,t_f) = \gamma^*(t_f ) \theta(0 \leq t_f \leq T) \delta(x_f-a)
 + P^*(x_f,T ) \theta(-\infty \leq x_f \leq a)\delta(t_f-T)
\label{Estar}
\end{eqnarray}
where the first contribution corresponds to the trajectories ending at the absorbing boundary $x_f=a$
at times $t_f \in [0,T] $, 
while the second contribution corresponds to the trajectories ending at time $t_f=T$
at the positions  $x_f \in ]-\infty,a[$.
The normalization of Eq. \ref{normaEstar} can be checked using Eqs \ref{survivalTstar}
and \ref{deadTstar}
\begin{eqnarray}
\int_{-\infty}^{+\infty} dx_f \int_0^{+\infty} dt_f E^*(x_f,t_f) 
= \int_{0}^T dt_f \gamma^*(t_f) + \int_{-\infty}^{a} dx_f P^*(x_f,T )
=  [1- S^*(T )] + S^*(T)=1
\label{normaEstar2}
\end{eqnarray}

%%%%%%%%%%%%%%%%%%%%%%%%%%%%%%%%%%%

\subsubsection{ Conditioned probability distribution $P^*(x,t ) $ at any intermediate time $t \in [0,T]$ }

At any intermediate time $t \in [0,T]$, the conditioned probability distribution $P^*(x,t ) $ to be surviving at position $x$ can be obtained via Eq. \ref{sumoverbridge}
using the joint distribution $E^*(x_f,t_f) $ of Eq. \ref{Estar}
to obtain 
\begin{eqnarray}
P^*(x,t) && =  \int_t^{T} dT_a \gamma^*( T_a) 
 \left[\oplim_{\epsilon \to 0} \frac{P(a-\epsilon,T_a \vert x,t) P(x,t \vert x_0,0)}{P(a-\epsilon,T_a \vert x_0,0)} \right]
 \nonumber \\ &&
 +  \int_{-\infty}^a dy P^*(y,T )  \frac{ P(y,T \vert x,t) P(x,t \vert x_0,0)}{P(y,T \vert x_0,0) }
\label{conditionaldef}
\end{eqnarray}
The first contribution involving the conditioned absorbing-time distribution 
$ \gamma^*( T_a) $ contains the bridge formula of Eq. \ref{bridge} ending at 
position $(a-\epsilon) \to a$ at time $T_a \in [t,T]$, 
while the second contribution involving the conditioned 
survival probability $P^*(y,T ) $ contains the bridge formula of Eq. \ref{bridge} ending at 
position $y$ at time $T$.

The normalization of Eq. \ref{conditionaldef}
over the possible positions $x \in ]-\infty,a[$ at time $t$
can be computed using the Chapman-Kolmogorov property of the unconditioned process
and Eq. \ref{survivalstarinter}
\begin{eqnarray}
\int_{-\infty}^a dx P^*(x,t) && =  \int_t^{T} dT_a \gamma^*( T_a) 
 \left[\oplim_{\epsilon \to 0} \frac{\int_{-\infty}^a dx P(a-\epsilon,T_a \vert x,t) P(x,t \vert x_0,0)}{P(a-\epsilon,T_a \vert x_0,0)} \right]
 \nonumber \\ &&
 +  \int_{-\infty}^a dy P^*(y,T )  \frac{ \int_{-\infty}^a dx P(y,T \vert x,t) P(x,t \vert x_0,0)}{P(y,T \vert x_0,0) }
 \nonumber \\
 && =\int_t^{T} dT_a \gamma^*( T_a) +  \int_{-\infty}^a dy P^*(y,T ) 
% \nonumber \\ && 
 = \left[ S^*(t ) - S^*(T )\right] +S^*(T ) = S^*(t )
\label{conditionalnorma}
\end{eqnarray}
i.e. one obtains, as it should, the conditioned survival probability $S^*(t ) $ that one imposes.

The initial condition at $t=0$ is the same as for the initial process
 $P(x,t=0 \vert x_0,0) = \delta(x-x_0) $ of Eq. \ref{initial}
as a consequence of Eqs \ref{survivalTstar}
and \ref{deadTstar}
\begin{eqnarray}
P^*(x,t=0) && =  \int_0^{T} dT_a \gamma^*( T_a) 
 \left[\oplim_{\epsilon \to 0} \frac{P(a-\epsilon,T_a \vert x,0) P(x,0 \vert x_0,0)}{P(a-\epsilon,T_a \vert x_0,0)} \right]
 \nonumber \\ &&
 +  \int_{-\infty}^a dy P^*(y,T )  \frac{ P(y,T \vert x,0) P(x,t \vert x_0,0)}{P(y,T \vert x_0,0) }
 =\delta(x-x_0)
\label{conditionaldefinitialt}
\end{eqnarray}

In the first contribution of Eq. \ref{conditionaldef}, 
the property of Eq. \ref{Absgamma} allows one to rewrite the limit involving $\epsilon \to 0$ in terms of 
the first-passage distributions $ \gamma(T_a \vert x,t)$ and $\gamma(T_a\vert x_0,0) $
\begin{eqnarray}
 \oplim_{\epsilon \to 0}  \frac{P(a-\epsilon,T_a \vert x,t) }{P(a-\epsilon,T_a \vert x_0,0)}
= \frac{\gamma(T_a \vert x,t)}{\gamma(T_a\vert x_0,0)}
\label{ratio}
\end{eqnarray}
In summary, the conditioned probability distribution of Eq. \ref{conditionaldef} can be rewritten in the product form of Eq. \ref{conditional},
where the function 
\begin{eqnarray}
Q_T(x,t)  \equiv 
\int_t^{T} dT_a  \frac{\gamma^*(T_a)}{\gamma(T_a\vert x_0,0)}  \gamma(T_a \vert x,t)
+   \int_{-\infty}^a dy  \frac{ P^*(y,T )  }{P(y,T \vert x_0,0) } P(y,T \vert x,t)
 \label{Qdef}
\end{eqnarray}
inherits the backward Fokker-Planck dynamics of Eq. \ref{gammabackward} concerning $\gamma(T_a \vert x,t) $ 
and of Eq. \ref{backward} concerning $P(y,T \vert x,t) $ with respect to their initial variables $(x,t)$
\begin{eqnarray}
- \partial_t  Q_T(x,t) ={\cal F}_x  Q_T(x,t) =     \mu(x) \partial_{x}   Q_T(x,t)  + D(x)\partial^2_{x} Q_T(x,t)
\label{Qbackward}
\end{eqnarray}
since the derivative with respect to the time $t$ appearing as the lower boundary of the integral of the first
contribution of Eq. \ref{Qdef} gives zero as a consequence of Eq. \ref{gammafirstinitial}
\begin{eqnarray}
-  \frac{\gamma^*(t)}{\gamma(t\vert x_0,0)}  \gamma(t \vert x,t) =0
\label{Qderiborneintegrale}
\end{eqnarray}

%%%%%%%%%%%%%%%%%%%%%%%%%%%%

\subsubsection{ Dynamics of the conditioned process $X^*(t)$  }

As explained in the Introduction, once the conditioned probability distribution $P^*(x,t ) $ 
has been written in terms of the conditioned constraints,
the next goal is to analyze the corresponding dynamics.
Using the forward dynamics of Eq. \ref{forward} satisfied by the unconditioned propagator $P(x,t \vert x_0,0) $
\begin{eqnarray}
\partial_t P(x,t \vert x_0,0)  =  -  \partial_{x} \left[  \mu(x) P(x,t \vert x_0,0)  \right]+ \partial^2_{x} 
\left[ D(x) P(x,t \vert x_0,0)  \right]
\label{forwardbis}
\end{eqnarray}
and the backward dynamics of Eq. \ref{Qbackward} satisfied by $Q_T(x,t)$,
one obtains that the time derivative of the conditioned probability distribution
of Eq. \ref{conditional}
reads
\begin{eqnarray}
  \partial_t P^*(x,t)  && = P(x,t \vert x_0,0)  \left[   \partial_t Q_T(x,t) \right]  
+ Q_T(x,t) \left[  \partial_t P(x,t \vert x_0,0)  \right]
\nonumber \\
&& = P(x,t \vert x_0,0) \left[  - \mu(x) \partial_{x}   Q_T(x,t)  - D(x) \partial^2_{x} Q_T(x,t) \right]  
\nonumber \\
&& + Q_T(x,t) \left[  -  \partial_{x} \left[  \mu(x) P(x,t \vert x_0,0)  \right]
+ \partial^2_{x} \left[ D(x) P(x,t \vert x_0,0)  \right] \right]
\label{conditionalderi}
\end{eqnarray}
Using Eq. \ref{conditional} to replace the unconditioned propagator
\begin{eqnarray}
P(x,t \vert x_0,0)  =  \frac{ P^*(x,t) } {Q_T(x,t) }
\label{eliminate}
\end{eqnarray}
into Eq. \ref{conditionalderi}, one obtains that the conditioned probability distribution $P^*(x,t)$
satisfies the following forward Fokker-Planck dynamics with respect to $(x,t)$
\begin{eqnarray}
  \partial_t P^*(x,t)  && 
 = - \frac{ P^*(x,t)  } {Q_T(x,t) } \mu(x) \partial_{x}   Q_T(x,t)  -  \frac{ P^*(x,t)  } {Q_T(x,t) } D(x) \partial^2_{x} Q_T(x,t)   
 \nonumber \\ &&
 - Q_T(x,t)   \partial_{x} \left[  \mu(x)  \frac{ P^*(x,t)  } {Q_T(x,t) }  \right]
+ Q_T(x,t) \partial^2_{x} \left( D(x)  \frac{ P^*(x,t)  } {Q_T(x,t) }  \right) 
\nonumber \\
&& = -  \partial_{x} \left[  \mu_T^*(x,t) P^*(x,t)  \right]
+ \partial^2_x \left[ D(x) P^*(x,t)  \right]
\label{conditionaldyn}
\end{eqnarray}
where the diffusion coefficient $D(x)  $ is the same as in the unconditional dynamics of Eq. \ref{forward}, while the conditioned drift $\mu^*_T(x,t)$ differs from the initial drift $\mu(x)$ 
and involves the function $Q_T(x,t)$ of Eq. \ref{Qdef}
\begin{eqnarray}
\mu^*_T(x,t) && =\mu(x) + 2 D(x) \partial_x \ln Q_T(x,t) 
\nonumber \\
&& = \mu(x) +  2 D(x) \partial_x \ln \left(
\int_t^{T} dT_a  \frac{\gamma^*(T_a)}{\gamma(T_a\vert x_0,0)}  \gamma(T_a \vert x,t)
+   \int_{-\infty}^a dy  \frac{ P^*(y,T )  }{P(y,T \vert x_0,0) } P(y,T \vert x,t)
\right)\label{driftdoobfp}
\end{eqnarray}
The corresponding Ito Stochastic Differential Equation for the conditioned process $X^*(t)$ 
of Eq. \ref{itostar}
can be then used to generate stochastic trajectories of the conditioned process $X^*(t)$
with absorption at $x=a$.

%%%%%%%%%%%%%%%%%%%%%%%%%%%%

\subsubsection{ Supplementary information
that can be obtained from the Schr\"odinger perspective  }

A natural question is how different is the conditioned process $X^*(t)$
with respect to the initial unconditioned process $X(t)$. 
However this question is usually not addressed within the Doob perspective 
that we have applied in the present main text,
while it plays a major role within the Schr\"odinger perspective
that we describe in the two Appendices,
in relation of the large deviation properties of a large number $N$ of unconditioned processes :

(i) In Appendix \ref{app_lessdetailed}, the relative entropy cost 
of the conditioning constraints $\left[ P^*(.,T) ; \gamma^*(.)  \right]$
is written in Eq. \ref{RateSanovstar}
and is used
to give some precise meaning to conditioning constraints that are less detailed than the whole distributions $\left[ P^*(.,T) ; \gamma^*(.)  \right] $ at the time horizon $T$,
with various illustrative examples.

(ii) In Appendix \ref{app_largedev}, we explain  
how the Schr\"odinger perspective provides an alternative construction of the conditioned process via the optimization of its dynamical relative entropy 
in the presence of the conditioning constraints, 
which allows one to make the link with the stochastic control theory.

%%%%%%%%%%%%%%%%%%%%%%%%%%%%%%%%%%%%%%%%%%%%

\subsection{ Cases $ S^*(T )=0$ where the conditioning is toward full absorption before 
the horizon time $T$ }

When the conditioning corresponds to full absorption before 
the horizon time $T$ in Eqs \ref{survivalTstar} and \ref{deadTstar}
\begin{eqnarray}
 P^*(y,T )   && = 0 \ \ {\rm for } \ \  y \in ]-\infty,a[
 \nonumber \\
 \int_{0}^T dT_a \gamma^*(T_a )   && =1
\label{normatstar0}
\end{eqnarray}
then the function $Q_T(x,t)$ contains only the first contribution of Eq. \ref{Qdef}
involving an integral over the time $T_a \in ]t,T[$
\begin{eqnarray}
Q_T^{[time]}(x,t) = 
\int_t^{T} dT_a \gamma^*(T_a) \frac{\gamma(T_a \vert x,t)}{\gamma(T_a\vert x_0,0)}  
 \label{Qtime}
\end{eqnarray}

%%%%%%%%%%%%%%%%%%%%%%%%%%%%%%%%%%%%%%%%%%%%

\subsection{ Cases $ S^*(\infty )=0$ where the conditioning is toward full absorption before 
the infinite horizon $T=+\infty$ }

The limit of the infinite horizon $T \to +\infty$ can be directly taken on Eqs \ref{normatstar0}
and \ref{Qtime}
to obtain the following conclusion: 
when the conditioning is toward some first-passage-time distribution 
$\gamma^*(T_a )$ normalized over $T_a \in ]0,+\infty[$
\begin{eqnarray}
  \int_{0}^{+\infty} dT_a \gamma^*(T_a )    =1
\label{normatstar0infinity}
\end{eqnarray}
the function $Q_{\infty}^{[time]}(x,t) $ reads
\begin{eqnarray}
Q_{\infty}^{[time]}(x,t) = 
\int_t^{+\infty} dT_a \gamma^*(T_a) \frac{\gamma(T_a \vert x,t)}{\gamma(T_a\vert x_0,0)}  
 \label{Qtimeinfinity}
\end{eqnarray}

%%%%%%%%%%%%%%%%%%%%%%%%%%%%%%%%%%%%%%%%%%

\subsection{ Cases $ S^*(T )=1$ where the conditioning is toward full survival 
at the horizon time $T$ }

When the conditioning corresponds to full survival at the horizon time $T$  
in Eqs \ref{survivalTstar} and \ref{deadTstar}
\begin{eqnarray}
 \int_{-\infty}^a dy P^*(y,T )   && = 1
 \nonumber \\
 \gamma^*(T_a )   && =0 \ \ {\rm for } \ \  T_a \in ]0,T]
\label{normatstar1}
\end{eqnarray}
then the function $Q_T(x,t)$ contains only the second contribution of Eq. \ref{Qdef}
involving an integral over the spatial position $y \in ]-\infty,a[$
\begin{eqnarray}
Q_T^{[space]}(x,t)  \equiv 
   \int_{-\infty}^a dy P^*(y,T ) \frac{  P(y,T \vert x,t)  }{P(y,T \vert x_0,0) }
 \label{Qspace}
\end{eqnarray}

Let us now consider the important specific choice
 where the conditioned probability distribution $ P^{*[surviving]}(y,T ) $ 
is simply given by the unconditioned probability distribution $P(y,T \vert x_0,0) $
normalized by the corresponding survival probability $S (T \vert x_0,0) $
of the unconditioned process
\begin{eqnarray}
 P^{*[surviving]}(y,T )    = \frac{P(y,T \vert x_0,0)}{\int_{-\infty}^a d y' P(y',T \vert x_0,0)}
 = \frac{P(y,T \vert x_0,0)}{ S (T \vert x_0,0)}
\label{renorma}
\end{eqnarray}
This choice can be justified via the optimization of the appropriate relative entropy,
as explained around Eq. \ref{pyoptfinal}
of Appendix \ref{app_lessdetailed}.
Then the function of Eq. \ref{Qspace}
\begin{eqnarray}
Q_T^{[surviving]}(x,t) =   \int_{-\infty}^a dy P^{*[surviving]}(y,T ) \frac{  P(y,T \vert x,t)  }{P(y,T \vert x_0,0) }
=   \frac{  \int_{-\infty}^a dy P(y,T \vert x,t) }{ S (T \vert x_0,0)} = \frac{ S (T \vert x,t)  }{ S (T \vert x_0,0)}
 \label{Qspacesurvival}
\end{eqnarray}
reduces to the ratio of the two survival probabilities $S (T \vert x,t)  $ and $S (T \vert x_0,0) $ of the unconditioned process.

%%%%%%%%%%%%%%%%%%%%%%%%%%%%%%%%%%%%%%%%%%

\subsection{ Cases $ S^*(\infty )=1$ where the conditioning is toward full survival 
at the infinite horizon $T=+\infty$ }

When one considers the limit of the infinite horizon $T \to +\infty$ of Eqs \ref{normatstar1}
and \ref{Qspace},
 one first needs to choose what spatial conditioning $P^*(y,T ) $ 
one should consider during the limit procedure $T \to +\infty $.

To be concrete, we will now focus only the specific choice of Eq.\ref{renorma},
where the limit of the infinite horizon $T \to +\infty$ can be taken in Eq. \ref{Qspacesurvival}
to obtain
\begin{eqnarray}
Q_{\infty}^{[surviving]}(x,t) = \lim_{T \to +\infty} \frac{ S (T \vert x,t)  }{ S (T \vert x_0,0)}
 \label{Qspacesurvivalinfinity}
\end{eqnarray}
The rewriting in terms of the forever-survival probabilities $S(\infty \vert .)$ 
and in terms of the first-passage-time distributions $\gamma(t_2 \vert .,.) $ as
\begin{eqnarray}
Q_{\infty}^{[surviving]}(x,t) = \lim_{T \to +\infty} \frac{ S(T \vert x,t) }{S(T \vert x_0,0) }
 = \lim_{T \to +\infty} \frac{ S(\infty \vert x) +  \int_{T}^{+\infty} dt_2 \gamma(t_2 \vert x,t)  }
 {S(\infty \vert x_0) +  \int_{T}^{+\infty} dt_2 \gamma(t_2 \vert x_0,0)  }
\label{Qspacesurvivalinfty}
\end{eqnarray}
shows that the evaluation of this limit will depend on whether the forever-survival probability $S(\infty \vert .)$ 
of the unconditioned process vanishes or not :

(a) If the forever-survival probability $S(\infty \vert .)$ of the unconditioned process is finite,
the limit of Eq. \ref{Qspacesurvivalinfty} will only involve the ratio of the two 
forever-survival probabilities $S(\infty \vert x)$ and $S(\infty \vert x_0)$
\begin{eqnarray}
Q_{\infty}^{[surviving]}(x,t) 
 = \lim_{T \to +\infty} \frac{ S(\infty \vert x) +  \int_{T}^{+\infty} dt_2 \gamma(t_2 \vert x,t)  }
 {S(\infty \vert x_0) +  \int_{T}^{+\infty} dt_2 \gamma(t_2 \vert x_0,0)  }
 = \frac{ S(\infty \vert x)   }
 {S(\infty \vert x_0)   }
\label{Qspacesurvivalinftycasea}
\end{eqnarray}

(b) If the forever-survival probability of the unconditioned process vanishes $S(\infty \vert .)=0$,
 the limit of Eq. \ref{Qspacesurvivalinfty} will involve the asymptotic behavior of the
 first-passage-time distributions $\gamma(t_2 \vert x,t) $ and $\gamma(t_2 \vert x_0,0) $ of the unconditioned process
\begin{eqnarray}
Q_{\infty}^{[surviving]}(x,t) 
 = \lim_{T \to +\infty} \frac{ S(\infty \vert x) +  \int_{T}^{+\infty} dt_2 \gamma(t_2 \vert x,t)  }
 {S(\infty \vert x_0) +  \int_{T}^{+\infty} dt_2 \gamma(t_2 \vert x_0,0)  }
 = \lim_{T \to +\infty} \frac{  \int_{T}^{+\infty} dt_2 \gamma(t_2 \vert x,t)  }
 { \int_{T}^{+\infty} dt_2 \gamma(t_2 \vert x_0,0)  }
\label{Qspacesurvivalinftycaseb}
\end{eqnarray}

%%%%%%%%%%%%%%%%%%%%%%%%%%%%%%%%%%%%%%%%%%

%%%%%%%%%%%%%%%%%%%%%%%%%%%%%%%%%%%%%%%%%%

\subsection{ Cases $ S^*(\infty )\in ]0,1[$ where the conditioning is toward partial survival 
at the infinite horizon $T=+\infty$ }

Let us now consider the limit of the infinite horizon $T \to +\infty$
 for the cases with partial forever-survival $ S^*(\infty ) \in ]0,1[$.
The normalization over $T_a \in [0,+\infty[$
 of the conditioned distribution $\gamma^*(T_a ) $
 is given by Eq. \ref{survivalTstar} for $T \to +\infty$
\begin{eqnarray}
 \int_{0}^{\infty} dT_a \gamma^*(T_a)   = 1- S^*(\infty ) 
\label{deadTstarpartial}
\end{eqnarray}
For the spatial component, let us consider the choice analogous to Eq. \ref{renorma}
with the additional normalization $S^*(\infty )  $ as prefactor to respect the normalization of Eq. \ref{survivalTstar}
\begin{eqnarray}
 P^{*[partial]}(y,T ) =S^*(\infty )  P^{*[surviving]}(y,T )
 =S^*(\infty )  \frac{P(y,T \vert x_0,0)}{ S (T \vert x_0,0)}
\label{renormapartial}
\end{eqnarray}
Using Eqs \ref{Qspacesurvival} and \ref{Qspacesurvivalinfty},
the function $ Q_{T \to + \infty} (x,t) $ of Eq. \ref{Qdef} becomes
\begin{eqnarray}
Q^{*[partial]}_{\infty}(x,t) && = 
\int_t^{\infty} dT_a \gamma^*(T_a) \frac{\gamma(T_a \vert x,t)}{\gamma(T_a\vert x_0,0)}  
+  S^*(\infty ) \left[ \lim_{T \to +\infty} \frac{ S (T \vert x,t)  }{ S (T \vert x_0,0)} \right]
\nonumber \\
&& = \int_t^{\infty} dT_a \gamma^*(T_a) \frac{\gamma(T_a \vert x,t)}{\gamma(T_a\vert x_0,0)}  
+  S^*(\infty ) \left[ \lim_{T \to +\infty} 
\frac{ S(\infty \vert x) +  \int_{T}^{+\infty} dt_2 \gamma(t_2 \vert x,t)  }
 {S(\infty \vert x_0) +  \int_{T}^{+\infty} dt_2 \gamma(t_2 \vert x_0,0)  }
 \right]
 \label{Qinfinitypartial}
\end{eqnarray}
where the evaluation of the last limit will depend 
on whether the forever-survival probability $S(\infty \vert .)$ 
of the unconditioned process vanishes or not, as already discussed in Eqs \ref{Qspacesurvivalinftycasea}
and \ref{Qspacesurvivalinftycaseb}.

As a final remark,
 let us stress again that the final result of Eq. \ref{Qinfinitypartial} for $ Q^{*[partial]}_{\infty}(x,t)$
 is based on the specific choice of Eq. \ref{renormapartial},
 which can be justified via the optimization of the appropriate relative entropy,
as explained around Eq. \ref{pyoptfinal}
of Appendix \ref{app_lessdetailed}.
 However if one considers another choice for $P^*(y,T ) $, then one can return to 
 the general expression of Eq. \ref{Qdef} and analyze its asymptotic behavior for $T \to +\infty$.

%%%%%%%%%%%%%%%%%%%%%%%%%%%%%%%%%%%

\section{ Application to the Brownian motion with drift $\mu$ for finite horizon $T$}

\label{sec_finitehorizon}

In this section, the conditioning for finite horizon $T$ described in section \ref{sec_general} 
is applied to the simplest case
where the unconditioned process is the Brownian motion with uniform drift $\mu$ 
starting at $x=0$ with absorbing condition at position $a>0$.

\subsection{ Unconditioned process $X(t)$ : Brownian motion with drift $\mu$ and absorbing condition at position $a>0$ }

\label{subsec_Brown}

The unconditioned process $X(t)$ satisfies the Stochastic Differential Equation \ref{ito} with 
$\mu(x)=\mu$ and $D(x)=1/2$
\begin{eqnarray}
dX(t) = \mu dt + dB(t)
\label{itoBrown}
\end{eqnarray}
with the initial condition $X(t=0)=0$ and the absorbing condition at position $a$.
The corresponding propagator $P^{[\mu]}(x_2,t_2 \vert x_1,t_1) $ 
obtained via the method of images
\begin{eqnarray}
P^{[\mu]}(x_2,t_2 \vert x_1,t_1) = \frac{1}{\sqrt{2 \pi (t_2-t_1)}} \left( e^{- \frac{(x_2-x_1-\mu(t_2-t_1) )^2}{2(t_2-t_1)}} 
- e^{ 2 \mu (a-x_1) } e^{- \frac{(x_2+x_1-2a-\mu(t_2-t_1))^2}{2(t_2-t_1)}} \right)
\label{images}
\end{eqnarray}
allows one to compute the distribution $\gamma^{[\mu]}(t_2 \vert x_1,t_1)$ of the absorption-time $t_2$ of Eq. \ref{gammafirst}
\begin{eqnarray}
\gamma^{[\mu]}(t_2 \vert x_1,t_1) 
&& = - \frac{1 }{2}  \left( \partial_{x_2} P^{[\mu]}(x_2,t_2 \vert x_1,t_1) \right)\bigg\vert_{x_2=a}
 =  \frac{(a-x_1) }{\sqrt{2 \pi} (t_2-t_1)^{\frac{3}{2}} } e^{- \frac{(a-x_1-\mu(t_2-t_1) )^2}{2(t_2-t_1)}} 
\nonumber \\
&& =  \frac{(a-x_1) e^{ \mu (a-x_1) }  }{\sqrt{2 \pi} (t_2-t_1)^{\frac{3}{2}} } 
 e^{- \frac{(a-x_1)^2}{2(t_2-t_1)} - \frac{\mu^2 }{2} (t_2-t_1)  } 
\label{gammafirstBrown}
\end{eqnarray}
The integral
\begin{eqnarray} 
\int_0^{+\infty} \frac{d\tau }{  \tau^{\frac{3}{2}} }  e^{- \frac{(a-x_1)^2}{2\tau}-  \frac{\mu^2}{2}\tau}
 = \frac{ \sqrt{2 \pi }}{ (a-x_1)} e^{ - \vert \mu \vert (a-x_1) }
\label{integral}
\end{eqnarray}
yields 
the normalization of $\gamma^{[\mu]}(t_2 \vert x_1,t_1)  $ over the possible finite times $t_2 \in [t_1,+\infty[$ 
\begin{eqnarray}
\int_{t_1}^{+\infty} d t_2 \gamma^{[\mu]}(t_2 \vert x_1,t_1) 
 =  \frac{(a-x_1) e^{ \mu (a-x_1) }  }{\sqrt{2 \pi}  } 
 \int_0^{+\infty} \frac{d\tau }{  \tau^{\frac{3}{2}} }  e^{- \frac{(a-x_1)^2}{2\tau}-  \frac{\lambda^2}{2}\tau}
 =   e^{ (\mu - \vert \mu \vert) (a-x_1) }  
=  \left\lbrace
  \begin{array}{lll}
    1  
    &~~\mathrm{if~~} \mu \ge 0
    \\
    e^{2 \mu (a-x_1) } 
    &~~\mathrm{if~~} \mu < 0  
  \end{array}
\right.
\label{gammafirstBrowninte}
\end{eqnarray}
So one recovers the well-known property that
 the forever-survival probability $ S^{[\mu]}(\infty \vert x_1)$ of Eq. \ref{gammanormalization}
vanishes only for positive drift $\mu \geq 0$
\begin{eqnarray}
S^{[\mu]}(\infty \vert x_1) = 1 - \int_{t_1}^{+\infty} d t_2  \gamma^{[\mu]}(t_2 \vert x_1,t_1) = 
 \left\lbrace
  \begin{array}{lll}
    0  
    &~~\mathrm{if~~} \mu \ge 0
    \\
    1- e^{2 \mu (a-x_1) } = 1- e^{-2 \vert \mu \vert (a-x_1) }
    &~~\mathrm{if~~} \mu < 0  
  \end{array}
\right.
\label{survivalBrown}
\end{eqnarray}
while for negative drift $\mu<0$, the particle starting at $x_1$ can escape toward $(-\infty)$ without touching the position $a$ with the finite probability $S^{[\mu]}(\infty \vert x_1)= 1- e^{2 \mu (a-x_1) }  $
satisfying Eq. \ref{survivalbackward}.

\subsection{ Conditioned process $X^*(t)$ with respect to the finite horizon $T$ }

For the Brownian motion with drift $\mu$ starting at position $0$ at time $t=0$ :
 
 (i) the probability distribution to be surviving at position $y \in ]-\infty,a[$ at time $T$ is given by Eq \ref{images}
\begin{eqnarray}
P^{[\mu]}(y,T \vert 0, 0) && = \frac{1}{\sqrt{2 \pi T}} \left( e^{- \frac{(y-\mu T )^2}{2T}} 
- e^{ 2 \mu a } e^{- \frac{(y-2a-\mu T)^2}{2T}} \right)
\nonumber \\
&& = 
\frac{1}{\sqrt{2 \pi T}}
e^{\mu y - \frac{(y-a)^2}{2T}- \frac{a^2}{2T} - \frac{\mu^2}{2}T}
\left( e^{ \frac{ a(a-y) }{T}} - e^{- \frac{ a (a-y) }{T}}\right)
\label{imagesBrown}
\end{eqnarray}

 (ii) the probability distribution of the absorption time $T_a $ 
 is given by Eq. \ref{gammafirstBrown}
  \begin{eqnarray}
\gamma^{[\mu]}(T_a \vert 0,0) && =\frac{a }{\sqrt{2 \pi} T_a^{\frac{3}{2}} } e^{- \frac{(a-\mu T_a )^2}{2T_a}} 
= \frac{a }{\sqrt{2 \pi} T_a^{\frac{3}{2}} } e^{\mu a - \frac{a^2}{2T_a}- \frac{\mu^2}{2}T_a} 
\label{gammaBrown}
\end{eqnarray}

As explained around Eqs \ref{survivalTstar} and \ref{deadTstar},
we now impose to the conditioned process the following properties instead :

(i) another probability distribution $P^*(y,T )$ to be surviving at position $y$ at time $T$;

 (ii) another probability distribution $\gamma^*(T_a ) $ of the absorption time $T_a $ 
 for $T_a \in [0,T]$.

The normalization of Eqs \ref{survivalTstar} and \ref{deadTstar}
involves the conditioned survival probability $S^*(T ) $ at the time $T$
\begin{eqnarray}
S^*(T ) = \int_{-\infty}^a dy P^*(y,T )   =
1- \int_{0}^T dT_a \gamma^*(T_a )   
\label{normatstarBrown}
\end{eqnarray}

The ratio of the first-passage time distributions computed using Eq. \ref{gammafirstBrown},
\begin{eqnarray}
\frac{\gamma^{[\mu]}(T_a \vert x,t)}{\gamma^{[\mu]}(T_a\vert 0,0)}
&& = \left(\frac{T_a}{T_a-t}  \right)^{\frac{3}{2} } \left( \frac{a-x}{a} \right)   
e^{\frac{(a-\mu T_a )^2}{2T_a} - \frac{(a-x-\mu (T_a-t) )^2}{2(T_a-t)}}
\nonumber \\
&& = \left(\frac{T_a}{T_a-t}  \right)^{\frac{3}{2} } \left( \frac{a-x}{a} \right)   
e^{ \frac{\mu^2}{2} t - \mu x  +\frac{a^2}{2T_a} - \frac{(a-x)^2}{2(T_a-t)}}
\label{ratioexpli}
\end{eqnarray}
and the ratio of the propagators computed using Eq. \ref{images}
\begin{eqnarray}
 \frac{ P^{[\mu]}(y,T \vert x,t)  }{P^{[\mu]}(y,T \vert 0,0) } 
&& = \frac{\frac{1}{\sqrt{2 \pi (T-t)}}
e^{\mu (y-x) - \frac{(y-a)^2}{2(T-t)}- \frac{(x-a)^2}{2(T-t)} - \frac{\mu^2}{2}(T-t)}
\left( e^{ \frac{ (a-y) (a-x)}{(T-t)}} - e^{- \frac{ (a-y) (a-x)}{(T-t)}}\right)}
{\frac{1}{\sqrt{2 \pi T}}
e^{\mu y - \frac{(y-a)^2}{2 T}- \frac{a^2}{2 T} - \frac{\mu^2}{2} T}
\left( e^{ \frac{ (a-y) a }{T}} - e^{- \frac{ (a-y) a}{T}}\right)}
\nonumber \\
&& = \sqrt{ \frac{T}{T-t} } e^{\frac{\mu^2}{2} t -\mu x + \frac{(a-y)^2}{2 T} - \frac{(a-y)^2}{2(T-t)}
+\frac{a^2}{2 T}- \frac{(a-x)^2}{2(T-t)}} 
\frac{ \sinh \left( \frac{(a-x) (a-y) }{T-t} \right)}
{\sinh \left( \frac{a (a-y) }{T} \right)}
\label{ratioexpliP}
\end{eqnarray}
can be plugged 
into Eq. \ref{Qdef}
to obtain that the dependence with respect to the initial drift $\mu$ can be factorized
\begin{eqnarray}
Q^{[\mu]}_T(x,t) && =
\int_t^{T} dT_a \gamma^*(T_a) \frac{\gamma^{[\mu]}(T_a \vert x,t)}{\gamma^{[\mu]}(T_a\vert 0,0)}  
+   \int_{-\infty}^a dy P^*(y,T )  \frac{ P^{[\mu]}(y,T \vert x,t)  }{P^{[\mu]}(y,T \vert 0,0) } 
\nonumber \\
&& =e^{ \frac{\mu^2}{2} t - \mu x} Q^{[0]}_T(x,t)
 \label{QBrownT}
\end{eqnarray}
while the remaining function $Q^{[0]}_T (x,t)$ corresponding to vanishing initial drift $\mu=0$ reads
\begin{eqnarray}
Q^{[0]}_T(x,t) && =
  \left( \frac{a-x}{a} \right)   
\int_t^{T} dT_a \gamma^*(T_a) 
  \left(\frac{T_a}{T_a-t}  \right)^{\frac{3}{2} }
e^{\frac{a^2}{2T_a} - \frac{(a-x)^2}{2(T_a-t)}}
\nonumber \\
&& +  e^{ \frac{a^2}{2 T}- \frac{(a-x)^2}{2(T-t)}} 
 \sqrt{ \frac{T}{T-t} } 
\int_{-\infty}^a dy P^*(y,T ) 
e^{ \frac{(a-y)^2}{2 T} - \frac{(a-y)^2}{2(T-t)}} 
\frac{ \sinh \left( \frac{(a-x) (a-y) }{T-t} \right)}
{\sinh \left( \frac{a (a-y) }{T} \right)}
 \label{Qzero}
\end{eqnarray}
As a consequence, 
the conditioned drift of Eq. \ref{driftdoobfp} 
\begin{eqnarray}
\mu^*_T(x,t)  = \mu+ \partial_x \ln Q^{[\mu]}_T(x,t) 
=  \mu+ \partial_x \ln \left[ e^{ \frac{\mu^2}{2} t - \mu x} Q^{[0]}_T(x,t) \right]    
 =  \partial_x \ln  Q^{[0]}_T(x,t)
\label{driftBrown}
\end{eqnarray}
is independent of the initial drift $\mu$.
The fact that conditioned processes can be independent of the unconditioned drift
has stressed a lot of interest recently \cite{Krapivsky,refMultiEnds} 
and we refer to these two references for detailed discussions.

%%%%%%%%%%%%%%%%%%%%%%%%%%%%%%%%%%%%%%

\subsection{ Cases $ S^*(T )=0$ where the conditioning is toward full absorption 
before the horizon $T$ }

When the conditioning corresponds to full absorption before the horizon  $T$ in Eq. \ref{normatstarBrown}
\begin{eqnarray}
 \int_{0}^T dT_a \gamma^*(T_a )   = 1- S^*(T ) =1
\label{normatstarBrowntime}
\end{eqnarray}
the function $Q^{[0]}_T(x,t)$ contains only the first contribution of Eq. \ref{Qzero}
\begin{eqnarray}
Q^{[0]}_T(x,t)  =
  \left( \frac{a-x}{a} \right)   
\int_t^{T} dT_a \gamma^*(T_a) 
  \left(\frac{T_a}{T_a-t}  \right)^{\frac{3}{2} }
e^{\frac{a^2}{2T_a} - \frac{(a-x)^2}{2(T_a-t)}}
 \label{QzeroDead}
\end{eqnarray}
and the conditioned drift of Eq. \ref{driftBrown} reads
\begin{eqnarray}
\mu^*_T(x,t) && = \partial_x \ln  Q^{[0]}_T(x,t)
=  \partial_x  \left[ \ln \left( \frac{a-x}{a} \right) + \ln \left(\int_t^{T} dT_a \gamma^*(T_a) 
  \left(\frac{T_a}{T_a-t}  \right)^{\frac{3}{2} }
e^{\frac{a^2}{2T_a} - \frac{(a-x)^2}{2(T_a-t)}} \right)  \right]
\nonumber \\ 
&& 
    = \frac{1}{x-a}+  (a-x) \ 
\frac{\int_t^{T} dT_a \gamma^*(T_a) 
 \left(\frac{T_a}{T_a-t}  \right)^{\frac{3}{2} } 
e^{\frac{a^2}{2T_a} - \frac{(a-x)^2}{2(T_a-t)}}
\left[   \frac{1}{T_a-t}  \right]}
{\int_t^{T} dT_a \gamma^*(T_a) 
 \left(\frac{T_a}{T_a-t}  \right)^{\frac{3}{2} } 
e^{\frac{a^2}{2T_a} - \frac{(a-x)^2}{2(T_a-t)}} }
\label{driftBrownDead}
\end{eqnarray}

%%%%%%%%%%%%%%%%%%%%%%%%%%%%%%%%

\subsubsection{ Case where the absorption-time $T_a$ takes the single value $T^* \in ]0,T[$ : $\gamma^*(T_a)=\delta(T_a-T^*) $ }

\label{sec_finitehorizondelta}

For the special case where the absorption-time $T_a$ takes the single value $T^* \in ]0,T[$
\begin{eqnarray}
\gamma^*(T_a)=\delta(T_a-T^* ) 
 \label{gammaDead1delta}
\end{eqnarray}
the function $Q^{[0]}_T(x,t)$ of Eq. \ref{QzeroDead} reads using the 
Heaviside function $\theta(.)$
\begin{eqnarray}
Q^{[0]}_T(x,t)  = 
\theta(T^* -t)    \left( \frac{a-x}{a} \right)  
     \left(\frac{T^*}{T^*-t}  \right)^{\frac{3}{2} }e^{\frac{a^2}{2T^*} - \frac{(a-x)^2}{2(T^*-t)}}
 \label{QzeroDead1delta}
\end{eqnarray}
and the corresponding conditioned drift of Eq. \ref{driftBrownDead}
reduces for $t \in [0,T^*[$ to
\begin{eqnarray}
\mu^*(x,t) =   \partial_x \ln  Q^{[0]}_T(x,t) = -\frac{1}{a-x} +  \frac{a-x}{T^*-t} 
\label{driftdoobdeltaT}
\end{eqnarray}

Equation \ref{driftdoobdeltaT} can be found in~\cite{refMultiEnds}. This equation is also found in the mathematical literature (with the usual convention $T^* = 1$) where it is obtained using enlargements of filtration techniques~\cite{refBaudoin,refbookMansuy}. 
Observe that when the final time $T^*$ becomes arbitrarily large ($T^* \to \infty$), the drift Eq.\ref{driftdoobdeltaT} reduces to that of the taboo process (with taboo state $a$), 
\begin{equation}
	\mu^{[taboo]}(x)  = -\frac{1}{a-x}  
\label{drifttaboo}
\end{equation}
which is the unique diffusion on $]-\infty,a[$ with a generator of the form~\cite{refKnight} 
\begin{eqnarray}
	\frac{1}{2}\frac{d^2}{dx^2} - \frac{1}{a-x} \frac{d}{dx} 
\label{taboostatea}
\end{eqnarray}
Loosely speaking, one can see the taboo process as a Bessel process~\cite{refnotesLawler} but for the present geometry $]-\infty,a[$. With such a drift, the boundary now corresponds to an entrance boundary~\cite{refbookKarlin}, which means that the boundary cannot be reached from the interior of the state space (here the interval $]-\infty,a[$). Originally introduced in the mathematical literature, and since then widely studied in this field~\cite{refKnight,refPinsky,refKorzeniowski} for both semi-infinite and finite domains, the taboo process and its later generalizations have recently found applications in physics~\cite{refGarbaczewski} where they are relevant for studying confined polymers~\cite{refAdorisio}. For a physicist-oriented survey, we refer to the recent article~\cite{refAlainTaboo}.

Also observe that as the level $a$ becomes large ($a \to \infty$), the first term in the r.h.s. of Eq.\ref{driftdoobdeltaT}  is small compared to the second, except when $x$ approaches $a$ near the final time $T^*$. In this case, the drift Eq.\ref{driftdoobdeltaT} becomes
\begin{equation}
	\mu^{[BB]}(x)  = \frac{a - x}{T^*-t}   
\label{driftBrownianbridge}
\end{equation}
which is the drift of a Brownian bridge ending at $a$ at the final time $T^*$~\cite{refbookKarlin,refMajumdarOrland,refMazzoloJstat}. This can be understood intuitively since, when $a$ is large, the process spends most of the time far from the boundary (recall that the process starts at $x_0 = 0 \ll a$) and thus it does not feel the boundary, except at the final time $T^*$ when the process is constrained to end at the level $a$. Apart from near-final times, the process therefore has a very low probability of reaching $a$. 

Finally, also observe that when $a = 0$, then the drift Eq.\ref{driftdoobdeltaT} becomes
\begin{eqnarray}
\mu_t(x) = \frac{1}{x} -  \frac{x}{T^*-t} 
\label{drift3DBesselbridge}
\end{eqnarray}
which is the drift of a three-dimensional Bessel bridge~\cite{refHernandez-del-Valle}, a process also known as Brownian excursion~\cite{refMajumdarOrland}. 

%%%%%%%%%%%%%%%%%%%%%%%%%%%%%%%

\subsubsection{ Case where the absorption-time $T_a$ takes only two values $T^*_-<T^*_+$ : 
$\gamma^*(T_a)=p \delta(T_a-T^*_-)+(1-p) \delta(T_a-T^*_+) $ }

For the special case where the absorption-time $T_a$ takes only two values $T^*_-<T^*_+$
\begin{eqnarray}
\gamma^*(T_a)=p \delta(T_a-T^*_-)+(1-p) \delta(T_a-T^*_+)
 \label{gammaDead2delta}
\end{eqnarray}
 the function $Q^{[0]}_T(x,t)$ of Eq. \ref{QzeroDead} reads 
\begin{eqnarray}
Q^{[0]}_T(x,t)  = p \theta(T^*_- -t)    \left( \frac{a-x}{a} \right)  
     \left(\frac{T^*_-}{T^*_--t}  \right)^{\frac{3}{2} }e^{\frac{a^2}{2T^*_-} - \frac{(a-x)^2}{2(T^*_--t)}}     
     +(1-p) \theta(T^*_+ -t)    \left( \frac{a-x}{a} \right)  
     \left(\frac{T^*_+}{T^*_+-t}  \right)^{\frac{3}{2} }e^{\frac{a^2}{2T^*_+} - \frac{(a-x)^2}{2(T^*_+-t)}}
 \label{QzeroDead2delta}
\end{eqnarray}

%%%%%%%%%%%%%%%%%%%%%%%%%%%%%%%%%%%%%%%%%%%%%%%%%%%%%%%%
%%                      FIGURE                        %%
%%%%%%%%%%%%%%%%%%%%%%%%%%%%%%%%%%%%%%%%%%%%%%%%%%%%%%%%                          
\begin{figure}[h]
\centering
\includegraphics[width=4.in,height=3.in]{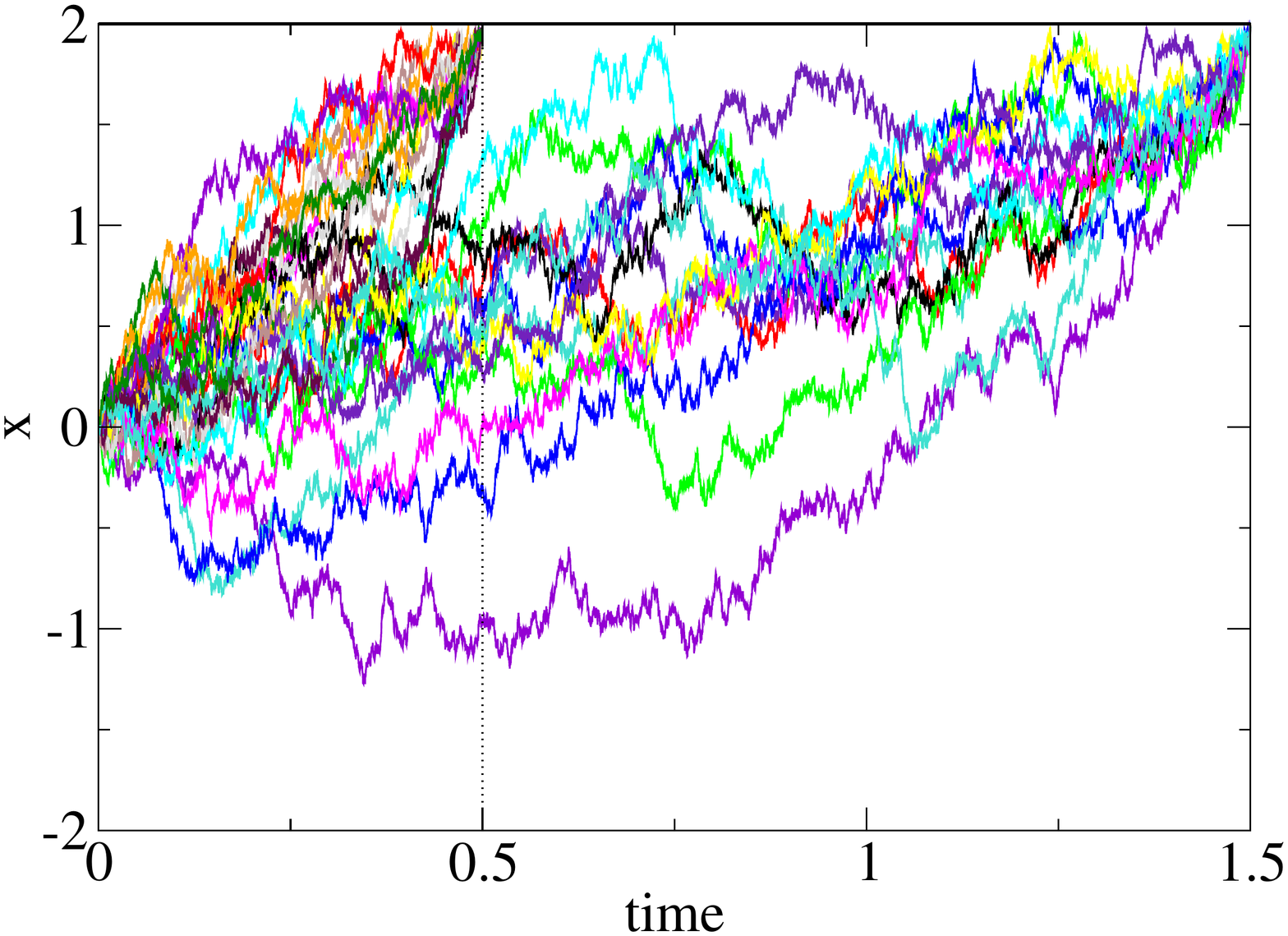}
\setlength{\abovecaptionskip}{15pt}  
\caption{A sample of 30 diffusions for the conditioned drift given by Eq. \ref{driftdoob2delta} and Eq. \ref{driftdoobdeltaplus} with parameters $a = 2$ , $T^*_- = 0.5$ and $T^*_+ = 1.5$. The time step used in the discretization is $dt = 10^{-4}$. All trajectories generated with different noise histories are statistically independent.}
\label{fig1}
\end{figure}
%%%%%%%%%%%%%%%%%%%%%%%%%%%%%%%%%%%%%%%%%%%%%%%%%%%%%%%% 

So here one needs to separate two regions for the conditioned drift of Eq. \ref{driftBrownDead}

(i)  in the region I corresponding to $0 \leq t \leq T^*_-$, Eq. \ref{driftBrownDead} yields
\begin{eqnarray}
\mu^*_{I}(x,t) 
 = \frac{1}{x-a} + (a-x)  
\frac{  p \left(\frac{T^*_-}{T^*_--t}  \right)^{\frac{3}{2} }   e^{ \frac{a^2}{2T^*_-}- \frac{(a-x)^2}{2(T^*_--t)}}\left[   \frac{1}{T^*_--t} \right]
+  (1-p) \left(\frac{T^*_+}{T^*_+-t}  \right)^{\frac{3}{2} }   e^{ \frac{a^2}{2T^*_+}- \frac{(a-x)^2}{2(T^*_+-t)}}\left[   \frac{1}{T^*_+-t} \right]  }
{ p \left(\frac{T^*_-}{T^*_--t}  \right)^{\frac{3}{2} }    e^{ \frac{a^2}{2T^*_-}- \frac{(a-x)^2}{2(T^*_--t)}}
+(1-p)  \left(\frac{T^*_+}{T^*_+-t}  \right)^{\frac{3}{2} }    e^{ \frac{a^2}{2T^*_+}- \frac{(a-x)^2}{2(T^*_+-t)}}}
\label{driftdoob2delta}
\end{eqnarray}

(ii) in the region II corresponding to $T^*_- \leq t \leq T^*_+$, the absorption at $T^*_-$ has already taken place, 
so the conditioned drift 
is similar to Eq. \ref{driftdoobdeltaT} with the replacement $T^* \to T^*_+$
\begin{eqnarray}
\mu^*_{II}(x,t)  =   \frac{1}{x-a} +  \frac{a-x}{T^*_+-t} 
\label{driftdoobdeltaplus}
\end{eqnarray}

Observe that 
\begin{eqnarray}
	\mu^*_{I}(x,T^*_-) = \lim_{t  \to T^*_-} \mu_{I}(x,t) =  \frac{1}{x-a} +  \frac{a-x}{T^*_+-T^*_-} = \mu^*_{II}(x,T^*_-) 
\end{eqnarray}
so that the conditioned drift is continuous on the whole interval $[0,T^*_+]$.

%%%%%%%%%%%%%%%%%%%%%%%%%%%%%%%%%%%%%%%%%%

\subsection{ Cases $ S^*(T )=1$ where the conditioning is toward full survival at the horizon $T$ }

When the conditioning corresponds to full survival at the horizon $T$ 
 in Eq. \ref{normatstarBrown}
\begin{eqnarray}
 \int_{-\infty}^a dy P^*(y,T )    = S^*(T )=1
\label{normatstar11}
\end{eqnarray}
the function $Q^{[0]}_T(x,t)$ contains only the second contribution of Eq. \ref{Qzero}
\begin{eqnarray}
Q^{[0]}_T(x,t)
 =  e^{ \frac{a^2}{2 T}- \frac{(a-x)^2}{2(T-t)}} 
 \sqrt{ \frac{T}{T-t} } 
\int_{-\infty}^a dy P^*(y,T ) 
e^{ \frac{(a-y)^2}{2 T} - \frac{(a-y)^2}{2(T-t)}} 
\frac{ \sinh \left( \frac{(a-x) (a-y) }{T-t} \right)}
{\sinh \left( \frac{a (a-y) }{T} \right)}
 \label{QzeroSurvival}
\end{eqnarray}
and the conditioned drift of Eq. \ref{driftBrown} reads
\begin{eqnarray}
\mu^*_T(x,t) && = \partial_x \ln  Q^{[0]}_T(x,t)
=  \partial_x  \left[ - \frac{(a-x)^2}{2(T-t)} + \ln \left( \int_{-\infty}^a dy P^*(y,T ) 
e^{ \frac{(a-y)^2}{2 T} - \frac{(a-y)^2}{2(T-t)}} 
\ \frac{ \sinh \left( \frac{(a-x) (a-y) }{T-t} \right)}
{\sinh \left( \frac{a (a-y) }{T} \right)}\right)  \right]
\nonumber \\ 
&& 
    = \frac{a-x}{T-t} 
    + \frac{\int_{-\infty}^a dy P^*(y,T ) 
e^{ \frac{(a-y)^2}{2 T} - \frac{(a-y)^2}{2(T-t)}} 
\frac{ \left( \frac{y-a}{T-t} \right) \cosh \left( \frac{(a-x) (a-y) }{T-t} \right)}
{\sinh \left( \frac{a (a-y) }{T} \right)}}
    {\int_{-\infty}^a dy P^*(y,T ) 
e^{ \frac{(a-y)^2}{2 T} - \frac{(a-y)^2}{2(T-t)}} 
\frac{ \sinh \left( \frac{(a-x) (a-y) }{T-t} \right)}
{\sinh \left( \frac{a (a-y) }{T} \right)}}
\label{driftBrownSurvival}
\end{eqnarray}

%%%%%%%%%%%%%%%%%%%%%%%%%%%%

\subsubsection{ Special case with a single position $y^*\in ]-\infty,a[ $ at time $T$ : $P^*(y,T )=\delta(y-y^*)   $  }

%%%%%%%%%%%%%%%%%%%%%%%%%%%%%%%%%%%%%%%%%%%%%%%%%%%%%%%%
%%                      FIGURE                        %%
%%%%%%%%%%%%%%%%%%%%%%%%%%%%%%%%%%%%%%%%%%%%%%%%%%%%%%%%                          
\begin{figure}[h]
\centering
\includegraphics[width=4.in,height=3.in]{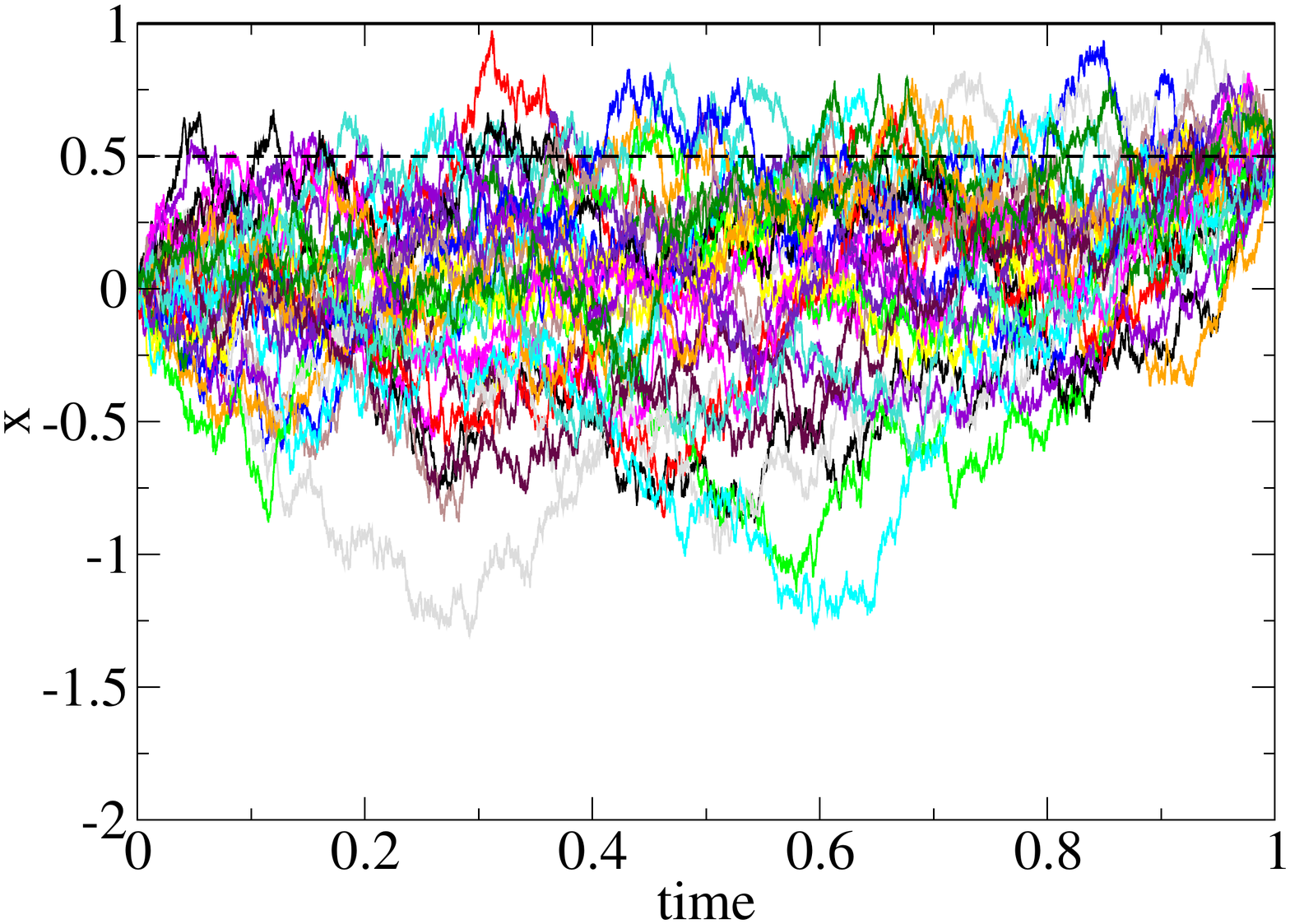}
\setlength{\abovecaptionskip}{15pt}  
\caption{A sample of 30 diffusions for the conditioned drift given by Eq. \ref{driftBrownSurvival1delta} with parameters $a = 1$ , $T = 1$ and $y^* = 0.5$. The time step used in the discretization is $dt = 10^{-4}$. All trajectories generated with different noise histories are statistically independent.}
\label{fig2}
\end{figure}
%%%%%%%%%%%%%%%%%%%%%%%%%%%%%%%%%%%%%%%%%%%%%%%%%%%%%%%% 

For the case with a single position $y^*\in ]-\infty,a[ $ at time $T$
\begin{eqnarray}
P^*(y,T )=\delta(y-y^*)
 \label{pspatial1delta}
\end{eqnarray}
 Eq. \ref{QzeroSurvival} reads
\begin{eqnarray}
Q^{[0]}_T(x,t)
 =  e^{ \frac{a^2}{2 T}- \frac{(a-x)^2}{2(T-t)}} 
 \sqrt{ \frac{T}{T-t} } 
e^{ \frac{(a-y^*)^2}{2 T} - \frac{(a-y^*)^2}{2(T-t)}} 
\ \frac{ \sinh \left( \frac{(a-x) (a-y^*) }{T-t} \right)}
{\sinh \left( \frac{a (a-y^*) }{T} \right)}
 \label{QzeroSurvivaly1delta}
\end{eqnarray}
and the conditioned drift of Eq. \ref{driftBrownSurvival} reduces to
\begin{eqnarray}
\mu^*_T(x,t) && = \partial_x \ln  Q^{[0]}_T(x,t)
= \partial_x \left(- \frac{(a-x)^2}{2(T-t)} + \ln \left[  \sinh \left( \frac{(a-x) (a-y^*) }{T-t} \right)\right]  \right)
\nonumber \\ 
&& 
    = \frac{a-x}{T-t} + \left( \frac{y^*-a}{T-t } \right) \coth \left( \frac{(a-x) (a-y^*) }{T-t} \right)
    \label{driftBrownSurvival1delta}
\end{eqnarray}
The drift of Eq. \ref{driftBrownSurvival1delta} corresponds to a Brownian bridge ending at $y^*$ at time $T$, conditioned to stay below the positive level $a$.
\noindent Observe that when $y^*$ is close to $a$ 
\begin{eqnarray}
 \lim_{y^* \to a} \mu^*_T(x,t) =  \frac{1}{a-x} +  \frac{a-x}{T-t}
\end{eqnarray}
corresponding to the drift of Eq. \ref{driftdoobdeltaT} as expected. Similarly, when the frontier $a$ becomes large, one get that  
\begin{eqnarray}
 \lim_{a  \to \infty} \mu^*_T(x,t) =   \frac{y^*-x}{T-t}
\end{eqnarray}
which is the drift of an unconstrained Brownian bridge ending at $y^*$ at time $T$.

\subsubsection{ Special case with two position $(y_1^*,y_2^*)$ at time $T$ : $P^*(y,T )=
p \delta(y-y_1^*) +  (1-p) \delta(y-y_2^*) $  }

%%%%%%%%%%%%%%%%%%%%%%%%%%%%%%%%%%%%%%%%%%%%%%%%%%%%%%%%
%%                      FIGURE                        %%
%%%%%%%%%%%%%%%%%%%%%%%%%%%%%%%%%%%%%%%%%%%%%%%%%%%%%%%%                          
\begin{figure}[h]
\centering
\includegraphics[width=4.in,height=3.in]{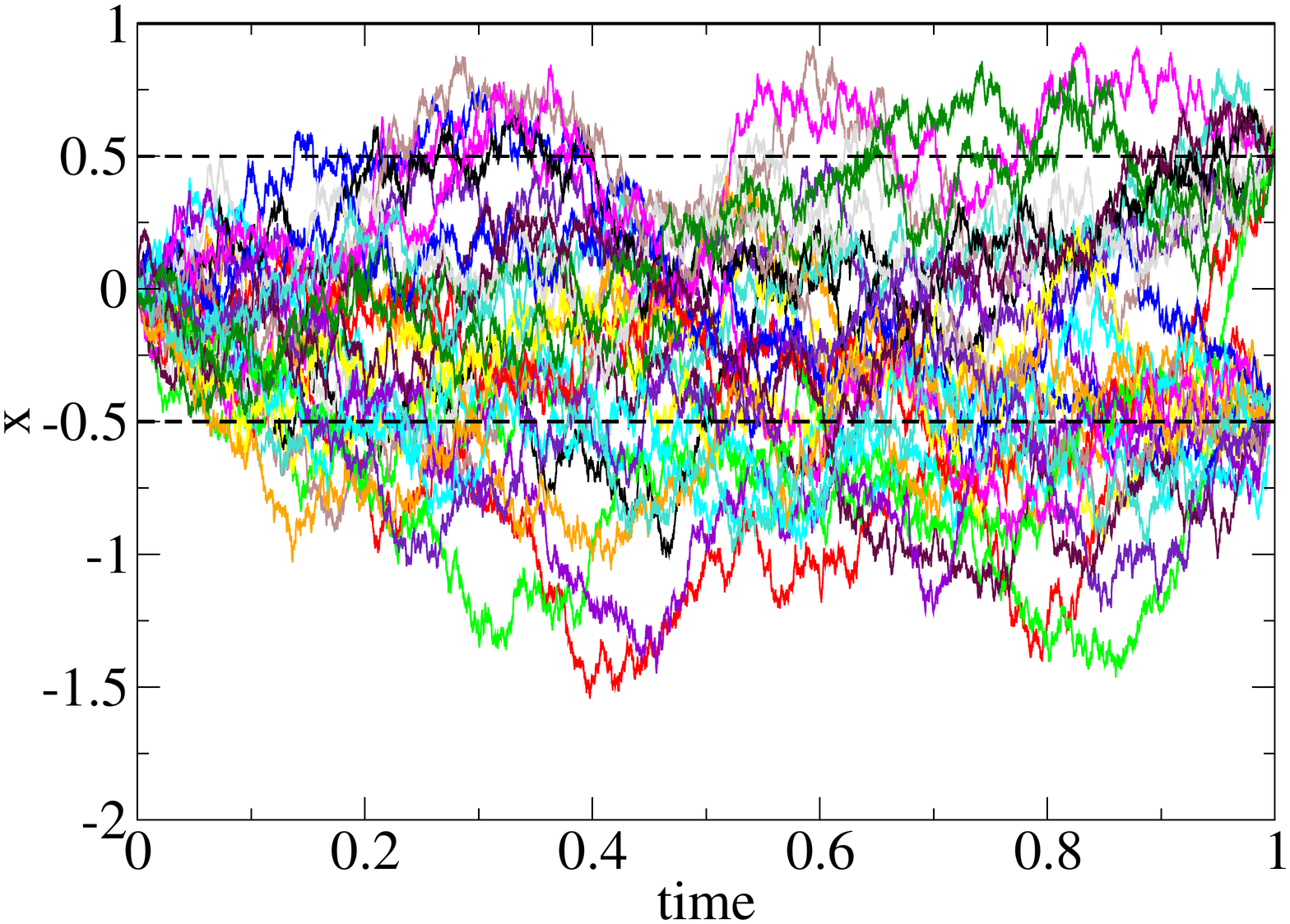}
\setlength{\abovecaptionskip}{15pt}  
\caption{A sample of 30 diffusions for the conditioned drift given by Eq. \ref{driftBrownSurvival2delta} with parameters $a = 1$ , $T = 1$, $y_1^* = 0.5$ and $y_2^* = -0.5$. The time step used in the discretization is $dt = 10^{-4}$. All trajectories generated with different noise histories are statistically independent.}
\label{fig3}
\end{figure}
%%%%%%%%%%%%%%%%%%%%%%%%%%%%%%%%%%%%%%%%%%%%%%%%%%%%%%%%

For the case with two positions $(y_1^*,y_2^*)$ at time $T$
\begin{eqnarray}
P^*(y,T)=p \delta(y-y_1^*) +  (1-p) \delta(y-y_2^*) 
 \label{pspatial2delta}
\end{eqnarray}
the conditioned drift of Eq. \ref{driftBrownSurvival} reads
\begin{eqnarray}
\mu^*_T(x,t)    = \frac{a-x}{T-t} 
    + \frac{ p 
e^{ \frac{(a-y_1^*)^2}{2 T} - \frac{(a-y_1^*)^2}{2(T-t)}} 
\frac{ \left( \frac{y_1^*-a}{T-t} \right) \cosh \left( \frac{(a-x) (a-y_1^*) }{T-t} \right)}
{\sinh \left( \frac{a (a-y_1^*) }{T} \right)}
+
(1-p) 
e^{ \frac{(a-y_2^*)^2}{2 T} - \frac{(a-y_2^*)^2}{2(T-t)}} 
\frac{ \left( \frac{y_2^*-a}{T-t} \right) \cosh \left( \frac{(a-x) (a-y_2^*) }{T-t} \right)}
{\sinh \left( \frac{a (a-y_2^*) }{T} \right)}
}
    { p 
e^{ \frac{(a-y_1^*)^2}{2 T} - \frac{(a-y_1^*)^2}{2(T-t)}} 
\frac{ \sinh \left( \frac{(a-x) (a-y_1^*) }{T-t} \right)}
{\sinh \left( \frac{a (a-y_1^*) }{T} \right)}
+ (1-p)
e^{ \frac{(a-y_2^*)^2}{2 T} - \frac{(a-y_2^*)^2}{2(T-t)}} 
\frac{ \sinh \left( \frac{(a-x) (a-y_2^*) }{T-t} \right)}
{\sinh \left( \frac{a (a-y_2^*) }{T} \right)} }
 \ \ \ \ 
\label{driftBrownSurvival2delta}
\end{eqnarray}

%%%%%%%%%%%%%%%%%%%%%%%%%%%%%%

\subsection{Simplest example with partial survival probability $S^*(T ) \in]0,1[ $ at time $T$   }

%%%%%%%%%%%%%%%%%%%%%%%%%%%%%%%%%%%%%%%%%%%%%%%%%%%%%%%%
%%                      FIGURE                        %%
%%%%%%%%%%%%%%%%%%%%%%%%%%%%%%%%%%%%%%%%%%%%%%%%%%%%%%%%                          
\begin{figure}[h]
\centering
\includegraphics[width=4.in,height=3.in]{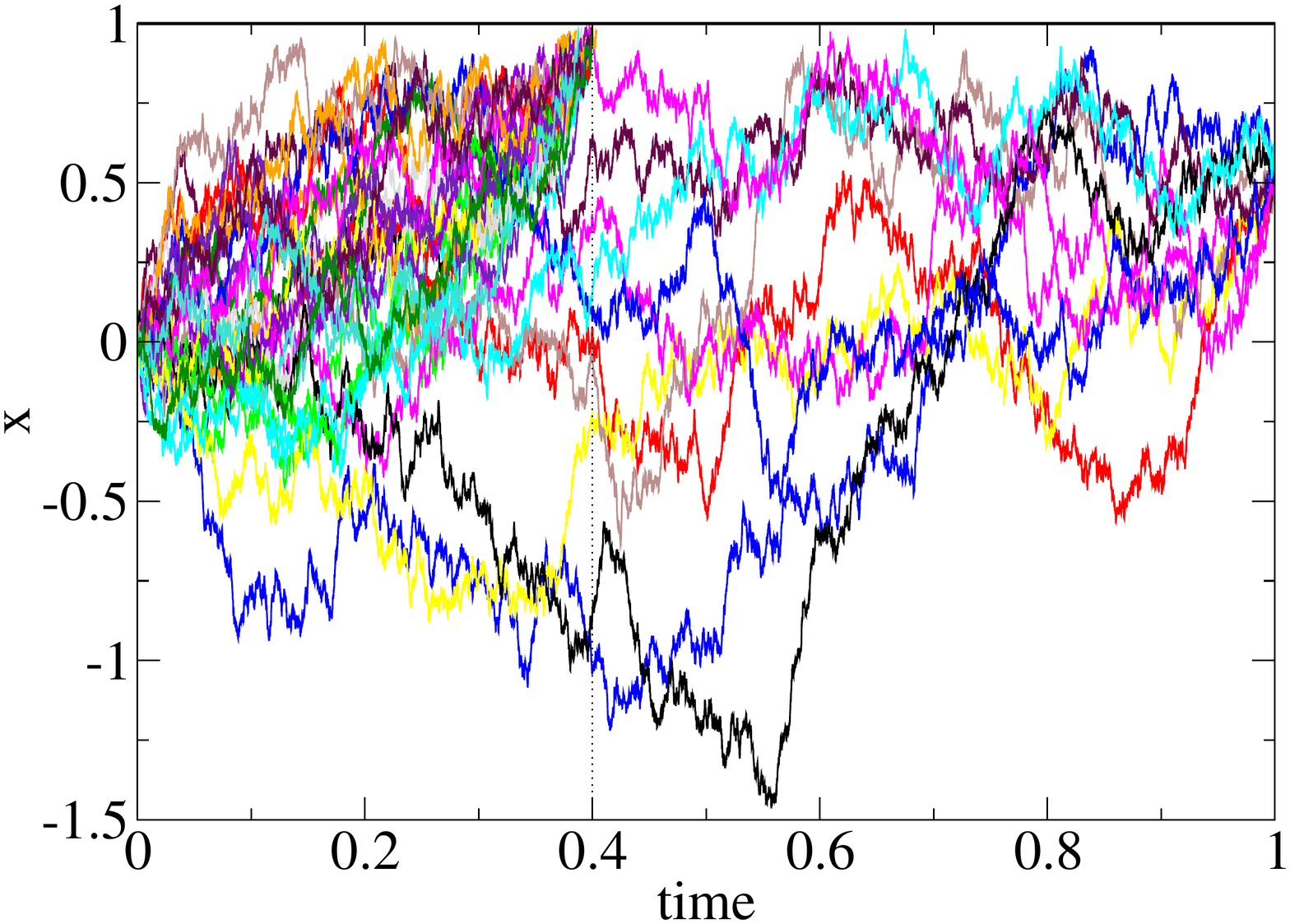}
\setlength{\abovecaptionskip}{15pt}  
\caption{A sample of 30 diffusions for the conditioned drift given by Eq. \ref{driftBrown2deltaintimespace} with parameters $a = 1$, $T^* = 0.4$, $T = 1$, $y^* = 0.5$ and $S^* = 1/3$. The time step used in the discretization is $dt = 10^{-4}$. All trajectories generated with different noise histories are statistically independent.}
\label{fig4}
\end{figure}
%%%%%%%%%%%%%%%%%%%%%%%%%%%%%%%%%%%%%%%%%%%%%%%%%%%%%%%%

As simplest example with partial survival probability $S^*(T )=S^* \in]0,1[ $ at time $T$
in Eq \ref{normatstarBrown},
let us consider the case with a single time $T^* \in ]0,T[$ and a single position $y^* \in ]-\infty, a[$
\begin{eqnarray}
\gamma^*(T_a)&& =(1-S^*) \delta(T_a-T^*)  
\nonumber \\
P^*(y,T )&& = S^* \delta(y-y^*)  
\label{2deltaintimespace}
\end{eqnarray}
Then Eq. \ref{Qzero} yields the function $Q^{[0]}_T (x,t)$  
\begin{eqnarray}
Q^{[0]}_T(x,t) && =(1-S^*) \theta(T^*-t)
  \left( \frac{a-x}{a} \right)   
  \left(\frac{T^*}{T^*-t}  \right)^{\frac{3}{2} }
e^{\frac{a^2}{2T^*} - \frac{(a-x)^2}{2(T^*-t)}}
 \nonumber \\ &&
 + S^* e^{ \frac{a^2}{2 T}- \frac{(a-x)^2}{2(T-t)}} 
 \sqrt{ \frac{T}{T-t} } 
e^{ \frac{(a-y^*)^2}{2 T} - \frac{(a-y^*)^2}{2(T-t)}} 
\frac{ \sinh \left( \frac{(a-x) (a-y^*) }{T-t} \right)}
{\sinh \left( \frac{a (a-y^*) }{T} \right)}
 \label{Qzero2deltaintimespace}
\end{eqnarray}
and its partial derivative with respect to $x$
\begin{eqnarray}
&& \partial_x Q^{[0]}_T(x,t)  =(1-S^*) \theta(T^*-t)
  \left( \frac{a-x}{a} \right)   
  \left(\frac{T^*}{T^*-t}  \right)^{\frac{3}{2} }
e^{\frac{a^2}{2T^*} - \frac{(a-x)^2}{2(T^*-t)}} \left[ \frac{1}{x-a} +  \frac{a-x}{T^*-t} \right]
 \nonumber \\ &&
 + S^* e^{ \frac{a^2}{2 T}- \frac{(a-x)^2}{2(T-t)}} 
 \sqrt{ \frac{T}{T-t} } 
e^{ \frac{(a-y^*)^2}{2 T} - \frac{(a-y^*)^2}{2(T-t)}} 
\frac{ \left[  \frac{a-x}{T-t} \sinh \left( \frac{(a-x) (a-y^*) }{T-t} \right)
+  \left( \frac{y^*-a}{T-t } \right) \cosh \left( \frac{(a-x) (a-y^*) }{T-t} \right)
\right] }
{\sinh \left( \frac{a (a-y^*) }{T} \right)} 
 \label{Qzero2deltaintimespacederi}
\end{eqnarray}
that allow to obtain the corresponding conditioned drift of Eq. \ref{driftdoobfp} 
\begin{eqnarray}
\mu^*_T(x,t)  
 =  \partial_x \ln  Q^{[0]}_T(x,t) = \frac{ \partial_x Q^{[0]}_T(x,t)}{Q^{[0]}_T(x,t)}
\label{driftBrown2deltaintimespace}
\end{eqnarray}

%%%%%%%%%%%%%%%%%%%%%%%%%%%%%%%%%%%%%%%%%%%%%%%%

\section{ Application to the Brownian motion with drift $\mu$ for infinite horizon }

\label{sec_infinitehorizon}

In this section, the conditioning for infinite horizon $T=+\infty$ described in section \ref{sec_general} 
is applied to the simplest case where the unconditioned process is the Brownian motion with uniform drift $\mu$ 
starting at $x=0$ with absorbing condition at position $a>0$,
whose properties have been already described in subsection \ref{subsec_Brown}.

%%%%%%%%%%%%%%%%%%%%%%%%%%%%%%%%%%%%%%%%%%%%

\subsection{ Cases $ S^*(\infty )=0$ where the conditioning is toward full absorption before the infinite horizon $T=+\infty$ }

When the conditioning corresponds to full absorption before the infinite horizon $T=+\infty$
in Eq. \ref{normatstarBrowntime}
\begin{eqnarray}
 \int_{0}^{+\infty} dT_a \gamma^*(T_a )   = 1- S^*(\infty ) =1
\label{normatstarBrowntimeinfty}
\end{eqnarray}
the function $Q^{[0]}_{\infty}(x,t)$ of Eq. \ref{QzeroDead} reads
\begin{eqnarray}
Q^{[0]}_{\infty}(x,t)  =
  \left( \frac{a-x}{a} \right)   
\int_t^{+\infty} dT_a \gamma^*(T_a) 
  \left(\frac{T_a}{T_a-t}  \right)^{\frac{3}{2} }
e^{\frac{a^2}{2T_a} - \frac{(a-x)^2}{2(T_a-t)}}
 \label{QzeroDeadinfty}
\end{eqnarray}
and the conditioned drift of Eq. \ref{driftBrownDead} 
can be rewritten as
\begin{eqnarray}
\mu^*_{\infty}(x,t)  =   
\frac{\int_t^{+\infty} dT_a \gamma^*(T_a)\left(\frac{T_a}{T_a-t}  \right)^{\frac{3}{2} }    
e^{\frac{a^2}{2T_a} - \frac{(a-x)^2}{2(T_a-t)}} \left[\frac{1}{x-a} +  \frac{a-x}{T_a-t}  \right]}
{\int_t^{+\infty} dT_a \gamma^*(T_a)\left(\frac{T_a}{T_a-t}  \right)^{\frac{3}{2} }   
e^{\frac{a^2}{2T_a} - \frac{(a-x)^2}{2(T_a-t)}} }
\label{driftdoobfiniteTchange}
\end{eqnarray}

%%%%%%%%%%%%%%%%%%%%%%%%%%%%%%%%%%%

\subsubsection{ Bounds on the conditioned drift $\mu^*_{\infty}(x,t)$ for an arbitrary normalized distribution $\gamma^*(T_a) $}

To get some physical intuition about the possible values of the 
conditioned drift $\mu^*_{\infty}(x,t)$ of Eq. \ref{driftdoobfiniteTchange}
for an arbitrary  normalized distribution $\gamma^*(T_a) $,
 one can make the change of variables from the first-passage time $T_a \in [t,+\infty[ $
toward the corresponding drift $\lambda(T_a) $ that appears when the value $T_a$ is alone (see Eq. \ref{driftdoobdeltaT})
\begin{eqnarray}
\lambda(T_a)=\frac{1}{x-a} +  \frac{a-x}{T_a-t}
\label{driftlambdaTa}
\end{eqnarray}
This change of variables is monotonic
\begin{eqnarray}
 \frac{d\lambda(T_a)}{dT_a} = - \frac{a-x}{(T_a-t)^2} <0
\label{monotonic}
\end{eqnarray}
The drift $\lambda(T_a) $ for $T_a =t$ gives the maximal possible drift
\begin{eqnarray}
\lambda_{max}=\lambda(T_a=t)=+\infty  
\label{lambdamax}
\end{eqnarray}
while the drift $\lambda(T_a) $ for $T_a =+\infty$ gives the minimal possible drift
\begin{eqnarray}
\lambda_{min}=\lambda(T_a=+\infty)= - \frac{1}{a-x} <0
\label{lambdamin}
\end{eqnarray}
where one recognizes the Bessel drift that will be discussed below in Eq. \ref{driftdoobforevermupositive}.
As a consequence, the conditioned drift $\mu^*_{\infty}(x,t)$ of Eq. \ref{driftdoobfiniteTchange}
belongs to the half-line
\begin{eqnarray}
\mu^*_{\infty}(x,t) \geq  \lambda_{min}=  - \frac{1}{a-x} 
\label{bound}
\end{eqnarray}
One could plug the change of variables of Eq. \ref{driftlambdaTa}
\begin{eqnarray}
T_a(\lambda) && =t+   \frac{(a-x)^2}{1+ (a-x) \lambda}
\label{driftlambdaTainversion}
\end{eqnarray}
into Eq. \ref{driftdoobfiniteTchange}, but the result will not be illuminating 
when the normalized distribution $\gamma^*(T_a) $ is arbitrary,
while there will be simplifications when the normalized distribution $\gamma^*(T_a) $
belongs to some special family, as described in the next subsection.

%%%%%%%%%%%%%%%%%%%%%%%%%%%%%%%%

\subsubsection{ Special family $\gamma^*(T_a )  = \int_0^{+\infty} d\lambda m(\lambda)  \frac{a }{\sqrt{2 \pi} T_a^{\frac{3}{2}} } e^{- \frac{(a-\lambda T_a )^2}{2T_a}}  $ with some normalized measure $ \int_{0}^{+\infty} d \lambda m(\lambda) =1 $ }

If one requires the conditioned first-passage-time distribution $\gamma^*(T_a )   $ to be
the normalized first-passage-time distribution $\gamma^{[\lambda]}(T_a \vert 0,0) $ of Eq. \ref{gammaBrown}
that would have the Brownian motion of uniform drift $\lambda \geq 0$
 \begin{eqnarray}
\gamma^*(T_a )  = \gamma^{[\lambda]}(T_a \vert 0,0)  
\equiv \frac{a }{\sqrt{2 \pi} T_a^{\frac{3}{2}} } e^{- \frac{(a-\lambda T_a )^2}{2T_a}} 
= \frac{a }{\sqrt{2 \pi} T_a^{\frac{3}{2}} } e^{\lambda a - \frac{a^2}{2T_a}-\frac{\lambda^2}{2}T_a} 
\label{gammalambda}
\end{eqnarray}
then the function of Eq. \ref{QzeroDeadinfty}
can be evaluated using the integral of Eq. \ref{integral} for $\lambda \geq 0$
\begin{eqnarray}
Q^{[0]}_{\infty}(x,t)  && = 
 \left( \frac{a-x}{a} \right)  
\int_t^{+\infty} dT_a  \frac{a }{\sqrt{2 \pi} T_a^{\frac{3}{2}} } e^{\lambda a - \frac{a^2}{2T_a}-\frac{\lambda^2}{2}T_a}
\left(\frac{T_a}{T_a-t}  \right)^{\frac{3}{2} } 
e^{\frac{a^2}{2T_a} - \frac{(a-x)^2}{2(T_a-t)}}
\nonumber \\
&& = \left( \frac{a-x}{\sqrt{2 \pi}} \right)   e^{\lambda a - \frac{\lambda^2}{2} t }
\int_t^{+\infty} dT_a  
\left(\frac{1}{T_a-t}  \right)^{\frac{3}{2} } 
e^{-\frac{\lambda^2}{2}(T_a-t)  - \frac{(a-x)^2}{2(T_a-t)}}
\nonumber \\
&& = \left( \frac{a-x}{\sqrt{2 \pi}} \right)   e^{\lambda a - \frac{\lambda^2}{2} t }
 \frac{ \sqrt{2 \pi }}{ (a-x)} e^{ - \lambda (a-x) }
 \nonumber \\
&& =   e^{ \lambda x  - \frac{\lambda^2}{2} t }
\label{QdoobfiniteTlambda}
\end{eqnarray}
The conditioned drift then reduces to $\lambda$
\begin{eqnarray}
\mu^*_{\infty}(x,t)   =  \partial_x \ln Q^{[0]}_{\infty}(x,t)   = \lambda
\label{driftdoobfiniteT1lambdafinal}
\end{eqnarray}
in agreement with the physical intuition that the conditioning toward 
the normalized first-passage time distribution of Eq. \ref{gammalambda}
that would have the Brownian motion of uniform drift $\lambda$
should simply produce the conditioned drift $\lambda$ independently of the initial drift $\mu$.

This simple result 
suggests to considering the more general case where
the conditioned first-passage time distribution $\gamma^*(T_a ) $
can be decomposed as an integral over $\lambda \in [0,+\infty[$
of the normalized distributions $ \gamma^{[\lambda]}(T_a \vert 0,0)$ of Eq. \ref{gammalambda} 
\begin{eqnarray}
\gamma^*(T_a )  = \int_0^{+\infty} d\lambda m(\lambda) \gamma^{[\lambda]}(T_a \vert 0,0) 
 = \int_0^{+\infty} d\lambda m(\lambda)  \frac{a }{\sqrt{2 \pi} T_a^{\frac{3}{2}} } e^{- \frac{(a-\lambda T_a )^2}{2T_a}} 
\label{mparametrization}
\end{eqnarray}
with some measure $m(\lambda)$ normalized over $\lambda \in [0,+\infty[$
\begin{eqnarray}
1 = \int_0^{+\infty} dT_a \gamma^*(T_a) = \int_{0}^{+\infty} d \lambda m(\lambda) 
\label{mparametrizationnorma}
\end{eqnarray}
Then the function of Eq. \ref{QzeroDeadinfty}
reads using the previous computation of Eq. \ref{QdoobfiniteTlambda}
\begin{eqnarray}
Q^{[0]}_{\infty}(x,t)  
  =   \int_0^{+\infty} d\lambda m(\lambda) e^{ \lambda  x - \frac{\lambda^2}{2} t }
\label{QdoobfiniteTm}
\end{eqnarray}
and the conditioned drift 
\begin{eqnarray}
\mu^*_{\infty}(x,t) = \partial_x \ln Q^{[0]}_{\infty}(x,t)     
  =  \frac{  \int_0^{+\infty} d\lambda m(\lambda) \lambda e^{ \lambda x - \frac{\lambda^2}{2} t } }
  {  \int_0^{+\infty} d\lambda m(\lambda) e^{ \lambda  x - \frac{\lambda^2}{2} t } }   
\label{driftdoobfiniteTmbaudoin}
\end{eqnarray}
is in agreement with the formula (3.20) given in \cite{refBaudoin}.

%%%%%%%%%%%%%%%%%%%%%%%%%%%%%%%%%%%%%%%%%%

\subsection{ Cases $ S^*(\infty )=1$ where the conditioning is toward full survival at the infinite horizon $T=+\infty$ }

Now we apply the formula of Eq. \ref{Qspacesurvivalinfty},
which is based on the specific choice of Eq. \ref{renorma}, 
 where the conditioned probability distribution $ P^{*[surviving]}(y,T ) $ 
is simply given by the unconditioned probability distribution $P^{[\mu]}(y,T \vert 0,0) $
normalized by the corresponding survival probability $S^{[\mu]} (T \vert 0,0) $
of the Brownian motion of drift $\mu$
\begin{eqnarray}
 P^{*[surviving]}(y,T )    
 =  \frac{P^{[\mu]}(y,T \vert 0,0)}{\int_{-\infty}^a d y' P^{[\mu]}(y',T \vert 0,0)}
 = \frac{P^{[\mu]}(y,T \vert 0,0)}{ S^{[\mu]} (T \vert 0,0)}
\label{renormamu}
\end{eqnarray}
As a consequence,
the conditioned process will then a priori depend on the initial drift $\mu$.

\subsubsection{ When the initial drift is vanishing or positive $\mu\geq 0$ }

 When the initial drift is positive $\mu\geq 0$, the forever-survival probability of Eq. \ref{survivalBrown}
vanishes $S(\infty \vert .) =0$, so
the function $Q_{\infty}^{[surviving]}(x,t)  $ of Eq. \ref{Qspacesurvivalinftycaseb} 
can be computed from the first-passage-time distribution $\gamma^{[\mu]}(t_2 \vert x,t) $ of Eq. \ref{gammafirstBrown}
\begin{eqnarray}
Q_{\infty}^{[surviving]}(x,t) 
&& = \lim_{T \to +\infty} \frac{  \int_{T}^{+\infty} dt_2 \gamma^{[\mu]}(t_2 \vert x,t)  }
 { \int_{T}^{+\infty} dt_2 \gamma^{[\mu]}(t_2 \vert 0,0)  }
 = \lim_{T \to +\infty}\frac{   \int_{T}^{+\infty} dt_2 
  \frac{(a-x) e^{ \mu (a-x) }  }{\sqrt{2 \pi} (t_2-t)^{\frac{3}{2}} } 
 e^{- \frac{(a-x)^2}{2(t_2-t)} - \frac{\mu^2 }{2} (t_2-t)  }    }
 {  \int_{T}^{+\infty} dt_2 
  \frac{a e^{ \mu a }  }{\sqrt{2 \pi} t_2^{\frac{3}{2}} } 
 e^{- \frac{a^2}{2t_2} - \frac{\mu^2 }{2} t_2  }   }
  \nonumber \\
 && = \frac{a-x }{a} e^{ - \mu x  + \frac{\mu^2 }{2} t} 
 \lim_{T \to +\infty}\frac{   \int_{0}^{+\infty} d\tau 
  \frac{ 1 }{ (T-t+\tau)^{\frac{3}{2}} } 
 e^{- \frac{(a-x)^2}{2(T-t+\tau)} - \frac{\mu^2 }{2} \tau  }    }
 {  \int_{0}^{+\infty} d\tau 
  \frac{ 1 }{ (T+\tau)^{\frac{3}{2}} } 
 e^{- \frac{a^2}{2 (T+\tau)} - \frac{\mu^2 }{2} \tau  }   }   = \frac{a-x }{a} e^{ - \mu x  + \frac{\mu^2 }{2} t} 
\label{survivalTamupos}
\end{eqnarray}
The corresponding conditioned drift
\begin{eqnarray}
\mu^*_{\infty}(x,t)  = \mu +  \partial_x  \ln Q_{\infty}^{[surviving]}(x,t) 
= \mu +  \partial_x \left[ \ln (a-x ) - \mu x \right] = - \frac{1}{a-x} 
\label{driftdoobforevermupositive}
\end{eqnarray}
does not depend on the value of the initial drift $\mu$ within the region $\mu \geq 0$
that we consider in this subsection.
The drift of Eq. \ref{driftdoobforevermupositive}
corresponds to the taboo process (with taboo state a) that we encountered in section \ref{sec_finitehorizondelta}, namely a Brownian motion conditioned to remain forever below the level $a$~\cite{refKnight}. 

%%%%%%%%%%%%%%%%%%%%%%%%%%%%%%%%%%%%

\subsubsection{ When the initial drift is strictly negative $\mu < 0$ }

When the initial drift is strictly negative $\mu<0$, the forever-survival probability of Eq. \ref{survivalBrown} is finite $S^{[\mu]}(\infty \vert x) >0$, so
the function $Q_{\infty}^{[surviving]}(x,t)  $ of Eq. \ref{Qspacesurvivalinftycasea}
can be computed in terms of the finite survival probability $S^{[\mu]}(\infty \vert x) $ of Eq. \ref{survivalBrown}
\begin{eqnarray}
Q_{\infty}^{[surviving]}(x,t) 
 = \frac{ S^{[\mu]}(\infty \vert x)   }
 {S^{[\mu]}(\infty \vert x_0)   }
 = \frac{1- e^{2 \mu (a-x) } }{1- e^{2 \mu a } }
\label{survivalTamuneg}
\end{eqnarray}
The corresponding conditioned drift
\begin{eqnarray}
\mu^*_{\infty}(x,t)  && = \mu +  \partial_x  \ln Q_{\infty}^{[surviving]}(x,t) 
  = \mu +  \partial_x \ln \left( 1- e^{2 \mu (a-x)} \right)
= \mu + \frac{2 \mu e^{2 \mu (a-x)} }{1- e^{2 \mu (a-x)}}
\nonumber \\
&& = \mu \left( \frac{1+  e^{2 \mu (a-x)}}{1- e^{2 \mu (a-x)}} \right)
= -\mu \coth(\mu (a-x))
\label{driftdoobforevermunegative}
\end{eqnarray}
depends on the value of the initial drift $\mu$ within the region $\mu < 0$
that we consider in this subsection. In the limit of vanishing drift $\mu \to 0^-$, one recovers Eq. \ref{driftdoobforevermupositive} at leading order
\begin{eqnarray}
\mu^*_{\infty}(x,t)  = -\frac{1}{a-x} +O(\mu)
\label{driftdoobforevermunegativelimit}
\end{eqnarray}
as expected. Also observe that as $x$ approaches the boundary $a$, the conditioned drift behaves as
\begin{eqnarray}
\mu^*_{\infty}(x,t)  \opsim_{x \to a^-}  -\frac{1}{a-x}  
\end{eqnarray}
which is the taboo drift and consequently, the conditioned process can never cross the barrier $a$. Moreover, 
when $x \to -\infty$ the conditioned drift behaves as
\begin{eqnarray}
\mu^*_{\infty}(x,t) \opsim_{x \to -\infty}      \mu 
\end{eqnarray}
which means that the conditioned process "converges" toward $-\infty$ since $\mu$ is negative. To the best of our knowledge,  Eq.~\ref{driftdoobforevermunegative} (with $\mu = -1$) first appears in Williams' paper~\cite{refWilliams}.

For semi-infinite domains, the process conditioned to never touch the barrier and thus to survive forever is different depending on whether the drift of the original process is positive or strictly negative. Of course, in both cases the barrier $a$ becomes an entrance boundary, but in the first case, the conditioned process is a universal taboo process (in the sense that it is independent of the original drift) while, in the second case, the process strongly depends on the original drift. In the last case, at large times and far from the boundary, the conditioned process behaves like a free Brownian motion with negative drift and therefore ends up going to - infinity.

%%%%%%%%%%%%%%%%%%%%%%%%%%%%%%%%%%%%%%%%%%

\subsection{ Cases $ S^*(\infty )\in]0,1[  $ where the conditioning is toward partial survival at the infinite horizon $T=+\infty$ }

Now we would like to apply the formula of Eq. \ref{Qinfinitypartial},
which is based on the specific choice of Eq. \ref{renormapartial},
which is the analog of Eq. \ref{renormamu} discussed above
\begin{eqnarray}
 P^{*[partial]}(y,T ) 
 =S^*(\infty )    \frac{P^{[\mu]}(y,T \vert 0,0)}{ S^{[\mu]} (T \vert 0,0)}
\label{renormapartialmu}
\end{eqnarray}
so that the conditioned process will then a priori depend on the initial drift $\mu$.

%%%%%%%%%%%%%%%%%%%%%%%%%%%%%%%

\subsubsection{ When the initial drift is vanishing or positive $\mu\geq 0$ }

When the initial drift is vanishing or positive $\mu\geq 0$,
the function $ Q^{*[partial]}_{\infty}(x,t) $ of Eq. \ref{Qinfinitypartial} 
can be evaluated using the building blocks of Eq. \ref{ratioexpli}
and \ref{survivalTamupos}
\begin{eqnarray}
Q^{*[partial]}_{\infty}(x,t) && = 
\int_t^{\infty} dT_a \gamma^*(T_a) \frac{\gamma^{[\mu]}(T_a \vert x,t)}{\gamma^{[\mu]}(T_a\vert 0,0)}  
+  S^*(\infty ) \left[ \lim_{T \to +\infty} \frac{ S^{[\mu]} (T \vert x,t)  }{ S^{[\mu]} (T \vert 0,0)} \right]
\nonumber
 \\
&& 
= \int_t^{+\infty} dT_a  \gamma^*(T_a)  \left(\frac{T_a}{T_a-t}  \right)^{\frac{3}{2} } \left( \frac{a-x}{a} \right)   
e^{- \mu x + \frac{\mu^2}{2} t +\frac{a^2}{2T_a} - \frac{(a-x)^2}{2(T_a-t)}}
+ S^*(\infty )  \left[  \frac{a-x }{a} e^{ - \mu x  + \frac{\mu^2 }{2} t}  \right]
\nonumber  \\
&& =   \left( \frac{a-x}{a} \right)  e^{ - \mu x  + \frac{\mu^2 }{2} t} 
\left[ \int_t^{+\infty} dT_a  \gamma^*(T_a)  \left(\frac{T_a}{T_a-t}  \right)^{\frac{3}{2} }   
e^{\frac{a^2}{2T_a} - \frac{(a-x)^2}{2(T_a-t)}}
+ S^*(\infty )  \right]
\label{Qmupos}
\end{eqnarray}
The corresponding conditioned drift 
\begin{eqnarray}
 \mu^*_{\infty}(x,t) && = \mu +  \partial_x \ln Q^{*[partial]}_{\infty}(x,t) 
 \nonumber  \\  
&& = \mu +  \partial_x \ln \left[ \left(  \frac{a-x }{a} \right) e^{ - \mu x  + \frac{\mu^2 }{2} t} 
\left[ \int_t^{+\infty} dT_a  \gamma^*(T_a)  \left(\frac{T_a}{T_a-t}  \right)^{\frac{3}{2} }   
e^{\frac{a^2}{2T_a} - \frac{(a-x)^2}{2(T_a-t)}}
+ S^*(\infty )  \right]
\right)
\nonumber  \\
&& = \frac{1}{x-a}  +  \partial_x \ln \left(  \int_t^{+\infty} dT_a  \gamma^*(T_a)  \left(\frac{T_a}{T_a-t}  \right)^{\frac{3}{2} }   
e^{\frac{a^2}{2T_a} - \frac{(a-x)^2}{2(T_a-t)}}
+ S^*(\infty )  
\right)
\nonumber  \\
&& = \frac{1}{x-a}  +  
(a-x) \frac{  \int_t^{+\infty} dT_a  \gamma^*(T_a)  \left(\frac{T_a}{T_a-t}  \right)^{\frac{3}{2} }   
e^{\frac{a^2}{2T_a} - \frac{(a-x)^2}{2(T_a-t)}} \left[ \frac{1}{T_a-t} \right] }
{  \int_t^{+\infty} dT_a  \gamma^*(T_a)  \left(\frac{T_a}{T_a-t}  \right)^{\frac{3}{2} }   
e^{\frac{a^2}{2T_a} - \frac{(a-x)^2}{2(T_a-t)}}+ S^*(\infty )   }
\label{driftdoobmupos}
\end{eqnarray}
does not depend on the value of the initial drift $\mu$ with the region $\mu \geq 0$
that we consider in this subsection.

As a simple example, let us consider the case where $\gamma^*(T_a) $ 
is a delta function at the value $T^*$ with the weight $\left[ 1-S^*  \right] $ 
complementary to the survival probability $S^*(\infty )=S^* $
\begin{eqnarray}
\gamma^*(T_a)= ( 1-S^* )  \delta(T_a-T^*)
\label{gammapartial1delta}
\end{eqnarray}
The conditioned drift of Eq. \ref{driftdoobmupos} then reads 
 \begin{eqnarray}
 \mu^*_{\infty}(x,t)  
  =  \frac{ 1 }{x-a}
+ (a-x) \frac{ ( 1-S^* ) \theta(T^*-t)  \left(\frac{T^*}{T^*-t}  \right)^{\frac{3}{2} }   
e^{\frac{a^2}{2T^*} - \frac{(a-x)^2}{2(T^*-t)}} \left[ \frac{1}{T^*-t} \right] }
{( 1-S^* )  \theta(T^*-t)  \left(\frac{T^*}{T^*-t}  \right)^{\frac{3}{2} }   
e^{\frac{a^2}{2T^*} - \frac{(a-x)^2}{2(T^*-t)}} + S^*  }
\label{driftmuposfiniteinfinite}
\end{eqnarray}

%%%%%%%%%%%%%%%%%%%%%%%%%%%%%%%%%%%%%%%%%%%%%%%%%%%%%%%%
%%                      FIGURE                        %%
%%%%%%%%%%%%%%%%%%%%%%%%%%%%%%%%%%%%%%%%%%%%%%%%%%%%%%%%                          
\begin{figure}[h]
\centering
\includegraphics[width=4.in,height=3.in]{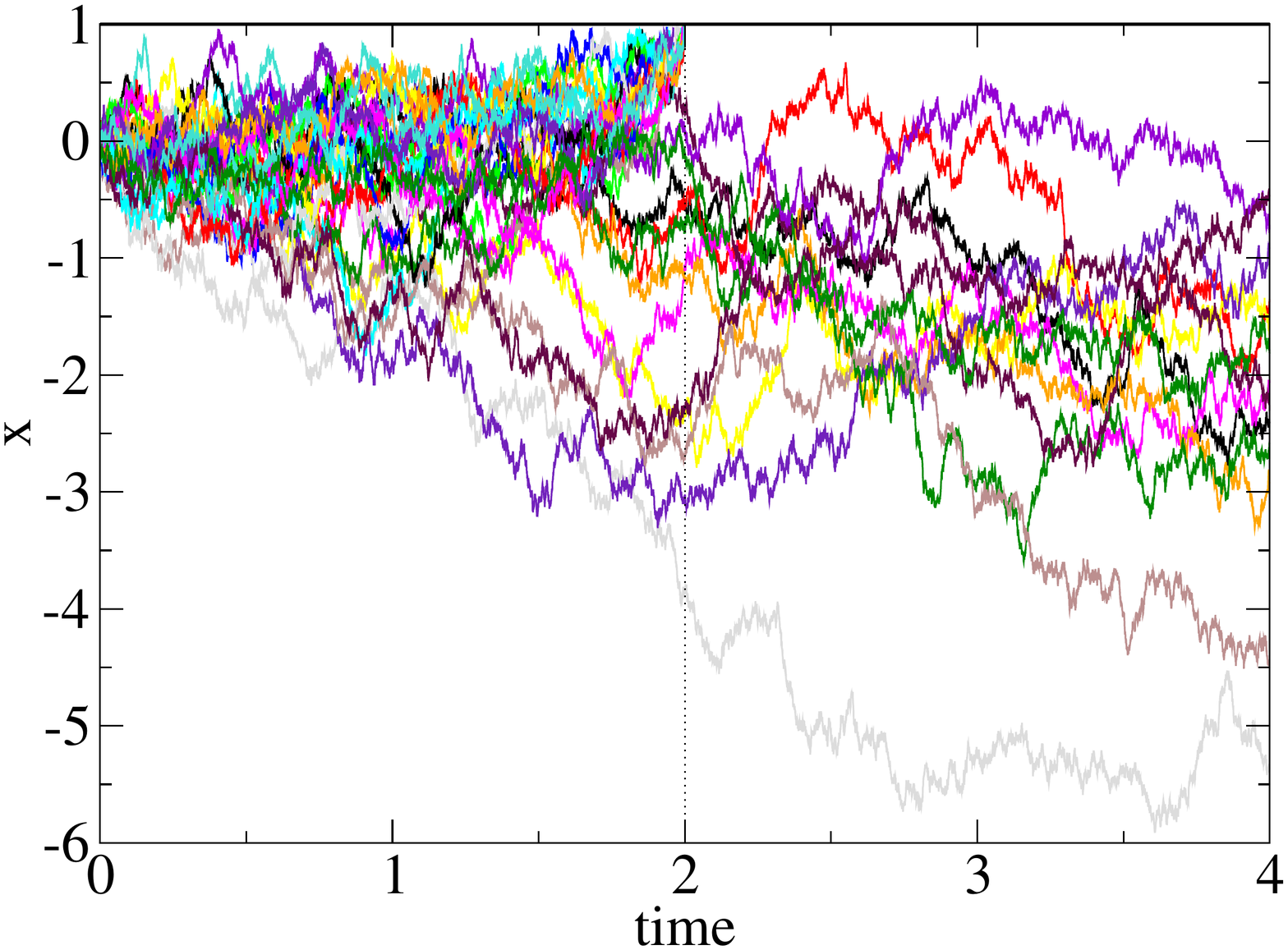}
\setlength{\abovecaptionskip}{15pt}  
\caption{A sample of 30 diffusions for the conditioned drift given by Eq. \ref{driftmuposfiniteinfinite} with parameters $a = 1$ , $T^* = 2$,  and $S^* = 0.5$. The time step used in the discretization is $dt = 10^{-4}$. All trajectories generated with different noise histories are statistically independent. The simulations of the surviving processes are stopped at time $t=4$.}
\label{fig5}
\end{figure}
%%%%%%%%%%%%%%%%%%%%%%%%%%%%%%%%%%%%%%%%%%%%%%%%%%%%%%%%

%%%%%%%%%%%%%%%%%%%%%%%%%%%%%%%%%%%%

\subsubsection{ When the initial drift is strictly negative $\mu < 0$ }

When the initial drift is strictly negative $\mu < 0$,
the function $ Q^{*[partial]}_{\infty}(x,t) $ of Eq. \ref{Qinfinitypartial} 
can be evaluated using the building blocks of Eq. \ref{ratioexpli}
and \ref{survivalTamuneg}
\begin{eqnarray}
Q^{*[partial]}_{\infty}(x,t) && = 
\int_t^{\infty} dT_a \gamma^*(T_a)
\frac{\gamma^{[\mu]}(T_a \vert x,t)}{\gamma^{[\mu]}(T_a\vert 0,0)}  
+  S^*(\infty ) \left[ \lim_{T \to +\infty} \frac{ S^{[\mu]} (T \vert x,t)  }{ S^{[\mu]} (T \vert 0,0)} \right]
\nonumber \\
&& = 
\int_t^{+\infty} dT_a  \gamma^*(T_a)  \left(\frac{T_a}{T_a-t}  \right)^{\frac{3}{2} } \left( \frac{a-x}{a} \right)   
e^{- \mu x + \frac{\mu^2}{2} t +\frac{a^2}{2T_a} - \frac{(a-x)^2}{2(T_a-t)}}
+ S^*(\infty )  \left[ \frac{1- e^{2 \mu (a-x) } }{1- e^{2 \mu a } } \right]
\ \ \ \ 
\label{Qmuneg}
\end{eqnarray}

The corresponding conditioned drift 
\begin{eqnarray}
&& \mu^*_{\infty}(x,t)  = \mu +  \partial_x \ln Q^{*[partial]}_{\infty}(x,t) 
 \nonumber  \\ 
&& = \mu +  
\frac{
\int_t^{+\infty} dT_a  \gamma^*(T_a)  \left(\frac{T_a}{T_a-t}  \right)^{\frac{3}{2} } \left( \frac{a-x}{a} \right)   
e^{- \mu x + \frac{\mu^2}{2} t +\frac{a^2}{2T_a} - \frac{(a-x)^2}{2(T_a-t)}}
\left[\frac{1}{x-a} - \mu +  \frac{a-x}{T_a-t}  \right]
+ S^*(\infty )  \left[ \frac{2 \mu e^{2 \mu (a-x) } }{1- e^{2 \mu a } } \right]
}
{
\int_t^{+\infty} dT_a  \gamma^*(T_a)  \left(\frac{T_a}{T_a-t}  \right)^{\frac{3}{2} } \left( \frac{a-x}{a} \right)   
e^{- \mu x + \frac{\mu^2}{2} t +\frac{a^2}{2T_a} - \frac{(a-x)^2}{2(T_a-t)}}
+ S^*(\infty )  \left[ \frac{1- e^{2 \mu (a-x) } }{1- e^{2 \mu a } } \right]
}
\nonumber
\\
&& =   
\frac{
\int_t^{+\infty} dT_a  \gamma^*(T_a)  \left(\frac{T_a}{T_a-t}  \right)^{\frac{3}{2} } \left( \frac{a-x}{a} \right)   
e^{- \mu x + \frac{\mu^2}{2} t +\frac{a^2}{2T_a} - \frac{(a-x)^2}{2(T_a-t)}}
\left[\frac{1}{x-a} +  \frac{a-x}{T_a-t}  \right]
+ S^*(\infty ) \mu \left[ \frac{ 1+ e^{2 \mu (a-x) } }{1- e^{2 \mu a } } \right]
}
{
\int_t^{+\infty} dT_a  \gamma^*(T_a)  \left(\frac{T_a}{T_a-t}  \right)^{\frac{3}{2} } \left( \frac{a-x}{a} \right)   
e^{- \mu x + \frac{\mu^2}{2} t +\frac{a^2}{2T_a} - \frac{(a-x)^2}{2(T_a-t)}}
+ S^*(\infty )   \left[ \frac{1- e^{2 \mu (a-x) } }{1- e^{2 \mu a } } \right]
}
\label{driftdoobmuneg}
\end{eqnarray}
depends on the value of the initial drift $\mu$ within the region $\mu < 0$
that we consider in this subsection.

As a simple example, let us consider again the case of Eq. \ref{gammapartial1delta} :
the conditioned drift of Eq. \ref{driftdoobmuneg} then reads  
\begin{eqnarray}
 \mu^*_{\infty}(x,t)   =   
\frac{
( 1-S^* ) \theta(T^*-t)   \left(\frac{T^*}{T^*-t}  \right)^{\frac{3}{2} } \left( \frac{a-x}{a} \right)   
e^{- \mu x + \frac{\mu^2}{2} t +\frac{a^2}{2T^*} - \frac{(a-x)^2}{2(T^*-t)}}
\left[\frac{1}{x-a} +  \frac{a-x}{T^*-t}  \right]
+ S^* \mu \left[ \frac{ 1+ e^{2 \mu (a-x) } }{1- e^{2 \mu a } } \right]
}
{
( 1-S^* ) \theta(T^*-t)   \left(\frac{T^*}{T^*-t}  \right)^{\frac{3}{2} } \left( \frac{a-x}{a} \right)   
e^{- \mu x + \frac{\mu^2}{2} t +\frac{a^2}{2T^*} - \frac{(a-x)^2}{2(T^*-t)}}
+  S^*  \left[ \frac{1- e^{2 \mu (a-x) } }{1- e^{2 \mu a } } \right]
}
\label{driftmunegfiniteinfinite}
\end{eqnarray}

%%%%%%%%%%%%%%%%%%%%%%%%%%%%%%%%%%%%%%%%%%%%%%%%%%%%%%%%
%%                      FIGURE                        %%
%%%%%%%%%%%%%%%%%%%%%%%%%%%%%%%%%%%%%%%%%%%%%%%%%%%%%%%%                          
\begin{figure}[h]
\centering
\includegraphics[width=4.in,height=3.in]{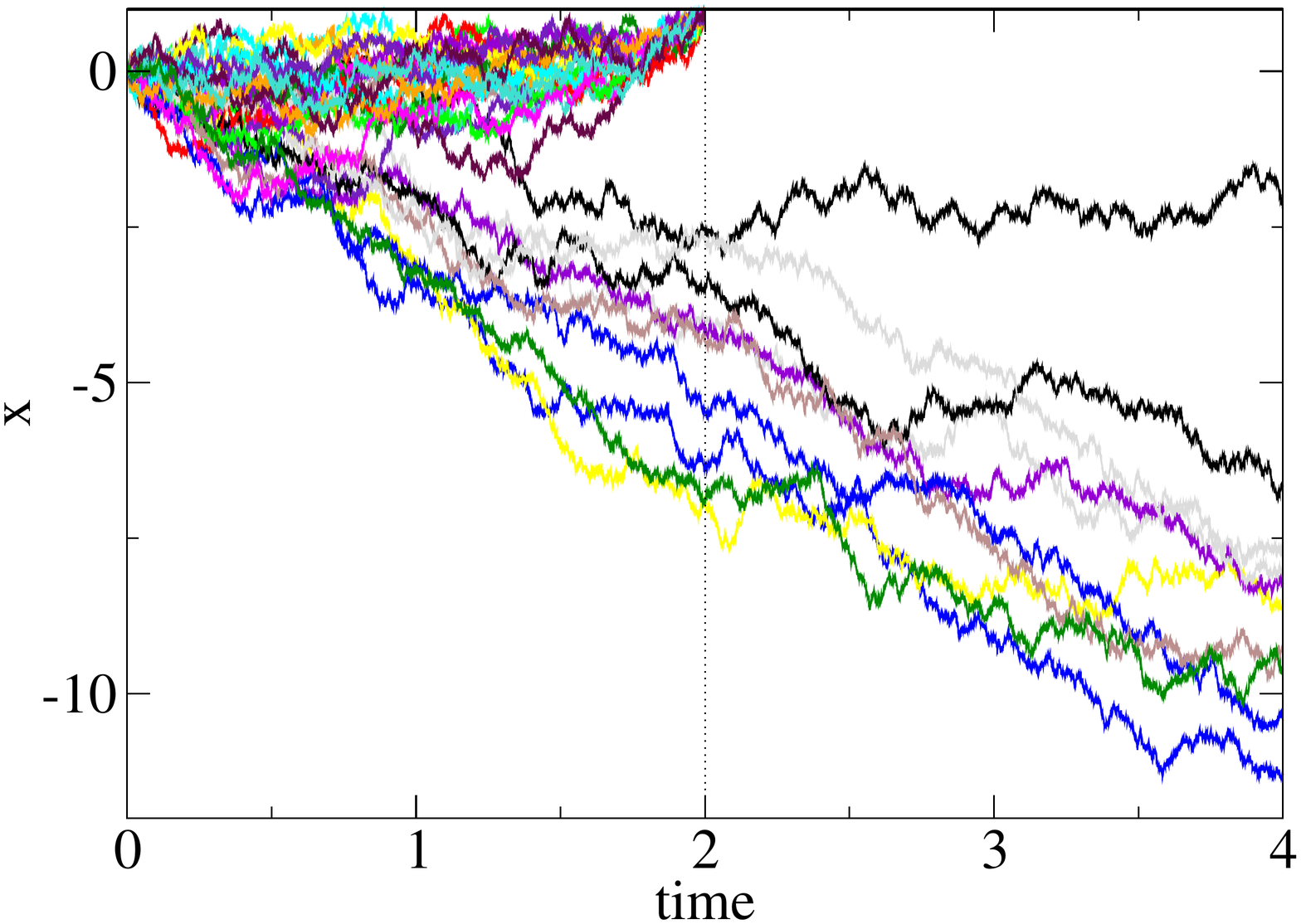}
\setlength{\abovecaptionskip}{15pt}  
\caption{A sample of 30 diffusions for the conditioned drift given by Eq. \ref{driftmunegfiniteinfinite} with parameters $a = 1$ ,  $T^* = 2$, $\mu = -2$ and $S^* = 1/3$. The time step used in the discretization is $dt = 10^{-4}$. All trajectories generated with different noise histories are statistically independent. The surviving processes asymptotically behave as Brownian motions with negative drift $\mu$. Simulations are stopped at time $t=4$.}
\label{fig6}
\end{figure}
%%%%%%%%%%%%%%%%%%%%%%%%%%%%%%%%%%%%%%%%%%%%%%%%%%%%%%%%

%%%%%%%%%%%%%%%%%%%%%%%%%%%%%%%%%%%%%%%%

\section{ Conclusions} 

\label{sec_conclusion}

In this paper, we have focused on the case where
 the unconditioned process is a diffusion living on the half-line $x \in ]-\infty,a[$ in the presence of an absorbing boundary condition at position $x=a$, to construct various conditioned processes corresponding to finite or infinite horizons. 
When the time horizon is finite $T<+\infty$, we have explained that the conditioning consists in imposing the probability distribution $P^*(y,T ) $ to be surviving at time $T$ and at the position $y \in ]-\infty,a[$, as well as the probability distribution $\gamma^*(T_a ) $ of the absorption time $T_a \in [0,T]$. When the time horizon is infinite $T=+\infty$, the conditioning consists in imposing the probability distribution $\gamma^*(T_a ) $ of the absorption time $T_a \in [0,+\infty[$, whose normalization $[1- S^*(\infty )]$ determines the conditioned probability $S^*(\infty ) \in [0,1]$ of forever-survival. This case of infinite horizon $T=+\infty$ can be thus reformulated as the conditioning of diffusion processes with respect to their first-passage-time properties at position $a$.  We have applied this general framework to the simplest case where the unconditioned process is the Brownian motion with uniform drift $\mu$ to generate stochastic trajectories satisfying various types of conditioning constraints. 
Finally, the Appendices describe the links with the Schr\"odinger perspective
that involve the dynamical large deviations of a large number
$N$ of independent unconditioned processes
and the stochastic control theory.

We thank an anonymous referee for suggesting the following two directions
to apply the present work in the future:

(i) in the field of population dynamical models subject to extinction/survival constraints,

(ii) in the field of absorbing-state phase transitions.

%%%%%%%%%%%%%%%%%%%%%%%%%%%%%%%%%%%%%%%%%%%

\appendix

\section{ Conditioning constraints that are less detailed than the distributions $\left[ P^*(.,T) ; \gamma^*(.)  \right] $ at $T$ }

\label{app_lessdetailed}

In Section \ref{sec_general} of main text, we have described the conditioning with respect
to the distributions $\left[ P^*(.,T) ; \gamma^*(.)  \right] $ at time $T$ and some of its consequences. In the present Appendix, we describe how to construct the appropriate conditioned processes when the conditioning constraints are less detailed than the distributions $\left[ P^*(.,T) ; \gamma^*(.)  \right] $ at $T$. 
In the present Appendix and in the next appendix, 
we adopt the Schr\"odinger perspective mentioned in the Introduction,
where one considers
 a large number $N$ of independent realizations  $X_n(t)$ of the unconditioned process
 labelled by $n=1,2,..,N$ starting all at the same initial condition $X_n(0)=x_0$, with the absorbing condition at position $x=a$ to analyze their large deviations properties with respect to $N$.

\subsection{ Empirical ensemble-averaged observables for $N$ independent unconditioned processes $X_n(t)$ }

The basic empirical observable is the ensemble-averaged density ${\hat P}(x,t) $ at position $x$ and at time $t$
 \begin{eqnarray}
 {\hat P}(x,t) \equiv  \frac{1}{N} \sum_{n=1}^N \delta( X_n(t) -x) 
\label{empiP}
\end{eqnarray}
In the bulk $x \in ]-\infty,a[$ where probability is conserved, this empirical density has to satisfy 
the continuity equation
 \begin{eqnarray}
 \partial_t {\hat P}(x,t) = - \partial_x {\hat J}(x,t)
\label{empidyn}
\end{eqnarray}
where the empirical current ${\hat J}(x,t) $ can be parametrized in terms of some empirical drift 
${\hat \mu}(x,t) $, while the diffusion coefficient $D(x)  $ is fixed
\begin{eqnarray}
{\hat J}(x,t) = {\hat \mu }(x,t)  {\hat P}(x,t) -  \partial_x  \left[ D(x) {\hat P}(x,t) \right]
\label{empiJ}
\end{eqnarray}
At the absorbing boundary condition $x=a$,
the empirical density $ {\hat P}(x,t) $ of Eq. \ref{empiP} vanishes at any time $t$
\begin{eqnarray}
 {\hat P}(x=a,t)   =0
\label{empiAbs}
\end{eqnarray}
The normalization of the empirical density $ {\hat P}(x,t) $ 
over the position $x \in ]-\infty,a[$ gives the empirical survival probability 
${\hat S}(t )$ at time $t$
\begin{eqnarray}
{\hat S}(t ) \equiv \int_{-\infty}^a dx {\hat P}(x,t)   
\label{empiS}
\end{eqnarray}
whose time-decay corresponds to the empirical distribution ${\hat \gamma}(t) $
of the absorption-time $t$ 
\begin{eqnarray}
{\hat \gamma}(t)  = - \frac{  d {\hat S}(t ) }{dt}  = - \int_{-\infty}^a dx \partial_t {\hat P}(x,t)   
\label{empigamma}
\end{eqnarray}
Using the continuity Eq. \ref{empidyn}, one obtains
that ${\hat \gamma}(t)  $ is directly related to the empirical current ${\hat J}(a,t)   $ entering the boundary $a$ 
\begin{eqnarray}
{\hat \gamma}(t) 
 =  \int_{-\infty}^a dx    \partial_x {\hat J}(x,t) =  \left[ {\hat J}(x,t)    \right]_{x=-\infty}^{x=a}
=  {\hat J}(a,t)  
\label{empigammaJ}
\end{eqnarray}
Using the parametrization of Eq. \ref{empiJ} and the absorbing boundary condition of Eq. \ref{empiAbs}, 
Eq. \ref{empigammaJ}
can be also rewritten in terms of the partial derivative of the density $\left( \partial_x {\hat P}(x,t) \right)\vert_{x=a} $ near the boundary as
\begin{eqnarray}
{\hat \gamma}(t) 
=  {\hat J}(a,t)   =  \left[  {\hat \mu }(x,t)  {\hat P}(x,t) -  \partial_x  \left( D(x) {\hat P}(x,t) \right)   \right]\vert_{x=a} = - D(a)  \left( \partial_x {\hat P}(x,t) \right)\vert_{x=a}
\label{empigammaderi}
\end{eqnarray}

In the thermodynamic limit $N \to +\infty$, all these empirical observables concentrate 
on their typical values given by the corresponding observables without hats described 
in subsection \ref{subsec_unconditioned} of the main text.
However for large finite $N$, dynamical fluctuations around these typical values are possible
and can be analyzed via the theory of large deviations.

%%%%%%%%%%%%%%%%%%%%%%%%%%%%%%%%%%%%%%%%%%%

\subsection{ Large deviations for the empirical observables associated to the finite time horizon $T$ }

\subsubsection{ Application of the Sanov theorem for $N$ independent processes observed at the finite time horizon $T$}

For each of the unconditioned process $X_n(t)$ starting at $x_0$ at time $t=0$,
its final state at the time horizon $T$ is characterized by :

(i) either its position $X_n(T)=y_n \in ]-\infty,a[$ if it is still surviving at $T$;

(ii) or its absorption time $T^{abs}_n \in ]0,T[$ if it did not survive up to time $T$.

The global normalization for the probabilities of these events read
\begin{eqnarray}
1 = \int_{-\infty}^a dy_n P(y_n,T \vert x_0, 0) + \int_0^T dT^{abs}_n \gamma(T_n^{abs} \vert x_0,0) 
\label{globalnormaT}
\end{eqnarray}

When one considers the $N$ independent processes $X_n(t) $ with $n=1,..,N$
at the finite time horizon $T$, the empirical histogram ${\hat P}(y,T) $ at time $T$ of the position $y \in ]-\infty,a[ $ 
\begin{eqnarray}
 {\hat P}(y,T) \equiv \frac{1}{N} \sum_{n=1}^N \delta( X_n(T) -y) = \frac{1}{N} \sum_{n=1}^N \delta( y_n -y)
\label{empiPTN}
\end{eqnarray}
and the empirical  histogram ${\hat \gamma}(T_a) $
of the absorption-time $T_a \in ]0,T[$ 
\begin{eqnarray}
{\hat \gamma}(T_a) \equiv \frac{1}{N} \sum_{n=1}^N \delta(T^{abs}_n -T_a )  
\label{empigammaN}
\end{eqnarray}
satisfy the global normalization analog to Eq. \ref{globalnormaT}
\begin{eqnarray}
1 = \int_{-\infty}^a dy {\hat P}(y,T) + \int_0^T dT_a {\hat \gamma}(T_a) 
\label{globalnormaTempi}
\end{eqnarray}

In the field of large deviations (see the reviews \cite{oono,ellis,review_touchette} and references therein),
 the Sanov theorem concerning the empirical histogram of independent identically distributed variables 
 is one of the most important result.
In our present case, its application gives the following conclusion :
The joint probability distribution to observe the empirical density  ${\hat P}(y,T) $ for $y \in ]-\infty,a[ $ 
and the empirical distribution ${\hat \gamma}(T_a) $ for $T_a \in ]0,T[$
satisfy the large deviation form for large $N$
\begin{eqnarray}
{\cal P}^{Sanov}_T \left[ {\hat P }(.,T) ; {\hat \gamma}(.)  \right]  \opsimeq_{N \to +\infty} 
\delta \left( \int_{-\infty}^a dy {\hat P}(y,T) + \int_0^T dT_a {\hat \gamma}(T_a) -1 \right) 
e^{- N {\cal I}^{Sanov}_T \left[ {\hat P }(.,T) ; {\hat \gamma}(.)  \right] }
\label{LevelSanov}
\end{eqnarray}
where the delta function imposes the normalization constraint of Eq. \ref{globalnormaTempi},
while the Sanov rate function 
\begin{eqnarray}
{\cal I}^{Sanov}_T \left[ {\hat P }(.,T) ; {\hat \gamma}(.)  \right]= 
  \int_{-\infty}^a dy   {\hat P}(y,T)  \ln \left( \frac{ {\hat P}(y,T)}{ P(y,T \vert x_0, 0)}  \right)
    + \int_0^T dT_a  {\hat \gamma}(T_a)   \ln \left(  \frac{{\hat \gamma}(T_a)  }{\gamma(T_a \vert x_0,0) }  \right)
\label{RateSanov}
\end{eqnarray} 
corresponds to the relative entropy
of the empirical distributions $\left[ {\hat P }(.,T) ; {\hat \gamma}(.)  \right] $ with respect to the true 
distributions $\left[ P(.,T \vert x_0, 0) ;  \gamma(. \vert x_0,0) \right] $.
The Sanov rate function of Eq. \ref{RateSanov}
vanishes only when the empirical distribution coincides the true 
distribution $\left[ {\hat P }(.,T) ; {\hat \gamma}(.)  \right] =\left[ P(.,T \vert x_0, 0) ;  \gamma(. \vert x_0,0) \right] $, and is strictly positive otherwise
\begin{eqnarray}
{\cal I}^{Sanov}_T \left[ {\hat P }(.,T) ; {\hat \gamma}(.)  \right] >0 
\ \ \ {\rm for } \ \ \left[ {\hat P }(.,T) ; {\hat \gamma}(.)  \right] \ne \left[ P(.,T \vert x_0, 0) ;  \gamma(. \vert x_0,0) \right]
\label{RateSanovpos}
\end{eqnarray} 

%%%%%%%%%%%%%%%%%%%%%%%%%%%%%%%%

\subsubsection{ Relative entropy cost of the conditioning constraints
$\left[ P^*(.,T) ; \gamma^*(.)  \right]$
imposed at the finite time horizon $T$}

The above framework involving $N$ independent unconditioned processes
provide the interesting alternative Schr\"odinger perspective on the 
conditioning constraints imposed at the finite horizon $T<+\infty$ in the main text:
one can interpret the imposed distributions $P^*(y,T ) $ for $y \in ]-\infty,a[$ and $\gamma^*(T_a ) $  for $T_a \in [0,T]$ of Eq. \ref{survivalTstar}
and Eq. \ref{deadTstar} as the empirical results $\left[ {\hat P }(.,T) ; {\hat \gamma}(.)  \right]$ 
obtained in an experiment concerning
$N$ independent unconditioned processes.
As a consequence, the Sanov rate function of Eq. \ref{RateSanov} evaluated for the imposed conditions
$\left[ P^*(.,T) ; \gamma^*(.)  \right]$ at the horizon $T$
\begin{eqnarray}
{\cal I}^{Sanov}_T \left[ P^*(.,T) ; \gamma^*(.)  \right]= 
  \int_{-\infty}^a dy   P^*(y,T)  \ln \left( \frac{ P^*(y,T)}{ P(y,T \vert x_0, 0)}  \right)
    + \int_0^T dT_a  \gamma^*(T_a)   \ln \left(  \frac{\gamma^*(T_a)  }{\gamma(T_a \vert x_0,0) }  \right)
\label{RateSanovstar}
\end{eqnarray} 
allows one to measure the relative entropy cost of the conditioning constraints $\left[ P^*(.,T) ; \gamma^*(.)  \right]$ with respect to the 'true' distributions $\left[ P(.,T \vert x_0, 0) ;  \gamma(. \vert x_0,0) \right]$. So this allows one to characterize how rare are the conditioning events one is interested in, and to compare the rarity of various conditioning constraints.

This Sanov rate function ${\cal I}^{Sanov}_T \left[ P^*(.,T) ; \gamma^*(.)  \right] $
  is also essential if one wishes
to give some precise meaning to conditioning constraints that are less detailed than the whole distributions $\left[ P^*(.,T) ; \gamma^*(.)  \right] $ at the time horizon $T$.
The idea is that one needs to optimize the Sanov rate function
in the presence of the conditioning constraints that one imposes.
Let us describe some simple examples in the following subsections.

%%%%%%%%%%%%%%%%%%%%%%%%%%%%%%%%%%%%%%%%%%%

\subsection{ Conditioning toward the surviving probability distribution 
$P^*(y,T) $ at time $T$ alone}

If one wishes to impose only the probability distribution $P^*(y,T) $ at time $T$, together
with its corresponding survival probability
\begin{eqnarray}
 S^*(T) \equiv \int_{-\infty}^a dy P^*(y,T) 
  \label{survivalspace}
\end{eqnarray}
one needs to optimize the Sanov rate function ${\cal I}^{Sanov}_T \left[ P^*(.,T) ; \gamma^*(.)  \right] $ of Eq. \ref{RateSanovstar}
over the absorption-time distribution $ \gamma^*(.) $ normalized to $\left[ 1-  S^*\right]$.
It is thus convenient to introduce the following Lagrangian involving the Lagrange multiplier $\alpha$ associated to the normalization constraint
\begin{eqnarray}
&& {\cal L}_T^{space} \left[  \gamma^*(.)  \right] 
 = {\cal I}^{Sanov}_T \left[ P^*(.,T) ; \gamma^*(.)  \right]
  + \alpha \left(  \int_0^T dT_a \gamma^*(T_a) - \left[ 1-  S^*(T) \right] \right) 
   \nonumber \\
 &&
 =  \int_{-\infty}^a dy   P^*(y,T)  \ln \left( \frac{ P^*(y,T)}{ P(y,T \vert x_0, 0)}  \right)
    + \int_0^T dT_a  \gamma^*(T_a)   \ln \left(  \frac{\gamma^*(T_a)  }{\gamma(T_a \vert x_0,0) }  \right)
  + \alpha \left(  \int_0^T dT_a \gamma^*(T_a) - \left[ 1-  S^*(T) \right] \right) 
  \ \ \ \ 
 \label{lagrangianspace}
\end{eqnarray}
The optimization of this Lagrangian 
over the distribution $\gamma^*(T_a) $ 
\begin{eqnarray}
0 = \frac{ \partial {\cal L}_T^{space} \left[  \gamma^*(.)  \right] }{ \partial  \gamma^*(T_a)} = 
  \ln \left(  \frac{\gamma^*(T_a)  }{\gamma(T_a \vert x_0,0) } \right)
+1  + \alpha
 \label{lagrangianspacederi}
\end{eqnarray}
leads to the optimal solution
\begin{eqnarray}
 \gamma^{*opt} (T_a)  =e^{-1-\alpha}  \gamma(T_a \vert x_0,0)  
 \label{gammaopt}
\end{eqnarray}
that should satisfy the normalization constraint
\begin{eqnarray}
1- S^*(T) = \int_0^T dT_a  \gamma^{*opt}(T_a) = e^{-1-\alpha} \int_0^T dT_a \gamma(T_a \vert x_0,0) = e^{-1-\alpha} \left[ 1- S(T \vert x_0,0)\right]
  \label{gammaoptnorm}
\end{eqnarray}
Plugging this value of the Lagrange multiplier $\alpha$ into Eq. \ref{gammaopt}
leads to the final 
optimal solution
\begin{eqnarray}
 \gamma^{*opt} (T_a)  = \left( \frac{1- S^*(T) }{1- S(T \vert x_0,0)} \right) \gamma(T_a \vert x_0,0)  
 \label{gammaoptfinal}
\end{eqnarray}
The contribution to the Lagrangian of Eq. \ref{lagrangianspace} of this optimal solution 
\begin{eqnarray}
\int_0^T dT_a  \gamma^{*opt}(T_a) 
  \ln \left(  \frac{\gamma^{*opt}(T_a)  }{\gamma(T_a \vert x_0,0) } \right)
&&  = \int_0^T dT_a \left( \frac{1- S^*(T) }{1- S(T \vert x_0,0)} \right) \gamma(T_a \vert x_0,0)
  \ln \left( \frac{1- S^*(T) }{1- S(T \vert x_0,0)} \right)
  \nonumber \\ &&
  =  \left( 1- S^*(T)  \right) 
  \ln \left( \frac{1- S^*(T) }{1- S(T \vert x_0,0)} \right)
 \label{lagrangianspaceopt}
\end{eqnarray}
leads to the relative entropy cost of the probability distribution $P^*(y,T) $ at time $T$
and of its corresponding survival probability $S^*(T) $ of Eq. \ref{survivalspace}
\begin{eqnarray}
 {\cal I}_T^{space} \left[ P^*(.,T) ; S^*(T)  \right] 
&& =  {\cal I}^{Sanov}_T \left[ P^*(.,T) ; \gamma^{*opt}(.)  \right]
 \nonumber \\
&&  =   \int_{-\infty}^a dy   P^*(y,T)  \ln \left( \frac{ P^*(y,T)}{ P(y,T \vert x_0, 0)}  \right)
  +  \left( 1- S^*(T)  \right) 
  \ln \left( \frac{1- S^*(T) }{1- S(T \vert x_0,0)} \right)
 \label{ratespacealone}
\end{eqnarray}

In conclusion, if one wishes to impose only the probability distribution $P^*(y,T) $ at time $T$,
one should use the optimal solution $\gamma^{*opt} (T_a) $ of Eq. \ref{gammaoptfinal}
in the formula given in the main text :
 the function of Eq. \ref{Qdef} becomes
\begin{eqnarray}
Q^{space}_T(x,t) && = 
\int_t^{T} dT_a  \frac{\gamma^{*opt}(T_a)}{\gamma(T_a\vert x_0,0)}  \gamma(T_a \vert x,t)
+   \int_{-\infty}^a dy  \frac{ P^*(y,T )  }{P(y,T \vert x_0,0) } P(y,T \vert x,t)
\nonumber \\
&&=(1- S^*(T))  \left( \frac{1- S (T \vert x,t) }{1- S(T \vert x_0,0)} \right) 
+   \int_{-\infty}^a dy P^*(y,T ) \frac{ P(y,T \vert x,t)  }{P(y,T \vert x_0,0) } 
 \label{Qdefspace}
\end{eqnarray}
and leads to the conditioned drift via Eq. \ref{driftdoobfp}
\begin{eqnarray}
\mu^{*space}_T(x,t) && =\mu(x) + 2 D(x) \partial_x \ln Q^{space}_T(x,t) 
\nonumber \\
&& = \mu(x) +  2 D(x) \partial_x \ln \left[
(1- S^*(T))  \left( \frac{1- S (T \vert x,t) }{1- S(T \vert x_0,0)} \right) 
+   \int_{-\infty}^a dy P^*(y,T ) \frac{ P(y,T \vert x,t)  }{P(y,T \vert x_0,0) }
\right]
\label{driftdoobfpspace}
\end{eqnarray}

%%%%%%%%%%%%%%%%%%%%%%%%%%%%%%%%%%%%%%%%%%%

\subsection{ Conditioning toward the absorption-time distribution $ \gamma^*(T_a) $ for $T_a \in ]0,T[$ alone}

If one imposes only the absorption-time distribution $\gamma^*(T_a) $ for $T_a \in ]0,T[$ alone, together
with its normalization 
\begin{eqnarray}
\int_0^T dT_a  \gamma^*(T_a) =1-  S^*(T) 
  \label{survivaltime}
\end{eqnarray}
then one needs to optimize the Sanov rate function ${\cal I}^{Sanov}_T \left[ P^*(.,T) ; \gamma^*(.)  \right] $
over the possible spatial distributions $P^*(y,T)  $ normalized to $S^*(T) $.
It is thus convenient to introduce the following Lagrangian involving the Lagrange multiplier $\beta$ associated to the normalization constraint
\begin{eqnarray}
&& {\cal L}_T^{time} \left[   P^*(.,T) \right]  
 = {\cal I}^{Sanov}_T \left[ P^*(.,T) ; \gamma^*(.)  \right]
   + \beta \left( \int_{-\infty}^a dy   P^*(y,T)  -  S^*(T) \right) 
   \nonumber \\
 &&
=\int_{-\infty}^a dy   P^*(y,T)  \ln \left( \frac{ P^*(y,T)}{ P(y,T \vert x_0, 0)}  \right)
    + \int_0^T dT_a  \gamma^*(T_a)   \ln \left(  \frac{\gamma^*(T_a)  }{\gamma(T_a \vert x_0,0) }  \right)
  + \beta \left( \int_{-\infty}^a dy   P^*(y,T)  -  S^*(T) \right) 
  \ \ 
 \label{lagrangiantime}
\end{eqnarray}
The optimization is thus very similar to the previous subsection
and leads to the optimal solution
\begin{eqnarray}
P^{*opt}(y,T)  = \left( \frac{ S^*(T) }{ S(T \vert x_0,0)} \right)  P(y,T \vert x_0, 0) 
 \label{pyoptfinal}
\end{eqnarray}
with the corresponding contribution to the Lagrangian of Eq. \ref{lagrangiantime}
\begin{eqnarray}  
\int_{-\infty}^a dy   P^{*opt}(y,T)  \ln \left( \frac{ P^{*opt}(y,T)}{ P(y,T \vert x_0, 0)} \right)
&&  = \int_{-\infty}^a dy 
\left( \frac{ S^*(T) }{ S(T \vert x_0,0)} \right)  P(y,T \vert x_0, 0) 
  \ln \left( \frac{ S^*(T) }{ S(T \vert x_0,0)} \right)
  \nonumber \\
&&  =  S^*(T) 
   \ln \left( \frac{ S^*(T) }{ S(T \vert x_0,0)} \right)
 \label{lagrangiantimeopt}
\end{eqnarray}

So the relative entropy cost of the absorption-time distribution $\gamma^*(T_a) $ for $T_a \in ]0,T[$
and of the corresponding survival probability $S^*(T) $ of Eq. \ref{survivaltime}
is given by
\begin{eqnarray}
 {\cal I}_T^{time} \left[  \gamma^*(.)  ; S^*(T) \right]
 && =  {\cal I}^{Sanov}_T \left[ P^{*opt}(.,T) ; \gamma^*(.)  \right]
 \nonumber \\
&&  =  \int_0^T dT_a  \gamma^*(T_a) 
  \ln \left(  \frac{\gamma^*(T_a)  }{\gamma(T_a \vert x_0,0) } \right) 
  +  S^*(T) 
   \ln \left( \frac{ S^*(T) }{ S(T \vert x_0,0)} \right)
   \label{ratetimealone}
\end{eqnarray}

In conclusion, if one wishes to impose only the absorption-time distribution 
$\gamma^*(T_a) $ for $T_a \in ]0,T[$,
one should use the optimal solution $P^{*opt}(y,T) $ of Eq. \ref{pyoptfinal}
as was done in Eqs \ref{renorma} and \ref{renormapartial} of the main text.
 The function of Eq. \ref{Qdef} becomes
\begin{eqnarray}
Q^{time}_T(x,t) 
&& = 
\int_t^{T} dT_a  \frac{\gamma^*(T_a)}{\gamma(T_a\vert x_0,0)}  \gamma(T_a \vert x,t)
+   \int_{-\infty}^a dy  \frac{ P^{*opt}(y,T )  }{P(y,T \vert x_0,0) } P(y,T \vert x,t)
\nonumber \\
&& = \int_t^{T} dT_a  \frac{\gamma^*(T_a)}{\gamma(T_a\vert x_0,0)}  \gamma(T_a \vert x,t)
+ S^*(T) \left( \frac{ S(T \vert x,t) }{ S(T \vert x_0,0)} \right)
 \label{Qdeftime}
\end{eqnarray}
and leads to the conditioned drift via Eq. \ref{driftdoobfp}
\begin{eqnarray}
\mu^{*time}_T(x,t) && =\mu(x) + 2 D(x) \partial_x \ln Q^{time}_T(x,t) 
\nonumber \\
&& = \mu(x) +  2 D(x) \partial_x \ln \left[
 \int_t^{T} dT_a  \frac{\gamma^*(T_a)}{\gamma(T_a\vert x_0,0)}  \gamma(T_a \vert x,t)
+ S^*(T) \left( \frac{ S(T \vert x,t) }{ S(T \vert x_0,0)} \right)
\right]
\label{driftdoobfptime}
\end{eqnarray}

%%%%%%%%%%%%%%%%%%%%%%%%%%%%%%

\subsection{ Conditioning toward zero survival $S^*(\infty)=0$ at $T=+\infty$ and the averaged absorption-time $T^*_{av} $ }

As a last example,
let us consider the infinite horizon $T=+\infty$ when the conditioned forever-survival probability vanishes $S^*(\infty)=0$,
i.e. when the conditioned absorption-time distribution $\gamma^*(T_a) $ is normalized over $T_a \in ]0,+\infty[$
\begin{eqnarray}
\int_0^{+\infty} dT_a  \gamma^*(T_a) =1 - S^*(\infty) =1
  \label{survivaltimeinfty}
\end{eqnarray}
The rate function of Eq. \ref{ratetimealone} for the infinite horizon $T \to +\infty$ then reduces to
\begin{eqnarray}
 {\cal I}_{\infty}^{time} \left[  \gamma^*(.)  ; S^*(\infty) =0\right]
   =  \int_0^{+\infty} dT_a  \gamma^*(T_a) 
  \ln \left(  \frac{\gamma^*(T_a)  }{\gamma(T_a \vert x_0,0) } \right) 
     \label{ratetimealoneinfty}
\end{eqnarray}

Let us assume that one imposes only the averaged absorption-time
\begin{eqnarray}
T^*_{av} \equiv   \int_0^{+\infty} dT_a  T_a  \gamma^*(T_a) 
   \label{timeav}
\end{eqnarray}
Then one needs to optimize the rate function $ {\cal I}_{\infty}^{time} \left[  \gamma^*(.)  ; S^*(\infty) =0\right] $ of Eq. \ref{ratetimealoneinfty}
over the absorption-time distribution $ \gamma^*(.) $ satisfying the two constraints
of Eqs \ref{survivaltimeinfty}
and \ref{timeav}.
Let us introduce the following Lagrangian
involving the two Lagrange multipliers $(\alpha,\omega) $ 
\begin{eqnarray}
{\cal L}_{\infty} \left[    \gamma^*(.)  \right] = 
\int_0^{+\infty} dT_a  \gamma^*(T_a) 
  \ln \left(  \frac{\gamma^*(T_a)  }{\gamma(T_a \vert x_0,0) } \right)
  + \alpha \left(  \int_0^{+\infty} dT_a \gamma^*(T_a) - 1 \right) 
    + \omega \left(  \int_0^{+\infty} dT_a T_a \gamma^*(T_a) - T^*_{av}  \right) 
 \label{lagrangianinfty}
\end{eqnarray}
The optimization of this Lagrangian 
over the distribution $\gamma^*(T_a) $ 
\begin{eqnarray}
0 = \frac{ \partial {\cal L}_{\infty} \left[   \gamma^*(.)  \right]  }{ \partial \gamma^*(T_a)} = 
  \ln \left(  \frac{\gamma^*(T_a)  }{\gamma(T_a \vert x_0,0) } \right)
+1  + \alpha +  \omega  T_a
 \label{lagrangianinftyderi}
\end{eqnarray}
leads to the optimal solution
\begin{eqnarray}
\gamma^{*opt} (T_a)  =e^{-1-\alpha - \omega  T_a}  \gamma(T_a \vert x_0,0)  
 \label{gammaoptinfty}
\end{eqnarray}
where the values of the two Lagrange multipliers $(\alpha,\omega)$ 
are determined by the two constraints
\begin{eqnarray}
1 && = \int_0^{+\infty} dT_a \gamma^{*opt}(T_a) 
= e^{-1-\alpha} \int_0^{+\infty} dT_a  e^{ - \omega  T_a}  \gamma(T_a \vert x_0,0) 
\nonumber \\
T^*_{av} && = \int_0^{+\infty} dT_a T_a \gamma^{*opt}(T_a) 
 = e^{-1-\alpha} \int_0^{+\infty} dT_a  T_a e^{ - \omega  T_a}  \gamma(T_a \vert x_0,0) 
  \label{2constraints}
\end{eqnarray}
The first constraint allows one to eliminate $\alpha$ via
\begin{eqnarray}
e^{1+\alpha} = \int_0^{+\infty} dT_a  e^{ - \omega  T_a}  \gamma(T_a \vert x_0,0) 
  \label{alphasol}
\end{eqnarray}
while $\omega $ has to be computed as the solution of the equation
\begin{eqnarray}
T^*_{av} = \frac{  \int_0^{+\infty} dT_a  T_a e^{ - \omega  T_a}  \gamma(T_a \vert x_0,0) }
{ \int_0^{+\infty} dT_a  e^{ - \omega  T_a}  \gamma(T_a \vert x_0,0) } 
= - \partial_{\omega} \ln \left[ \int_0^{+\infty} dT_a  e^{ - \omega  T_a}  \gamma(T_a \vert x_0,0) \right]
  \label{eqforomega}
\end{eqnarray}
So both equations involve the Laplace transform $ \left[ \int_0^{+\infty} dT_a  e^{ - \omega  T_a}  \gamma(T_a \vert x_0,0) \right] $ of the distribution $\gamma(T_a \vert x_0,0) $.

To be more concrete, let us now focus on the example where the unconditioned process
is the Brownian motion with uniform drift $\mu $ starting at $x_0=0$, with the 
absorption-time distribution $\gamma^{[\mu]}(T_a \vert 0,0)  $ of Eq. \ref{gammaBrown}
corresponding to the simple Laplace transform
  \begin{eqnarray}
\int_0^{+\infty} dT_a   e^{ - \omega  T_a} \gamma^{[\mu]}(T_a \vert 0,0)  =
 \frac{a }{\sqrt{2 \pi}  } e^{\mu a} \int_0^{+\infty} dT_a  T_a^{-\frac{3}{2} } 
 e^{ - \frac{a^2}{2T_a}-  \frac{\mu^2 + 2 \omega}{2}T_a} 
 = e^{ a \left( \mu - \sqrt{\mu^2 + 2 \omega } \right) }
\label{gammaBrownlaplace}
\end{eqnarray}
One can plug this Laplace transform 
into Eq. \ref{gammaoptnorm} to obtain the Lagrange multiplier
\begin{eqnarray}
\omega = \frac{a^2}{2 [T^*_{av}  ]^2 } - \frac{\mu^2}{2}
  \label{omegasol}
\end{eqnarray}
and then into 
Eq. \ref{alphasol} to obtain the Lagrange multiplier
\begin{eqnarray}
1+\alpha =  a \left( \mu - \sqrt{\mu^2 + 2 \omega } \right) = a \left( \mu  - \frac{a}{ T^*_{av} } \right)
  \label{alphasolbrown}
\end{eqnarray} 
The corresponding optimal solution of Eq. \ref{gammaoptinfty}
\begin{eqnarray}
\gamma^{*opt} (T_a) && =e^{-1-\alpha - \omega  T_a}  \gamma^{[\mu]}(T_a \vert x_0,0)  
= e^{-a  \mu + \frac{a^2}{ T^*_{av}  }   - \frac{a^2}{2 [T^*_{av}  ]^2 }  T_a + \frac{\mu^2 }{2}T_a}
 \frac{a }{\sqrt{2 \pi T_a^3 }  } e^{\mu a}    
 e^{ - \frac{a^2}{2T_a}-  \frac{\mu^2 }{2}T_a} 
\nonumber \\
&& =   \frac{a }{\sqrt{2 \pi T_a^3 }  }  
 e^{ \frac{a^2}{ T^*_{av} }   - \frac{a^2}{2 [T^*_{av}  ]^2 }  T_a- \frac{a^2}{2T_a}} 
 = \gamma^{[\lambda]}(T_a \vert 0,0)
 \label{gammaoptinftybrown}
\end{eqnarray}
is independent of the unconditioned drift $\mu$ and coincides with the distribution $\gamma^{[\lambda]}(T_a \vert 0,0) $ associated to the Brownian of drift
\begin{eqnarray}
\lambda \equiv \frac{a}{ T^*_{av} }
  \label{mueff}
\end{eqnarray} 
As a consequence, one recovers exactly the conditioning problem of Eq. \ref{gammalambda}
discussed in the main text, where the corresponding conditioned drift of Eq. \ref{driftdoobfiniteT1lambdafinal}
was simply $\lambda$. 
As a consequence here,
 the conclusion is that the conditioning based only on the averaged absorption-time
$T^*_{av} $ of Eq. \ref{timeav}
produces the constant conditioned drift
\begin{eqnarray}
\mu^*_{\infty}(x,t)   = \lambda = \frac{a}{ T^*_{av} }
\label{driftstarforavta}
\end{eqnarray}

%%%%%%%%%%%%%%%%%%%%%%%%%%%%%

%%%%%%%%%%%%%%%%%%%%%%%%%%%%%%%%%%%%%%%%%

\section{Links with the dynamical large deviations and the stochastic control theory}

\label{app_largedev}

As already stressed in the Introduction, the idea to analyze
the dynamics of
 a large number of independent identical diffusion processes
ending in an atypical distribution, has been introduced in 1931 by E. Schr\"odinger in his famous paper \cite{Schrodinger},
and is known nowadays as the 'Schr\"odinger bridge' problem
when both the initial distribution and the final distribution are given,
while in the present paper we focus 
on the much simpler case where the initial distribution is a delta function at $x_0$.
As explained in detail in the recent commentary \cite{CommentSchrodinger} accompanying its English translation
and in the two reviews \cite{ControlSchrodinger,MongeSchrodinger}), 
the analysis of this 'Schr\"odinger bridge' problem in terms of 'large deviations', of 'Doob conditioning',
of 'stochastic control' and of 'optimal transport' is actually already present in the Schr\"odinger paper \cite{Schrodinger},
even if this modern terminology did of course not yet exist in 1931!
In the present Appendix, we describe how the conditioned processes 
obtained via Doob's method in the main text can be alternatively constructed
via the Schr\"odinger perspective.
The idea is that the conditioned process $X^*(t)$
for the intermediate times $t \in ]0,T[$ 
can be interpreted as the most probable empirical dynamics 
that one can infer once the conditioning constraints are given.
 This interpretation is based on
the analysis of the relative entropy cost of the empirical dynamics 
during the whole time-window $t \in [0,T]$, as we now describe.

%%%%%%%%%%%%%%%%%%%%%%%%%%%%%%%%%%%%%%%%%%%

\subsection{ Large deviations at Level 2.5 for the empirical dynamics during the time-window 
$t \in [0,T]$}

In the field of dynamical large deviations for Markov processes 
(see the reviews \cite{oono,ellis,review_touchette} and references therein),
the initial standard classification into Levels 1,2,3
has turned out to be inappropriate :
Indeed, the Level 2 concerning the empirical density alone cannot be written explicitly in most cases,
while the Level 3 concerning the whole empirical process is actually far too general for many purposes.
As a consequence, a new Level has been introduced between the Level 2 and the Level 3
and has been called "Level 2.5", even if it is actually much closer in spirit to the Level 2,
since the "Level 2.5" describes the large deviations properties
of the joint distribution of the empirical density and of the empirical flows.
In contrast to the Level 2, the Level 2.5 can be written explicitly for general Markov processes,
including discrete-time Markov chains
 \cite{fortelle_thesis,fortelle_chain,review_touchette,c_largedevdisorder,c_reset,c_inference},
continuous-time Markov jump processes
\cite{fortelle_thesis,fortelle_jump,maes_canonical,maes_onandbeyond,wynants_thesis,chetrite_formal,BFG1,BFG2,chetrite_HDR,c_ring,c_interactions,c_open,c_detailed,barato_periodic,chetrite_periodic,c_reset,c_inference,c_runandtumble,c_jumpdiff,c_skew,c_metastable,c_east,c_exclusion}
and Diffusion processes 
\cite{wynants_thesis,maes_diffusion,chetrite_formal,engel,chetrite_HDR,c_reset,c_lyapunov,c_inference,c_metastable}.
In summary, the Level 2.5 plays an essential role because it is the smallest Level that is explicit
in full generality. Let us now describe the particular application to our present setting.

\subsubsection{ Large deviations at Level 2.5 for the empirical density 
$ {\hat P}(x,t) $ and the empirical current ${\hat J}(x,t) $ for $t \in [0,T]$}

In our present setting, the large deviations at Level 2.5 concerning the 
time-dependent empirical ensemble-averaged density and current
 associated to the $N$ independent unconditioned processes $X_n(t)$ 
 yields the following conclusion:
the joint probability distribution ${\cal P}^{[2.5]}_{[0,T]} \left[ {\hat P }(.,.) ; {\hat J}(.,.)  \right] $ to see the empirical density 
$ {\hat P}(x,t) $ and the empirical current ${\hat J}(x,t) $ 
on the half-line $x \in ]-\infty,a[$
during the time window $0 \leq t \leq T$
follows the large deviation form for large $N$
\begin{eqnarray}
{\cal P}^{[2.5]}_{[0,T]} \left[ {\hat P }(.,.) ; {\hat J}(.,.)  \right]  \opsimeq_{N \to +\infty} 
{\cal C}^{[2.5]}_{[0,T]} \left[ {\hat P }(.,.) ; {\hat J}(.,.)\right]
e^{- N {\cal I}^{[2.5]}_{[0,T]} \left[ {\hat P }(.,.) ; {\hat J}(.,.)  \right] }
\label{Level2.5}
\end{eqnarray}
with the following notations :

(i)  the rate function ${\cal I}^{[2.5]}_{[0,T]} \left[ {\hat P }(.,.) ; {\hat J}(.,.)  \right] $ at Level 2.5 
is given by the usual explicit form for diffusion processes 
in terms of the diffusion coefficient $D(x)$ and the unconditioned Ito drift $\mu(x) $
\begin{eqnarray}
{\cal I}^{[2.5]}_{[0,T]} \left[ {\hat P }(.,.) ; {\hat J}(.,.)  \right]
 = \int_0^T dt
\int_{-\infty}^a  \frac{d x}{ 4 D(x)  {\hat P}(x,t) } \left[  {\hat J}(x,t) - \mu (x) {\hat P}(x,t) +\partial_x \left( D(x) {\hat P}(x,t) \right) \right]^2
\label{rate2.5diff}
\end{eqnarray} 
This rate function is obviously positive and
vanishes only when the empirical density and current $\left[ {\hat P }(.,.) ; {\hat J}(.,.)  \right] $ 
coincide with their typical values $\left[ P(.,.) ; J(.,.)  \right] $  described in the subsection \ref{subsec_unconditioned}.

(ii) the constitutive constraints ${\cal C}^{[2.5]}_{[0,T]} \left[ {\hat P }(.,.) ; {\hat J}(.,.) \right] $ 
at Level 2.5 can be decomposed 
\begin{eqnarray}
{\cal C}^{[2.5]}_{[0,T]} \left[ {\hat P }(.,.) ; {\hat J}(.,.)  \right]
= \delta( {\hat P}(x,0) - \delta(x-x_0) ) \ 
 {\cal C}^{Bulk}_{[0,T]} \left[ {\hat P }(.,.) ; {\hat J}(.,.)  \right]
 {\cal C}^{Boundary}_{[0,T]} \left[ {\hat P }(.,.) ; {\hat J}(,.) \right]
  \label{C2.5}
\end{eqnarray}
into the initial condition $ {\hat P}(x,t=0) = \delta(x-x_0)$, 
the empirical dynamics of Eq. \ref{empidyn}
in the bulk $x \in ]-\infty,a[ $ during the time-window $t \in [0,T] $
\begin{eqnarray}
{\cal C}^{Bulk}_{[0,T]} \left[ {\hat P }(.,.) ; {\hat J}(.,.)  \right]
= \prod_{t \in [0,T]} \prod_{x \in ]-\infty,a[} \left( \partial_t {\hat P}(x,t) = - \partial_x {\hat J}(x,t)
\right)
  \label{C2.5diff}
\end{eqnarray}
and the boundary conditions of Eqs \ref{empiAbs} and \ref{empigammaderi} at the position $x=a$
during the time-window $t \in [0,T] $
\begin{eqnarray}
 {\cal C}^{Boundary}_{[0,T]} \left[ {\hat P }(.,.) ;  {\hat J}(.,.) \right]
 = \prod_{t \in [0,T]}  \delta \left[ {\hat P}(a,t) \right]  \delta \left[ 
 {\hat J}(a,t) + D(a)  \left( \partial_x {\hat P}(x,t) \right)\vert_{x=a}
  \right]
  \label{C2.5abs}
\end{eqnarray}

%%%%%%%%%%%%%%%%%%%%%%%%%

\subsubsection{ Large deviations for the empirical density 
$ {\hat P}(x,t) $, the empirical drift ${\hat \mu}(x,t) $, and the empirical distribution ${\hat \gamma}(t) $
}

The parametrization of Eq. \ref{empiJ} allows one to replace the empirical current ${\hat J}(x,t) $
in the bulk $x \in ]-\infty,a[$ by the empirical drift
\begin{eqnarray}
 {\hat \mu }(x,t) = \frac{ {\hat J}(x,t) + \partial_x  \left[ D(x) {\hat P}(x,t) \right] }{ {\hat P}(x,t) }
\label{empiJtomu}
\end{eqnarray}
while the empirical current ${\hat J}(a,t) $ at the boundary $x=a$ corresponds to the
empirical absorption-time distribution ${\hat \gamma}(t) = {\hat J}(a,t)$ as discussed in Eq. \ref{empigammaJ}.

As a consequence, the large deviations at Level 2.5 of Eq. \ref{Level2.5}
can be directly translated into the joint probability distribution
${\cal P}^{[2.5]}_{[0,T]} \left[ {\hat P }(.,.) ; {\hat \mu}(.,.) ; {\hat \gamma}(.) \right] $
to see the empirical density 
$ {\hat P}(x,t) $, the empirical drift ${\hat \mu}(x,t) $, and the empirical absorption-time distribution ${\hat \gamma}(t) $
\begin{eqnarray}
{\cal P}^{[2.5]}_{[0,T]} \left[ {\hat P }(.,.) ; {\hat \mu}(.,.) ; {\hat \gamma}(.) \right] \opsimeq_{N \to +\infty} 
{\cal C}^{[2.5]}_{[0,T]} \left[ {\hat P }(.,.)  ; {\hat \mu}(.,.) ;  {\hat \gamma}(.) \right]
e^{- N {\cal I}^{[2.5]}_{[0,T]} \left[ {\hat P }(.,.) ; {\hat \mu}(.,.)  \right] }
\label{Level2.5mu}
\end{eqnarray}
The rate function translated from Eq. \ref{rate2.5diff} 
reduces to the simpler Gaussian form
for the empirical drift $  {\hat \mu }(x,t)$ 
\begin{eqnarray}
{\cal I}^{[2.5]}_{[0,T]} \left[ {\hat P }(.,.) ; {\hat \mu}(.,.)  \right]
 = \int_0^T dt
\int_{-\infty}^a dx {\hat P}(x,t) \frac{ \left[  {\hat \mu }(x,t)  - \mu (x)  \right]^2}{ 4 D(x)   }
\label{rate2.5diffmu}
\end{eqnarray} 
The constitutive constraints translated from Eq. \ref{C2.5}
\begin{eqnarray}
{\cal C}^{[2.5]}_{[0,T]} \left[ {\hat P }(.,.) ; {\hat \mu}(.,.) ;  {\hat \gamma}(.)  \right]
= \delta( {\hat P}(x,0) - \delta(x-x_0) ) \ 
 {\cal C}^{Bulk}_{[0,T]} \left[ {\hat P }(.,.); {\hat \mu}(.,.)   \right]
 {\cal C}^{Boundary}_{[0,T]} \left[ {\hat P }(.,.) ; {\hat \gamma}(.) \right]
  \label{C2.5mu}
\end{eqnarray}
involve the contribution of the bulk $x \in ]-\infty,a[ $ during the time-window $t \in [0,T] $
translated from Eq. \ref{C2.5diff}
\begin{eqnarray}
{\cal C}^{Bulk}_{[0,T]} \left[ {\hat P }(.,.) ; {\hat \mu}(.,.)  \right]
= \prod_{t \in [0,T]} \prod_{x \in ]-\infty,a[} \left( \partial_t {\hat P}(x,t) 
+ \partial_x \left[ {\hat \mu }(x,t)  {\hat P}(x,t) \right] -  \partial_x^2  \left[ D(x) {\hat P}(x,t) \right]
\right)
  \label{C2.5diffmu}
\end{eqnarray}
and the contribution of the boundary  $x=a$
during the time-window $t \in [0,T] $ translated from Eq. \ref{C2.5abs}
\begin{eqnarray}
 {\cal C}^{Boundary}_{[0,T]} \left[ {\hat P }(.,.) ;  {\hat \gamma}(.,.) \right]
 = \prod_{t \in [0,T]}  \delta \left[ {\hat P}(a,t) \right]  \delta \left[ 
 {\hat \gamma}(t) + D(a)  \left( \partial_x {\hat P}(x,t) \right)\vert_{x=a}
  \right]
  \label{C2.5absmu}
\end{eqnarray}

%%%%%%%%%%%%%%%%%%%%%%%%%

\subsection{ Link with the stochastic control theory }

Let us now describe the link with the stochastic control theory 
(see the two reviews \cite{ControlSchrodinger,MongeSchrodinger} and references therein).
In this subsection, one assumes that the empirical density $ {\hat P}(x,T) $ at time $T$ is given
for $x \in ]-\infty,a[ $ 
and where the empirical distribution ${\hat \gamma}(t) $ is given for $0 \leq t \leq T$
\begin{eqnarray}
  {\hat P}(x,T) && = P^*(x,T) \ \ \ {\rm for } \ \ \ \ \ x \in ]-\infty,a[
  \nonumber \\
  {\hat \gamma}(t) && = \gamma^*(t)  \ \ \ \ \ \ \ {\rm for } \ \ \ \ \ \ t \in [0,T]
  \label{empiT}
\end{eqnarray}
The goal is then to optimize the rate function ${\cal I}^{[2.5]}_{[0,T]} \left[ {\hat P }(.,.) ; {\hat \mu}(.,.)  \right] $ at Level 2.5 of Eq. \ref{rate2.5diffmu}
over the empirical density ${\hat P}(x,t) $ and over the empirical drift ${\hat \mu }(x,t)  $
at all the intermediate times $t \in ]0,T[$, 
in the presence of the constitutive constraints of Eq. \ref{C2.5mu}
and the supplementary constraints of Eq. \ref{empiT}.

\subsubsection{ Optimization for a given density $ {\hat P}(x,T) =P^*(x,T) $ at time $T$ and a given distribution
 ${\hat \gamma}(t) = \gamma^*(t)$ for $0 \leq t \leq T$}

It is convenient to separate the constraints of Eqs \ref{C2.5mu} and \ref{empiT}
into :

(i) the time-boundary-conditions for the empirical density $ {\hat P }(.,.) $
at the initial time $t=0$ and at the final time $t=T$ for $x \in ]-\infty,a[$
\begin{eqnarray}
 {\hat P}(x,t=0) && = \delta(x-x_0) 
 \nonumber \\
   {\hat P}(x,t=T) &&  = P^*(x,T)  
 \label{timeboundaries}
\end{eqnarray} 

(ii) the space-boundary-conditions for the empirical density $ {\hat P }(.,.) $
and its spatial derivative at position $x=a$ for $t \in ]0,T[$
\begin{eqnarray}
  {\hat P}(x=a,t) && =0
  \nonumber \\
    D(a)  \left( \partial_x {\hat P}(x,t) \right)\vert_{x=a}  && = - {\hat \gamma}^*(t)
 \label{spaceboundaries}
\end{eqnarray} 

(iii) the bulk constraint ${\cal C}^{Bulk}_{[0,T]} \left[ {\hat P }(.,.) ; {\hat \mu}(.,.)  \right] $ of Eq. \ref{C2.5diffmu} concerning the
empirical dynamics for $x \in ]-\infty,a[$ and $t \in ]0,T[$
\begin{eqnarray}
  \partial_t {\hat P}(x,t) = - \partial_x \left[ {\hat \mu }(x,t)  {\hat P}(x,t) \right] +  \partial_x^2  \left[ D(x) {\hat P}(x,t) \right]
\label{bulkempi}
\end{eqnarray} 

As a consequence, in the space-time-bulk region $\left(x \in ]-\infty,a[ ; t \in ]0,T[\right) $,
one only needs to optimize the rate function ${\cal I}^{[2.5]}_{[0,T]} \left[ {\hat P }(.,.) ; {\hat \mu}(.,.)  \right] $ at Level 2.5 of Eq. \ref{rate2.5diffmu}
in the presence of the bulk constraint (iii). 
This optimization can be done via the introduction of the Lagrangian
\begin{eqnarray}
{\cal L}^{Bulk}\left[ {\hat P }(.,.) ; {\hat \mu}(.,.)  \right] 
 = {\cal I}^{[2.5]}_{[0,T]} \left[ {\hat P }(.,.) ; {\hat \mu}(.,.)  \right]
+ {\cal L}^{Empi}\left[ {\hat P }(.,.) ; {\hat \mu}(.,.)  \right] 
\label{lagrangiantot}
\end{eqnarray} 
where the contribution 
\begin{eqnarray}
{\cal L}^{Empi}\left[ {\hat P }(.,.) ; {\hat \mu}(.,.)  \right] 
\equiv  \int_0^T dt
\int_{-\infty}^a dx \ \psi(x,t) \left( \partial_t {\hat P}(x,t) 
+ \partial_x \left[ {\hat \mu }(x,t)  {\hat P}(x,t) \right] -  \partial_x^2  \left[ D(x) {\hat P}(x,t) \right]
\right)
\label{lagrangianpsi}
\end{eqnarray} 
involves the Lagrange multiplier $\psi(x,t)$ introduced to impose
the constraint of Eq. \ref{bulkempi} concerning the empirical dynamics.

%%%%%%%%%%%%%%%%%%%%%%%%%%%%%%%%%%%%%%

\subsubsection{ The adjoint-equation method to analyze the optimization problem }

As usual in stochastic control theory (see the reviews \cite{ControlSchrodinger,MongeSchrodinger}),
it is useful to make some transformation of the Lagrangian of Eq. \ref{lagrangiantot} before its optimization. 
In our present case, this amounts to rewrite the three terms of Eq. \ref{lagrangianpsi}
via integrations by parts,
 either over time $t \in ]0,T[$ using the time-boundary-conditions of Eq. \ref{timeboundaries}
\begin{eqnarray}
\int_0^T dt \ \psi(x,t) \partial_t {\hat P}(x,t) 
&& = \left[ \psi(x,t)  {\hat P}(x,t)\right]_{t=0}^{t=T} - \int_0^T dt {\hat P}(x,t) \partial_t  \psi(x,t)
\nonumber \\
&& =  \psi(x,T)  P^*(x,T) -  \psi(x,0) \delta(x-x_0)- \int_0^T dt {\hat P}(x,t) \partial_t  \psi(x,t)
\label{integtime}
\end{eqnarray} 
or over space $x \in ]-\infty,a[$ using the space-boundary-conditions of Eq. \ref{spaceboundaries},
both for the contribution involving the empirical drift ${\hat \mu }(x,t) $
\begin{eqnarray}
 \int_{-\infty}^a dx \ \psi(x,t) 
 \partial_x \left( {\hat \mu }(x,t)  {\hat P}(x,t)  \right)
&&  = \left[ \psi(x,t)  {\hat \mu }(x,t)  {\hat P}(x,t)   \right]_{x=-\infty}^{x=a}
 - \int_{-\infty}^a dx   {\hat \mu }(x,t)  {\hat P}(x,t)  \partial_x\psi(x,t) 
 \nonumber \\
 && =  - \int_{-\infty}^a dx    {\hat \mu }(x,t)  {\hat P}(x,t)   \partial_x\psi(x,t) 
\label{integspace}
\end{eqnarray} 
and for the contribution involving the diffusion coefficient $D(x)$
\begin{eqnarray}
 && - \int_{-\infty}^a dx \ \psi(x,t)   \partial^2_x  \left[ D(x) {\hat P}(x,t) \right] 
 = -  \left[ \psi(x,t)   \partial_x  \left( D(x) {\hat P}(x,t) \right)  \right]_{x=-\infty}^{x=a}
 + \int_{-\infty}^a dx 
  \left(  \partial_x  \left[ D(x) {\hat P}(x,t) \right] \right) \partial_x\psi(x,t) 
  \nonumber \\
  && = \psi(a,t)  {\hat \gamma}^*(t) 
  +  \left[   D(x) {\hat P}(x,t)   \partial_x\psi(x,t)\right]_{x=-\infty}^{x=a}
  - \int_{-\infty}^a dx 
D(x) {\hat P}(x,t)  \partial^2_x\psi(x,t) 
  \nonumber \\
  && = \psi(a,t)  {\hat \gamma}^*(t)   - \int_{-\infty}^a dx D(x) {\hat P}(x,t)  \partial^2_x\psi(x,t) 
\label{integspace2}
\end{eqnarray} 

Putting everything together, the contribution of Eq. \ref{lagrangianpsi} reads
\begin{eqnarray}
 {\cal L}^{Empi}\left[ {\hat P }(.,.) ; {\hat \mu}(.,.)  \right] =
&& \int_{-\infty}^a dx \left[  \psi(x,T)  P^*(x,T) -  \psi(x,0) \delta(x-x_0)- \int_0^T dt {\hat P}(x,t) \partial_t  \psi(x,t)\right]
\nonumber \\
&& + \int_0^T dt \left[  - \int_{-\infty}^a dx    {\hat \mu }(x,t)  {\hat P}(x,t)   \partial_x\psi(x,t) \right]
\nonumber \\
&& + \int_0^T dt \left[ \psi(a,t)  {\hat \gamma}^*(t)   - \int_{-\infty}^a dx D(x) {\hat P}(x,t)  \partial^2_x\psi(x,t) 
 \right]
 \nonumber \\
 && =- \int_0^T dt  \int_{-\infty}^a dx {\hat P}(x,t) \left[ \partial_t  \psi(x,t) 
 + {\hat \mu }(x,t)     \partial_x\psi(x,t) + D(x)  \partial^2_x\psi(x,t) 
 \right]
 \nonumber \\
 && +  \int_{-\infty}^a dx  \psi(x,T)  P^*(x,T)  -  \psi(x_0,0)   + \int_0^T dt  \psi(a,t)  {\hat \gamma}^*(t)
\label{lagrangianpsiinteg}
\end{eqnarray} 
so that the bulk lagrangian of Eq. \ref{lagrangiantot} becomes 
using the explicit rate function at Level 2.5 of Eq. \ref{rate2.5diffmu} 
\begin{eqnarray}
&& {\cal L}^{Bulk}\left[ {\hat P }(.,.) ; {\hat \mu}(.,.)  \right] 
 = \int_{-\infty}^a dx  \psi(x,T)  P^*(x,T)  -  \psi(x_0,0)   + \int_0^T dt  \psi(a,t)  {\hat \gamma}^*(t)
\nonumber \\
 &&
+ \int_0^T dt \int_{-\infty}^a dx {\hat P}(x,t) 
 \left( \frac{ \left[  {\hat \mu }(x,t)  - \mu (x)  \right]^2}{ 4 D(x)   }
  -  \left[ \partial_t  \psi(x,t) 
 + {\hat \mu }(x,t)     \partial_x\psi(x,t) + D(x)  \partial^2_x\psi(x,t) 
 \right] \right)
\label{lagrangianbulk}
\end{eqnarray} 
The optimization of Eq. \ref{lagrangianbulk}
over the empirical drift ${\hat \mu }(x,t) $
\begin{eqnarray}
0 && = \frac{ {\cal L}^{Bulk}\left[ {\hat P }(.,.) ; {\hat \mu}(.,.)  \right] }{ \partial {\hat \mu }(x,t) } = 
 {\hat P}(x,t)  \left( \frac{   {\hat \mu }(x,t)  - \mu (x)  }{ 2 D(x)   }
 -   \partial_x\psi(x,t) \right)
 \label{lagrangianbulkderimu}
\end{eqnarray} 
allows one to 
evaluate the optimal empirical drift ${\hat \mu }^{opt}(x,t) $ in terms of the Lagrange multiplier $\psi(x,t)  $
\begin{eqnarray}
  {\hat \mu }^{opt}(x,t)  = \mu (x) + 2 D(x)   \partial_x\psi(x,t)
 \label{mupsi}
\end{eqnarray} 
The further optimization of Eq. \ref{lagrangianbulk}
over the empirical density ${\hat P}(x,t) $ reads using the optimal drift of Eq. \ref{mupsi}
\begin{eqnarray}
0 && = - \frac{ {\cal L}^{Bulk}\left[ {\hat P }(.,.) ; {\hat \mu}(.,.)  \right] }{ \partial {\hat P }(x,t) } = 
 - \frac{ \left[  {\hat \mu }^{opt}(x,t)  - \mu (x)  \right]^2}{ 4 D(x)   }
 +  \partial_t  \psi(x,t) 
 + {\hat \mu }^{opt}(x,t)     \partial_x\psi(x,t) + D(x)  \partial^2_x\psi(x,t) 
 \nonumber \\
&& =  - D(x)  \left[  \partial_x\psi(x,t)  \right]^2
 +  \partial_t  \psi(x,t) 
 + \left(\mu (x) + 2 D(x)   \partial_x\psi(x,t) \right)    \partial_x\psi(x,t) + D(x)  \partial^2_x\psi(x,t) 
     \nonumber \\
&& =   \partial_t  \psi(x,t)  + \mu (x)       \partial_x\psi(x,t) + D(x)  \partial^2_x\psi(x,t) 
  + D(x) \left[    \partial_x\psi(x,t) \right]^2
\label{lagrangianbulkderiP}
\end{eqnarray} 
This Hamilton-Jacobi-Bellman equation for $\psi(x,t)$ can be transformed via
the change of variables
\begin{eqnarray}
 \psi(x,t)  = \ln q(x,t)
 \label{psiQ}
\end{eqnarray} 
into the linear backward Fokker-Planck equation 
involving the unconditioned generator ${\cal F}_x $ of Eq. \ref{generator}
\begin{eqnarray}
- \partial_t  q(x,t)    =   \mu (x)   \partial_x  q(x,t) + D(x)   \partial^2_x  q(x,t) ={\cal F}_x q(x,t)
\label{backwardsmallq}
\end{eqnarray} 
for the function $q(x,t)$.
Using Eq. \ref{psiQ},
the optimal empirical drift ${\hat \mu }^{opt}(x,t) $ of Eq. \ref{mupsi} becomes
\begin{eqnarray}
  {\hat \mu }^{opt}(x,t)  = \mu (x) + 2 D(x)   \partial_x  \ln q(x,t)
   \label{mupsiQ}
\end{eqnarray} 
while the optimal empirical density ${\hat P }^{opt}(x,t) $ should be the solution of the corresponding empirical forward dynamics of Eq. \ref{bulkempi}
\begin{eqnarray}
 \partial_t {\hat P}^{opt}(x,t) && =
  -  \partial_x \left[ {\hat \mu }^{opt}(x,t)  {\hat P}^{opt}(x,t) \right] +  \partial_x^2  \left[ D(x) {\hat P}^{opt}(x,t) \right]
  \nonumber \\
 && =
  -  \partial_x \left[ \left( \mu (x) + 2 D(x)   \partial_x  \ln q(x,t)\right) {\hat P}^{opt}(x,t) \right] 
  +  \partial_x^2  \left[ D(x) {\hat P}^{opt}(x,t) \right] 
\label{forwardhat}
\end{eqnarray} 

Using the backward unconditioned dynamics of Eq. \ref{backwardsmallq} for the function $q(x,t)$
and the forward optimal dynamics of Eq. \ref{forwardhat} for ${\hat P}^{opt}(x,t) $,
one obtains that the ratio
\begin{eqnarray}
  p(x,t)  \equiv \frac{{\hat P}^{opt}(x,t)}{ q(x,t)}
   \label{defsmallp}
\end{eqnarray} 
satisfies the forward unconditioned dynamics involving the adjoint operator $ {\cal F}^{\dagger}_x$ of Eq. \ref{adjoint}
\begin{eqnarray}
  \partial_t p(x,t)  && = \frac{ 1}{ q(x,t)} \partial_t {\hat P}^{opt}(x,t) - \frac{{\hat P}^{opt}(x,t)}{ q^2(x,t)} \partial_t q(x,t)
  \nonumber \\
  && =   \frac{ 1}{ q(x,t)} \left[  -  \partial_x \left( \left[ \mu (x) + 2 D(x)   \partial_x  \ln q(x,t)\right] {\hat P}^{opt}(x,t) \right) 
  +  \partial_x^2  \left( D(x) {\hat P}^{opt}(x,t) \right) \right] 
\nonumber \\
&&    + \frac{{\hat P}^{opt}(x,t)}{ q^2(x,t)} \left[  \mu (x)   \partial_x  q(x,t) + D(x)   \partial^2_x  q(x,t)\right]
  \nonumber \\
  && = -  \partial_x \left[ \mu(x) p(x,t)  \right] + \partial^2_{x} \left[ D(x) p(x,t)  \right] = {\cal F}^{\dagger}_x p(x,t)
   \label{forwardsmallp}
\end{eqnarray} 

%%%%%%%%%%%%%%%%%%%%%%%%%%%%%%%%%%%%%%

\subsubsection{ Taking into account the space-time boundary conditions to obtain the final optimal solution }

In summary, the optimal solution ${\hat P}^{opt}(x,t) $ is given the product of Eq. \ref{defsmallp}
\begin{eqnarray}
 {\hat P}^{opt}(x,t) =  q(x,t) p(x,t)
   \label{optimalprod}
\end{eqnarray} 
where $q(x,t)$ satisfies the backward unconditioned dynamics of Eq. \ref{backwardsmallq},
while $p(x,t)$ satisfies the forward unconditioned dynamics of Eq. \ref{forwardsmallp}.
In addition, we have to take into account
the time-boundary-conditions of Eq. \ref{timeboundaries}
 for $x \in ]-\infty,a[$ at the initial time $t=0$ and at the final time $t=T$
\begin{eqnarray}
 \delta(x-x_0)  && = {\hat P}^{opt}(x,t=0) =  q(x,0) p(x,0)
 \nonumber \\
 P^*(x,T) &&  = {\hat P}^{opt}(x,t=T) =  q(x,T) p(x,T)
 \label{timeboundariessmall}
\end{eqnarray} 
as well as the space-boundary-conditions of Eq. \ref{spaceboundaries}
 for $t \in ]0,T[$ at the position $x=a$
\begin{eqnarray}
  0 && =  {\hat P}(x=a,t) =  q(a,t) p(a,t)
  \nonumber \\
  - {\hat \gamma}^*(t) && =   D(a)  \left( \partial_x {\hat P}^{opt}(x,t) \right)\vert_{x=a}  = 
  D(a)  q(a,t) \left( \partial_x p(x,t) \right)\vert_{x=a} 
  + D(a) p(a,t)  \left( \partial_x q(x,t) \right)\vert_{x=a} 
 \label{spaceboundariessmall}
\end{eqnarray} 

For the function $p(x,t)$, it is natural to choose the unconditioned propagator $P(x, t \vert x_0,0)$
that would be the solution if one were not imposing atypical constraints
\begin{eqnarray}
p(x,t) = P(x,t \vert x_0,0)
 \label{smallpBigP}
\end{eqnarray} 
Plugging this choice into Eq. \ref{timeboundariessmall}, one obtains that the function $q(x,t)$ should satisfy 
time-boundary-conditions of Eq. \ref{timeboundariessmall}
 for $x \in ]-\infty,a[$ at the initial time $t=0$ and at the final time $t=T$
\begin{eqnarray}
 q(x,t=0) && =1
 \nonumber \\
 q(x,t=T) && = \frac{P^*(x,T) }{ P(x,T \vert x_0,0)} 
 \label{timeboundariessmallq}
\end{eqnarray} 
as well as the space-boundary-condition of Eq. \ref{spaceboundaries}
 for $t \in ]0,T[$ at the position $x=a$, using Eq. \ref{gammafirst}
\begin{eqnarray}
 q(a,t) = \frac{  {\hat \gamma}^*(t) }
 { \left[ - D(a)   \left( \partial_x P(x,t \vert x_0,0) \right)\vert_{x=a} \right]} 
 = \frac{  {\hat \gamma}^*(t) } { \gamma (t \vert x_0,0) }
 \label{spaceboundariessmallq}
\end{eqnarray} 
The solution $q(x,t)$
of the backward unconditioned dynamics of Eq. \ref{backwardsmallq}
that satisfies the boundary conditions of Eqs \ref{timeboundariessmallq}
and \ref{spaceboundariessmallq}
reads
\begin{eqnarray}
q(x,t) &&  = \int_t^{T} dT_a  q(a,T_a) \gamma(T_a\vert x,t)+   \int_{-\infty}^a dy  q(y,T) P(y,T \vert x,t) 
\nonumber \\
&& =  \int_t^{T} dT_a   \frac{  {\hat \gamma}^*(T_a) } { \gamma (T_a \vert x_0,0) } \gamma(T_a\vert x,t)
+   \int_{-\infty}^a dy   \frac{P^*(y,T) }{ P(y,T \vert x_0,0)}  P(y,T \vert x,t) = Q_T(x,t)
 \label{smallqsolution}
\end{eqnarray}
and thus coincides with the function $Q_T(x,t)$ introduced in Eq. \ref{Qdef} of the main text.

%%%%%%%%%%%%%%%%%%%%%%%%%%%%%%%%%%%%%%

\subsubsection{ Corresponding optimal value of the Lagrangian }

The corresponding optimal value of the Lagrangian of Eq. \ref{lagrangianbulk} 
reduces to the boundary terms, since the bulk contribution vanishes as a consequence of the optimization
Eq. \ref{lagrangianbulkderiP}
\begin{eqnarray}
 {\cal L}^{Bulk}\left[ {\hat P }^{opt}(.,.) ; {\hat \mu}^{opt}(.,.)  \right] 
 = \int_{-\infty}^a dx  \psi(x,T)  P^*(x,T)  -  \psi(x_0,0)   + \int_0^T dt  \psi(a,t)  {\hat \gamma}^*(t)
\label{lagrangianbulkopt}
\end{eqnarray} 
Using Eq. \ref{psiQ} and the solution of Eq. \ref{smallqsolution}, the Lagrange multiplier $\psi(x,t)$ 
\begin{eqnarray}
\psi(x,t) = \ln  q(x,t)= \ln Q_T(x,t) = \ln \left( \frac{P^*(x,t) }{P(x,t \vert x_0,0)} \right)
\label{smallqbigQ}
\end{eqnarray} 
and its particular values
\begin{eqnarray}
\psi(x_0,0) && = \ln q(x_0,0) = \ln \left( \frac{P^*(x_0,0) }{P(x_0,0 \vert x_0,0)} \right) = \ln ( 1 )=0
   \nonumber \\
\psi(a,t) && = \ln q(a,t)      = \ln \left( \frac{\gamma^*(t) }{\gamma(t\vert x_0,0)} \right)
\label{lagrangianoptimalpsi}
\end{eqnarray} 
can be plugged into Eq. \ref{lagrangianbulkopt} to obtain that the optimal value of the Lagrangian
\begin{eqnarray}
 {\cal L}^{Bulk}\left[ P^*(.,.) ; \mu^*(.,.)  \right] 
 = \int_{-\infty}^a dx   P^*(x,T) \ln \left(  \frac{P^*(x,T) }{P(x,T \vert x_0,0)} \right)   
    + \int_0^T dt   {\hat \gamma}^*(t)  \ln \left(  \frac{\gamma^*(t) }{\gamma(t\vert x_0,0)} \right)
\label{lagrangianbulkoptqfin}
\end{eqnarray} 
coincides with the Sanov rate function ${\cal I}^{Sanov}_T \left[ P^*(.,T) ; \gamma^*(.)  \right] $
 of Eq. \ref{RateSanovstar} as it should for consistency.
The physical interpretation is thus that the conditioned dynamics described in the main text
is the optimal dynamics satisfying the imposed constraints from the point of view of the dynamical relative entropy cost 
as measured by the rate function at Level 2.5.

%%%%%%%%%%%%%%%%%%%%%%%%%%%%%%%%

%%%%%%%%%%%%%%%%%%%%%%%%%%%%%%%%%%%%%%%%%%%%

\end{document}